\documentclass[preprint,aps,prd,floatfix,nofootinbib,11pt]{revtex4-2}
\pdfoutput=1
\usepackage{graphicx}
\usepackage{amsfonts}
\usepackage{amssymb}
\usepackage{xcolor}
\usepackage{natbib}
\usepackage{braket}

\usepackage{hyperref}

\usepackage{amsmath}
\usepackage{rotating}
\usepackage{amssymb,amsthm}
\usepackage{soul}

\usepackage{xcolor}

\usepackage{latexsym}
\usepackage{graphics}

\usepackage{amsfonts}
\usepackage{amsmath}
\usepackage{rotating}
\usepackage{amssymb}

\usepackage{amsmath}
\usepackage{rotating}
\usepackage{amssymb}
\usepackage{soul}

\usepackage{xcolor}

\usepackage{xcolor}
\def\beq{\begin{eqnarray}}  
	\def\eeq{\end{eqnarray}}

\begin{document}
	
	\title{Scalar Field Kantowski--Sachs Solutions in Teleparallel $F(T)$ Gravity}

	\author{A. Landry}
	\email{a.landry@dal.ca}
	\affiliation{Department of Mathematics and Statistics, Dalhousie University, Halifax, Nova Scotia, Canada, B3H 3J5}


\begin{abstract}
In this paper, we investigate time-dependent Kantowski--Sachs spherically symmetric teleparallel $F(T)$ gravity with a scalar field source. We begin by setting the exact field equations to be solved and solve conservation laws for possible scalar field potential, $V\left(\phi\right)$, solutions. Then, we find new non-trivial teleparallel $F(T)$ solutions by using power-law and exponential ansatz for each potential case arising from conservation laws, such as linear, quadratic, or logarithmic, to name a few. We find a general formula allowing us to compute all possible new teleparallel $F(T)$ solutions applicable for any scalar field potential and ansatz. Then, we apply this formula and find a large number of exact and approximate new teleparallel $F(T)$ solutions for several types of cases. Some new $F(T)$ solution classes may be relevant for future cosmological applications, especially concerning dark matter, dark energy quintessence, phantom energy leading to the Big Rip event, and quintom models of physical processes.
\end{abstract}

\maketitle

\newpage

\tableofcontents


\section{Introduction}\label{sect1}


{ 
	Teleparallel $F(T)$ gravity is an alternative theory to General Relativity (GR) of the frame-based type~\cite{Lucas_Obukhov_Pereira2009,Aldrovandi_Pereira2013,Bahamonde:2021gfp,Krssak:2018ywd,MCH,Coley:2019zld,Krssak_Pereira2015}. This theory is very promising and is in great expansion. It is defined in terms of spacetime torsion tensor $T^a_{~bc}$ and torsion scalar $T$, and its physical quantities are all defined in terms of {the coframe} 
	${\bf h}^a$ and the spin-connection $\omega^a_{~bc}$ (and their derivatives). This is in contrast to GR being defined in terms of the metric $g_{\mu\nu}$ and the spacetime curvature $R^a_{~b\mu\nu}$. One of the features of teleparallel gravity and its frame-based approach is the new possible spacetime symmetries, especially for non-trivial linear isotropy groups, and its Lorentz-invariant geometries~\cite{chinea1988symmetries, estabrook1996moving, papadopoulos2012locally}. There is thus an approach to determine these symmetries for any independent coframe/spin-connection pairs while treating spacetime curvature and torsion as geometric objects~\cite{MCH,Coley:2019zld,olver1995equivalence}. From~this point, any geometry described by an even coframe/spin-connection whose curvature and non-metricity are both zero ($R^a_{~b\mu\nu}=0$ and $Q_{a\mu\nu}=0$ conditions) is a teleparallel geometry. Covariantly, this type of geometry is defined as being gauge invariant (valid for any $g_{ab}$). In~the orthonormal gauge $g_{ab} =\eta_{ab}= Diag[-1,1,1,1]$, the~relations to satisfy for such a geometry are~\cite{olver1995equivalence}:
	\begin{align}
		\mathcal{L}_{{\bf X}} {\bf h}^a =& \lambda^a_{~b} {\bf h}^b , \label{intro1}
		\\
		\mathcal{L}_{{\bf X}} {\omega}^a_{~bc} =& 0, \label{Intro2}
	\end{align}
	where ${\bf h}^a$ is the orthonormal coframe basis, $\mathcal{L}_{{\bf X}}$ is the Lie derivative in terms of Killing Vectors (KV) ${\bf X}$, and $\lambda^a_{~b}$ is the generator of the Lorentz transformations $\Lambda^a_{~b}$. For~a pure teleparallel geometry, one must also satisfy the zero Riemann curvature criterion (see refs.~\cite{Lucas_Obukhov_Pereira2009,Aldrovandi_Pereira2013,Bahamonde:2021gfp,MCH,Krssak:2018ywd,Coley:2019zld,Krssak_Pereira2015} and references therein):
	\begin{align}\label{zerocurvatureeqsol}
		R^a_{~b\mu\nu}  =& \partial_{\mu}\omega^a_{~b\nu} -\partial_{\nu}\omega^a_{~b\mu}+\omega^a_{~e\mu}\omega^e_{~b\nu}-\omega^a_{~e\nu}\omega^e_{~b\mu} =0 ,
		\nonumber\\
		\Rightarrow\quad &\omega^a_{~b\mu} = \Lambda^a_{~c}\partial_{\mu}\Lambda_b^{~c}  .
	\end{align} 
	
	{The solution} 
	of Equation~\eqref{zerocurvatureeqsol} leads to the teleparallel spin-connection defined in terms of a Lorentz transformation $\Lambda^a_{~b}$. Note also that $\omega^a_{~b\mu}=0$ for all proper frames and that $\omega^a_{~b\mu}\neq 0$ for all non-proper frames. Moreover, a~proper frame can be defined in several ways in terms of spin-connections; this can make it difficult to determine the~symmetries.

	There is a direct equivalent to GR defined under teleparallel gravity: the Teleparallel Equivalent to General Relativity (TEGR or $F(T)=T$ theory) \cite{Aldrovandi_Pereira2013}. However, TEGR easily generalizes to teleparallel $F(T)$ gravity, with $F$ as a function of the scalar torsion   {$T$ \cite{Ferraro:2006jd,Ferraro:2008ey,Linder:2010py}}. Teleparallel $F(T)$ gravity is known to be locally invariant under the definition of covariant Lorentz~\cite{Krssak_Pereira2015}. In~addition to the teleparallel $F(T)$-type gravity, all the considerations mentioned so far have been adapted for extended theories like New General Relativity (NGR) (see refs.~\cite{kayashi,beltranngr,bahamondengr} and references therein) and~symmetric teleparallel $F(Q)$-type gravity (see refs.~\cite{heisenberg1,heisenberg2,faithman1,hohmannfq} and references therein), as well as several intermediate theories like $F(T,Q)$-type, $F(R,Q)$-type, $F(R,T)$-type, and others (see refs.~\cite{jimeneztrinity,nakayama,ftqgravity,ftqspecial,frqspecial,frtspecial,frttheory} and references therein). Therefore, the best approach is to go with the teleparallel $F(T)$ gravity for the present~approach.

	There have been several works and papers on spherically symmetric spacetimes and its solutions in teleparallel $F(T)$ gravity using various approaches and for several types of uses~\cite{golov1,golov2,golov3,debenedictis,SSpaper,TdSpaper,nonvacSSpaper,nonvacKSpaper,roberthudsonSSpaper,coleylandrygholami,scalarfieldTRW,baha1,bahagolov1,awad1,baha6,nashed5,pfeifer2,elhanafy1,benedictis3,baha10,baha4,calza}. However, solving FEs in the orthonormal gauge has been favored for reasons of determining coframe/spin-connection pairs and also to avoid the extra degrees of freedom (DoF) problem associated with using proper frames to solve FEs (see   {refs.~\cite{SSpaper,nonvacSSpaper,nonvacKSpaper,roberthudsonSSpaper}} for detailed discussions). However, the~symmetric FEs and their $F(T)$ solutions will be very similar between different gauges (orthonormal or not); this further confirms the gauge invariance of general teleparallel $F(T)$ gravity FEs. But~by opting for the orthonormal gauge, one will solve the non-trivial antisymmetric and symmetric parts of FEs covering all DoFs for a diagonal coframe and a non-trivial spin-connection (see, for example,   {refs.~\cite{SSpaper,nonvacSSpaper,nonvacKSpaper,roberthudsonSSpaper})}. In~the literature, many prefer to use tetrads $e^a_{~\mu}$ instead of coframes $h^a_{~\mu}$, because~they are defined in an orthonormal gauge by default~\cite{golov1,golov2,golov3,debenedictis,baha1,bahagolov1,awad1,baha6,nashed5,pfeifer2,elhanafy1,benedictis3,baha10,baha4,calza}. However, the~tetrad approach prevents the use of frames other than orthonormal, and this is the greatest weakness of the latter approach. It is to avoid this type of limitation, to~respect gauge invariance as well as to work on any type of frame that is not necessarily orthonormal, that the coframe notation $h^a_{~\mu}$ will be used here and has been used in refs.~\cite{SSpaper,nonvacSSpaper,nonvacKSpaper,roberthudsonSSpaper}.

	One of the major scopes and aims in the study of teleparallel $F(T)$ solutions concerns the study of cosmological models involving the different forms of dark energy, such as quintessence, phantom (or negative) energy, quintom, and also dark matter. To~achieve this aim, some recent studies in teleparallel $F(T)$ gravity deal with Kantowski--Sachs (KS) spacetimes and the possible $F(T)$ solutions by using various ansatz~\cite{SSpaper,nonvacKSpaper,roberthudsonSSpaper,Rodrigues2015,Amir2015}. This type of spacetime is characterized by the radial derivative $\partial_r$ as the fourth symmetry (KV), leading directly to the pure time dependence for coframes, spin-connections, and FEs~\cite{SSpaper,nonvacKSpaper,roberthudsonSSpaper}. Previously, there have been some detailed dynamical studies concerning the evolution of critical points, boundary values, asymptotes, and $\overset{\ \circ}{R}$ curvature in KS spacetimes under the GR and/or the $f(\overset{\ \circ}{R})$ gravity frameworks~\cite{leon1,KScurvature,KSanisotropic}. Very similar studies using the same types of dynamical methods for KS spacetimes and their generalizations have also been carried out, not only in $F(\overset{\ \circ}{R})$ gravity but~also in $F(T,B)$-type,~LRS Bianchi III Universe, $F(T,R)$-type, and $F(Q)$-type gravity theories~\cite{leon2,leon3,frtkssol1,frtkssol2,fqks1,fqks2}. Recent works on cosmological solutions in teleparallel KS spacetimes with a quantized scalar field source should also be mentioned~\cite{palia2022KS,palia2023KS}. All these studies confirm the relevance and the accuracy of cosmological KS spacetimes solutions, especially for models involving different forms of dark~energy.

	However, most of these previous studies were not carried out in terms of teleparallel $F(T)$ gravity and do not really provide new $F(T)$ solutions. In particular, there has been a recent study on perfect fluid KS teleparallel $F(T)$ solutions, revealing relevant solutions for the dark energy quintessence, but~also for the phantom dark energy models~\cite{nonvacKSpaper}. It is briefly mentioned in this study that the different forms of dark energy mentioned above are at the same time explainable by fundamental scalar fields. It is therefore quite logical to study cosmological KS $F(T)$ solutions for different types of scalar fields as a logical follow-up to this last study. All this will allow one to not only confirm the $F(T)$ solutions obtained in this ref.~\cite{nonvacKSpaper} but also~to better study the $F(T)$ solutions and the fundamental scalar fields associated with dark matter, dark energy quintessence, phantom energy, and also quintom cosmological models using the teleparallel gravity framework and without the quantization of the scalar field source. Most recently, this type of study has been carried out for Teleparallel Robertson--Walker (TRW) spacetimes in $F(T)$ and $F(T,B)$ gravity (spacetime geometry at six KVs instead of four) with various scalar field sources~\cite{scalarfieldTRW,FTBcosmogholamilandry}. Beyond~spacetime and symmetry considerations, this type of study will mainly serve purely physical~motivations.
	
}


A significant physical motivation for this type of study concerns the dark energy quintessence models~\cite{steinhardt1,steinhardt2,steinhardt3,carroll1,quintessencecmbpeak}. This new type of model was first developed by Paul J. Steinhardt~et~al. in the late 1990s~\cite{steinhardt1,steinhardt2,steinhardt3}. There have been various phenomenological studies performed via the GR frameworks in order to obtain a model explaining the accelerating universe expansion (see refs.~\cite{steinhardt1,steinhardt2,steinhardt3,carroll1,quintessencecmbpeak,quintessenceholo,steinhardt2024,quintchakra2024} and several other ones). In~these models, a~perfect linear fluid of negative pressure is assumed ($P=\alpha_Q\,\rho$, where $-1<\alpha_Q<-\frac{1}{3}$ models), supported and induced by a scalar field. From~this starting point, there have been various dynamical and other studies on the possibilities of scalar fields and possible potentials that most faithfully describe the dark energy quintessence process and thus the accelerating universe expansion~\cite{quintessencecmbpeak,quintessenceholo}. Even recently, there are increasingly advanced studies on this very important subject in cosmology, specifically for observational constraints~\cite{steinhardt2024,quintchakra2024}. However, several years before Paul Steinhardt's works on quintessence processes, there had been attempts at models of accelerated universe expansion through the use of scalar field-based models and their symmetries explaining accelerating universe expansion~\cite{cosmofate,rollingscalarfield}. The~same precursor models therefore allowed all of P. Steinhardt's work on dark energy quintessence, generalizations, and other subsequent studies to emerge and arrive at the current~models.

Among the set of recent models are the teleparallel gravity-based dark energy quintessence models and their extensions ( in particular, $F(T)$-type, $F(T,B)$-type, scalar-torsion-based) \cite{attractor,scalar1,scalar2,cosmoftbmodels,cosmoftbenergy,ftphicosmo,leonftbcosmo1,leonftbcosmo2,Kofinas,ljsaidftb}. However, these papers have mainly concentrated on dynamics and especially on stability studies on FEs using simple and predefined $F(T)$ and $F(T,B)$ functions as superpositions of power-law and/or logarithm terms. Even if these latter types of function are often used in teleparallel cosmology, this somewhat simplistic approach does not truly and fundamentally solve the FEs of teleparallel $F(T)$ gravity and/or its extensions. These studies mainly only primarily study the impacts on some physical parameters of teleparallel gravity. At~the same time, this type of approach makes the teleparallel $F(T)$ exact, and correct solutions of FEs that can really model the dark energy quintessence processes go under the radar. There is therefore room to go further by solving the teleparallel $F(T)$ gravity FEs for scalar field sources, as was the case in refs.~\cite{nonvacKSpaper,roberthudsonSSpaper} for teleparallel $F(T)$ solutions for perfect fluid sources. Most recently, there have been detailed studies of Teleparallel Robertson--Walker (TRW) $F(T)$ and $F(T,B)$ solutions with perfect fluids and scalar field sources, allowing a larger number of possible solutions for teleparallel dark energy quintessence models~\cite{coleylandrygholami,FTBcosmogholamilandry}. In~this manner, we will find the most fundamental solutions for teleparallel KS spacetimes, allowing us to study the best and most fundamental possible models explaining dark energy quintessence by using the frameworks and FEs of teleparallel $F(T)$ gravity.

In addition, the~first and primarily dark energy quintessence model studies also led to the scenarios of cosmological models involving strong negative pressure dark energy perfect fluids (where $\alpha_Q<-1$) \cite{quintessencephantom,strongnegative}. This is also a new physical motivation for studying more extreme scenarios of strong and fast accelerating universe expansion, such as phantom (or negative) dark energy models~\cite{caldwell1,baumframpton,farnes,phantomdivide}. Even in teleparallel gravity, some studies have been considered in phantom dark energy models using the same approaches as for dark energy quintessence processes studies~\cite{cosmoftbmodels,cosmoftbenergy,phantomteleparallel1,ripphantomteleparallel2,phantomteleparallel3}. These works still focus on dynamic and stability studies using very specific teleparallel $F(T)$ functions (and/or $F(T,B)$) to study the evolution of some specific physical parameters without really directly solving  the FEs of teleparallel gravity. Obviously, there has been less work on this type of scenario, because~it is more hypothetical. This type of model often involves nonlinear fluids, but~the determining feature is mostly the energy condition violation of such a cosmological system (i.e. $P+\rho \ngeq 0$ situation), which makes that hypothetical. Recent works in teleparallel $F(T)$ and $F(T,B)$ types of gravity for TRW spacetimes provide a larger amount of material and new solutions to further study the phantom energy models in addition to quintessence processes~\cite{coleylandrygholami,FTBcosmogholamilandry}. There is also a way and a reason here to go further by directly solving the teleparallel $F(T)$ gravity FEs with a scalar field source, especially for the KS~spacetimes.

Furthermore, the~works mentioned show the existence of at least three possible types of dark energy: quintessence, cosmological constant, phantom energy. There are also the quintom dark energy models defined as a mixture of quintessence dark energy ($-1 < \alpha_Q < -\frac{1}{3}$) and phantom dark energy ($\alpha_Q<-1$), having as the intermediate limit the cosmological constant~\cite{quintom1,quintom2,quintom3,quintom4,quintomcoleytot}. This type of hybrid model is often defined by spacetime geometry and two scalar fields: the first field can describe quintessence and the second phantom energy or one field for the unified process and the other for coupling between the two types of dark energy~\cite{quintom1,quintom2,quintomcoleytot}. Some works specifically address this problem through the oscillation between the quintessence and phantom states~\cite{quintom3}. However, in teleparallel gravity there have been very few works dealing with this type of more complex physical problem~\cite{quintomteleparallel1}. This is also an additional reason favoring the present approach to solve the teleparallel $F(T)$ gravity FEs with a scalar field source. The~results arising from the current work will provide significantly more materials and tools for better developing the teleparallel quintom dark energy models in the future. There are also anisotropic cosmological models that can be added to the list of possibilities emerging from the present approach~\cite{teleparallelcosmosolanisotropic1}. There is no lack of possibilities, but~here we focus on the KS teleparallel $F(T)$ solutions with scalar fields to open up these possibilities for future~development.

For this paper, we assume a time-coordinate dependent spherically symmetric teleparallel geometry, in~particular Kantowski--Sachs teleparallel geometry, in~an orthonormal gauge as defined and used in refs.~\cite{SSpaper,nonvacKSpaper,roberthudsonSSpaper}. We will find $F(T)$ solutions for a power-law defined scalar field and then for an exponential scalar field source. After~a summary of the teleparallel FEs and Kantowski--Sachs class of geometries in Section~\ref{sect2}, we will find in   {Section~\ref{sect3}} the power-law scalar field exact and approximated $F(T)$ solutions. We then repeat the exercise in Section \ref{sect4} with an exponential scalar field source for exact and approximated $F(T)$ solutions. In~both cases, we will use a power-law and an exponential coframe component ansatz for a better comparison between $F(T)$ solutions. { We will finish the investigation by discussing in Section~\ref{sect5} the other possible scalar field sources and solving methods, especially by solving the special case of logarithmic source in Section~\ref{sect51}.} This paper also has common features, aims, and a similar structure to some recent papers on perfect fluid teleparallel $F(T)$ solutions (see refs.~\cite{nonvacSSpaper,nonvacKSpaper,roberthudsonSSpaper} for details).

The notation is defined as follows: coordinate indices $\mu, \nu, \ldots$, tangent space indices $a,b,\ldots$ (see ref.~\cite{MCH}), spacetime coordinates $x^\mu$, coframe ${\bf h}_a$ or ${\bf h}^a$, vierbein $h_a^{~\mu}$ or $h^a_{~\mu}$, metric $g_{\mu \nu}$, gauge metric $g_{ab}$, spin-connection $\omega^a_{~b} = \omega^a_{~bc} {\bf h}^c$, curvature tensor $R^a_{~bcd}$, torsion tensor $T^a_{~bc}$. The~derivatives with respect to (w.r.t.) $t$, $F_{t} = F'$ (with a prime).


\section{Summary of Teleparallel Gravity and Scalar Field Kantowski--Sachs Field~Equations}\label{sect2}
\unskip

\subsection{Teleparallel $F(T)$ Gravity Theory}\label{sect21}

{As in several recent works,} {the teleparallel $F(T)$-type gravity action integral is~\cite{Aldrovandi_Pereira2013,Bahamonde:2021gfp,Krssak:2018ywd,MCH,Coley:2019zld,SSpaper,nonvacSSpaper,nonvacKSpaper,roberthudsonSSpaper}}:
\begin{equation}\label{1000}
	S_{F(T)} = \int\,d^4\,x\,\left[\frac{h}{2\kappa}\,F(T)+\mathcal{L}_{Source}\right],
\end{equation}
where $h$ is the coframe determinant and $\kappa$ is the coupling constant. { We apply the least-action principle to Equation~\eqref{1000}, and we find the general FEs in terms of coframe and spin-connection~\cite{Aldrovandi_Pereira2013,Bahamonde:2021gfp,Krssak:2018ywd,MCH,Coley:2019zld}:
	\begin{align}\label{1000a}
		\kappa\,\Theta_a^{~~\mu} =&\, h^{-1}\,F_T\left(T\right)\,\partial_{\nu}\left(h\,S_a^{~~\mu\nu}\right) + F_{TT}\left(T\right)\,S_a^{~~\mu\nu}\,\partial_{\mu} T+\frac{F\left(T\right)}{2}\,h_a^{~~\mu}-F_T\left(T\right)\,T^b_{~~a\nu}S_b^{~~\mu\nu}
		\nonumber\\
		&\;-F_T\left(T\right)\,\omega^b_{~~a\nu}S_b^{~~\mu\nu} ,
	\end{align}
	where $\Theta_a^{~~\mu}$ is the energy--momentum, $T$ the torsion scalar, $T^b_{~~a\nu}$ the torsion tensor, $h_a^{~~\mu}$ the coframe (or tetrad for orthonormal frames), $\omega^b_{~~a\nu}$ the spin-connection, and $S_a^{~~\mu\nu}$ the superpotential (torsion dependent). We can separate Equation~\eqref{1000a} into} the symmetric and antisymmetric parts of FEs as~\cite{SSpaper,nonvacSSpaper,nonvacKSpaper,roberthudsonSSpaper}:
\begin{eqnarray}
	\kappa\,\Theta_{\left(ab\right)} &=& F_T\left(T\right) \overset{\ \circ}{G}_{ab}+F_{TT}\left(T\right)\,S_{\left(ab\right)}^{~~\mu}\,\partial_{\mu} T+\frac{g_{ab}}{2}\,\left[F\left(T\right)-T\,F_T\left(T\right)\right],  \label{1001a}
	\\
	0 &=& F_{TT}\left(T\right)\,S_{\left[ab\right]}^{~~\mu}\,\partial_{\mu} T, \label{1001b}
\end{eqnarray}
with $\overset{\ \circ}{G}_{ab}$ being the Einstein tensor and $g_{ab}$ the gauge metric. {Equations \eqref{1000a}--\eqref{1001b} are valid for any type of gauge (they satisfy the gauge invariance principle), coframe, and spin-connection, including the orthonormal frames and gauges~\cite{Krssak:2018ywd,Coley:2019zld,SSpaper}. However, there are some papers where the coframe $h^a_{~\mu}$ will be replaced for orthonormal frames by the tetrads $e^a_{~\mu}$ (see refs.~\cite{golov1,golov2,golov3,debenedictis,baha1,bahagolov1,awad1,baha6,nashed5,pfeifer2,elhanafy1,benedictis3,baha10,baha4,calza}), but~this last notation is not as general as the coframe $h^a_{~\mu}$ \cite{Coley:2019zld,SSpaper}.}

\subsection{{ General Conservation~Laws}}\label{sect22}

The canonical energy--momentum{ and its conservation laws are} obtained from the $\mathcal{L}_{Matter}$ term of Equation~\eqref{1000} as~\cite{Aldrovandi_Pereira2013,Bahamonde:2021gfp,Krssak:2018ywd,SSpaper,nonvacSSpaper,nonvacKSpaper,roberthudsonSSpaper}:
\begin{align}
	&\Theta_a^{\;\;\mu}=\frac{1}{h} \frac{\delta \mathcal{L}_{Source}}{\delta h^a_{\;\;\mu}}, \label{1001ca}
	\\
	{	\Rightarrow\,}& \overset{\ \circ}{\nabla}_{\nu}\left(\Theta^{\mu\nu}\right)=0, \label{1001e}
\end{align}
{with $\overset{\ \circ}{\nabla}_{\nu}$ being the covariant derivative and $\Theta^{\mu\nu}$ the conserved energy--momentum tensor. Equation~\eqref{1001e} is also the GR conservation of energy--momentum expression. The Equation~\eqref{1001ca} antisymmetric and symmetric parts are~\cite{SSpaper,nonvacSSpaper,nonvacKSpaper,roberthudsonSSpaper}:}
\begin{equation}\label{1001c}
	\Theta_{[ab]}=0,\qquad \Theta_{(ab)}= T_{ab},
\end{equation}
where $T_{ab}$ is the symmetric part of the energy--momentum tensor. Equation~\eqref{1001c} is valid only when the matter field interacts with the metric $g_{\mu\nu}$, defined from the coframe $h^a_{\;\;\mu}$ and the gauge $g_{ab}$, and~is not directly coupled to the $F(T)$ gravity. Equation~\eqref{1001e} also imposes the symmetry of $\Theta^{\mu\nu}$ and hence the condition stated by Equations \eqref{1001c}. All { the previous considerations are valid only} when the hypermomentum is zero (i.e., $\mathfrak{T}^{\mu\nu}= 0 $), as discussed in refs.~\cite{golov3,nonvacSSpaper,nonvacKSpaper,roberthudsonSSpaper}. { The hypermomentum condition is defined from Equations \eqref{1001a} and \eqref{1001b} as~\cite{golov3}:}
\begin{align}\label{1001h}
	\mathfrak{T}_{ab}=\kappa\Theta_{ab}-F_T\left(T\right) \overset{\ \circ}{G}_{ab}-F_{TT}\left(T\right)\,S_{ab}^{\;\;\;\mu}\,\partial_{\mu} T-\frac{g_{ab}}{2}\,\left[F\left(T\right)-T\,F_T\left(T\right)\right] = 0.
\end{align}

{Therefore, there exists a general non-zero hypermomentum (i.e., $\mathfrak{T}^{\mu\nu}\neq 0 $) conservation law condition for any teleparallel gravity theory which generalizes the Equation~\eqref{1001e} condition~\cite{hypermomentum1,hypermomentum2,hypermomentum3,golov3}. This general hypermomentum conservation law is trivially satisfied for the $\mathfrak{T}^{\mu\nu}= 0$ situation (the current situation and any GR solution).}

\subsection{Teleparallel Kantowski--Sachs { Coframe and Spin-Connection~Solutions}}\label{sect23}

The orthonormal time-dependent Kantowski--Sachs resulting vierbein { in $\left(t,r,\theta,\phi\right)$ coordinates} is~\cite{SSpaper,nonvacKSpaper,roberthudsonSSpaper}:
\beq h^a_{~\mu} = Diag\left[  1,\, A_2(t),\,A_3(t),\,A_3(t) \sin(\theta) \right], \label{VB:SS} \eeq

\noindent where we are able to choose new coordinates such that $A_1(t)=1$ without any loss of generality. This will allow us to find cosmological-like~solutions.

Another possible orthonormal coframe { solution} is taking the choice $A_3(t)=t$ and then solve for $A_1(t)$ and $A_2(t)$ \cite{roberthudsonSSpaper}:
\beq h^a_{~\mu} = Diag\left[  A_1(t),\, A_2(t),\,t,\,t\,\sin(\theta) \right]. \label{VB:SS2} \eeq

In refs.~\cite{SSpaper,nonvacKSpaper}, we find that the pure vacuum KS coframe solution implies that $A_3=t$. This is another justification for the Equation~\eqref{VB:SS2} coframe expression. However, Equation~\eqref{VB:SS2} is not as practical and appropriate as Equation~\eqref{VB:SS} for cosmological-like solutions. {Equations \eqref{VB:SS} and \eqref{VB:SS2} can also be expressed in principle in terms of tetrads $e^a_{~\mu}$ only for orthonormal gauges. But~the tetrad notation, as seen in the literature, limits the scope of the gauge choices for physical quantity expression.  Because~the teleparallel $F(T)$ gravity is gauge invariant and the FEs are defined for any gauge choice (gauge invariant), it is preferable to express the frame components in terms of $h^a_{~\mu}$ instead of $e^a_{~\mu}$ \cite{Krssak:2018ywd,Coley:2019zld,SSpaper}. By~using the $h^a_{~\mu}$ coframe notation, we are able to make the right gauge choice, and we are only limited to orthonormal gauges.}

The spin-connection $\omega_{abc}$ non-null components for time-dependent spacetimes are expressed using the Equation~\eqref{VB:SS} coframe components as $\omega_{abc}=\omega_{abc} (\psi(t),\chi(t))$, where $\chi$ and $\psi$ are arbitrary functions and $a \neq b$ indexes (see refs.~\cite{SSpaper,nonvacKSpaper,roberthudsonSSpaper} for detailed discussion and calculations). { In refs.~\cite{SSpaper,nonvacKSpaper,roberthudsonSSpaper}, we found from the antisymmetric part of the $F(T)$ FEs that the solution is $\psi = 0$ and $\chi = \frac{\pi}{2}$ ($\frac{3\pi}{2}$ is also a solution). In~terms of spin-connection, the~only non-zero $\omega_{abc}$ components are~\cite{SSpaper,nonvacKSpaper,roberthudsonSSpaper}:}
\begin{align}\label{1021}
	\omega_{234} = -\omega_{243} = \delta, \quad \omega_{344} = - \frac{\cos(\theta)}{A_3 \sin(\theta)} ,
\end{align}
where $\delta=\pm 1$. Fundamentally, Equations \eqref{VB:SS} (or Equation~\eqref{VB:SS2}) and \eqref{1021} are also the solutions of Equations \eqref{intro1}--\eqref{Intro2} using the Cartan--Karlhede (CK) algorithm method and also Equation~\eqref{zerocurvatureeqsol} for zero curvature criteria (see refs.~\cite{Coley:2019zld,MCH,SSpaper} for details and justifications). 


\subsection{Scalar Field Source Conservation~Laws}\label{sect24}

For the scalar field source action integral, the~$\mathcal{L}_{Source}$ term will be described by the Lagrangian density~\cite{Bahamonde:2021gfp,coleylandrygholami,scalar1,scalar2,leonftbcosmo1,leonftbcosmo2,FTBcosmogholamilandry}:
\begin{align}\label{300aa}
	\mathcal{L}_{Source} = \frac{h}{2}{\overset{\ \circ}{\nabla}}\,_{\nu}\phi\,\overset{\ \circ}{\nabla}\,^{\nu}\phi -h\,V\left(\phi\right).
\end{align}

We use, in Equation~\eqref{300a}, the covariant derivative ${\overset{\ \circ}{\nabla}}\,_{\nu}$ for satisfying the GR conservation law defined by Equation~\eqref{1001e}.

For a time-dependent $\phi=\phi(t)$ scalar field source and a cosmological-like spacetime (i.e., $A_1=1$), Equation~\eqref{300aa} will be simplified as:
\begin{align}\label{300a}
	\mathcal{L}_{Source} = h\frac{\phi'^2}{2} -h\,V\left(\phi\right) ,
\end{align}
where $\phi'=\partial_t\,\phi$. By~applying the least-action principle to Equation~\eqref{300a}, we find that the energy--momentum tensor $T_{ab}$ is~\cite{hawkingellis1,coleybook,nonvacSSpaper,SSpaper,nonvacKSpaper,roberthudsonSSpaper,cosmofluidsbohmer}:
\begin{align}\label{1001d}
	T_{ab}= \left(P_{\phi}+\rho_{\phi}\right)\,u_a\,u_b+g_{ab}\,P_{\phi},
\end{align}
where $u_a=(-1,\,0,\,0,\,0)$ and~$P_{\phi}$ and $\rho_{\phi}$ are, respectively, the pressure and density equivalent for the scalar field { defined by} \cite{coleylandrygholami}:
\begin{align}\label{300b}
	P_{\phi}= \frac{\phi'^2}{2}-V\left(\phi\right) \quad \text{and}\quad   \rho_{\phi}=\frac{\phi'^2}{2}+V\left(\phi\right), 
\end{align}
where $\phi'=\phi_t$ and $V=V\left(\phi\right)$ is the scalar field potential. Then, the conservation law for scalar fields with density and pressure defined by Equations \eqref{300b} for time-dependent spacetimes is~\cite{SSpaper,nonvacKSpaper,roberthudsonSSpaper}:
\begin{align}
	\phi'\,\left(\ln(A_2\,A_3^2)\right)'+\phi''+\frac{dV}{d\phi} =0. \label{302d}
\end{align}

Equation~\eqref{302d} is the most general scalar field conservation law equation for any $A_2$ and $A_3$ ansatz components with any $\phi(t)$ scalar field expression. For~the coming steps, we will first solve Equation~\eqref{302d} for its possible potential for study in this investigation. Then, we will solve the FEs to find the corresponding classes of teleparallel $F(T)$ solutions for each potential case satisfying Equation~\eqref{302d}. We will do these derivation steps for power-law and exponential scalar field $\phi(t)$ definitions.{ Equation~\eqref{300b} pressure and density definitions allow us to use the perfect fluid energy--momentum equivalence, as seen in Equation~\eqref{1001d}, for solving the FEs with a scalar field source. The~Equations \eqref{1001d}--\eqref{302d} also show that the change from a pure linear perfect fluid of $P=\alpha\,\rho$ to a scalar field pressure $P_{\phi}$ and density $\rho_{\phi}$ is clearly a natural and a straightforward generalization of the physical problem. This transformation will allow us to more easily study the fundamental causes of the dark energy physical processes by only using the mathematical analogy between the two types of physical system.}

To complete the discussion on physical implications from~Equation \eqref{300b} and to { better} make the link with a linear perfect fluid EoS equivalent $P_{\phi}=\alpha_Q\rho_{\phi}$, we need to find the { dark energy} coefficient index $\alpha_Q$ { (also called the quintessence coefficient)} for making the parallel between $\phi$ and the dark energy processes. { The $\alpha_Q$ index will be defined from Equation \eqref{300b} as~\cite{steinhardt1,steinhardt2,steinhardt3,steinhardt2024}:
	\begin{align}\label{Quintessenceindex}
		\alpha_Q =& \frac{P_{\phi}}{\rho_{\phi}} = \frac{\phi'^2-2V\left(\phi\right)}{\phi'^2+2V\left(\phi\right)}=-1+\frac{2\phi'^2}{\phi'^2+2V\left(\phi\right)}.
	\end{align}
	
	We can use this Equation~\eqref{Quintessenceindex} for the dark energy perfect fluid equivalent for every new teleparallel $F(T)$ solution, any potential $V\left(\phi\right)$, and any ansatz. But~there are physical and especially dark energy state significations for $\alpha_Q$, such as:
	\begin{enumerate}
		\item \textbf{{Quintessence} 
		} $-1<\alpha_Q <-\frac{1}{3}$: This is the most relevant and possible dark energy state explaining the controlled universe accelerating expansion by using an associated scalar field $\phi$, called the quintessence scalar field~\cite{steinhardt1,steinhardt2,steinhardt3,steinhardt2024}.
		
		\item \textbf{{Phantom Energy}} $\alpha_Q<-1$: This case leads to phantom dark energy processes characterized by an uncontrolled accelerating universe expansion going to the Big Rip at the end of the physical process~\cite{farnes,baumframpton,caldwell1}. There is also a fundamental scalar field $\phi$ describing the phantom energy~processes.
		
		\item \textbf{{Cosmological constant}} $\alpha_Q=-1$: This is the first and most fundamental dark energy state. It is also the boundary between quintessence and phantom dark energy states. A~constant scalar field leads to the cosmological constant according to Equation~\eqref{Quintessenceindex}, a~GR solution~\cite{FTBcosmogholamilandry}.
		
		\item \textbf{{Quintom models}}: This is a mix of dark energy quintessence and phantom dark energy physical processes and~is usually described by a two-scalar-fields model~\cite{quintom1,quintom2,quintom3,quintom4,quintomcoleytot}. 
	\end{enumerate}
	
	This classification is also to express the new teleparallel $F(T)$ solutions in terms of dark energy quintessence and phantom and quintom~processes.

}

\subsection{Symmetric and Unified Field~Equations}\label{sect25}

The torsion scalar and the symmetric FE components for the $\chi = \frac{\pi}{2}$ ($\delta=+1$) case in Equation~\eqref{1021} are~\cite{SSpaper,nonvacKSpaper,roberthudsonSSpaper}:
\begin{align}
	T =& 2\left(\ln(A_3)\right)'\left(\left(\ln(A_3)\right)'+2\left(\ln(A_2)\right)'\right) - \frac{2}{A_3^2} , \label{301a}
	\\
	B' =&  -\left(\ln\left(A_2\,A_3^2\right)\right)'+\frac{\frac{1}{A_3^2} - \left(\ln\left(\frac{A_2}{A_3}\right)\right)''}{\left(\ln\left(\frac{A_2}{A_3}\right)\right)' }, \label{301b} 
	\\
	\kappa\,\phi'^2+2\kappa\,V\left(\phi\right) =& -F(T) + 2\left( T+\frac{2}{A_3^2} \right) F_T(T), \label{301c}
	\\ 
	-\kappa \phi'^2 =& \left[\left(\ln(A_3)\right)'\left(B'-\left(\ln\left(\frac{A_2}{A_3}\right)\right)'\right)+ \left(\ln(A_3)\right)''\right] F_T(T), \label{301d}
\end{align}
\noindent where $F_T(T) \neq$ constant and $B'=\partial_t\left(\ln\,F_T(T)\right)$. Compared with the Kantowski--Sachs $F(T)$-gravity FEs in the literature~\cite{Rodrigues2015,Amir2015}, the~FEs are different. For the~$\delta=-1$ FEs set, there are some small minor differences for some terms in Equations \eqref{301b}--\eqref{301d}, mainly some different signs at very specific terms. For~the rest, the~general form of   {Equations \eqref{301b}--\eqref{301d}} remains identical, regardless of $\delta$. However, we can simplify Equations \eqref{301b}--\eqref{301d} by adding Equations \eqref{301c} and \eqref{301d} and then by substituting Equations \eqref{301a} and \eqref{301b}:
\vspace{-12pt}

	\begin{align}
		2\kappa\,V\left(\phi(T)\right) =& -F(T) +\left[\frac{3}{2}\left( T+\frac{2}{A_3^2} \right)+\frac{\left(\ln(A_3)\right)'}{\left(\ln\left(\frac{A_2}{A_3}\right)\right)' }\left(\frac{1}{A_3^2} - \left(\ln\left(\frac{A_2}{A_3}\right)\right)''\right) + \left(\ln(A_3)\right)''\right] F_T(T). \label{305}
	\end{align}

Equation~\eqref{305} will be the DE to solve for $F(T)$. We will substitute the right ansatz and again use Equation~\eqref{301a} as a characteristic equation, as in ref.~\cite{nonvacKSpaper}. In~addition, we will have to calculate for each used ansatz the corresponding $V\left(\phi(T)\right)$ by using Equation~\eqref{302d} for conservation laws. The~Equation~\eqref{301a} characteristic equation will also be useful for finding the $\phi(T)$ scalar field and then the $V\left(\phi(T)\right)$ functions { in terms of $T$.

	Another method based on Equation~\eqref{305} consists of setting coframe components, Equation~\eqref{301a} and a $F(T)$ function ansatz to find the exact scalar field potential $V(\phi)$, and then $\phi(t)$ by using Equation~\eqref{302d}. However, this method does not allow one to control the source and/or to take into account the exact homogeneous parts of an $F(T)$.}


\section{Power-Law Scalar Field~Solutions}\label{sect3}

In this section, we will set a power-law scalar field $\phi(t)=p_0\,t^p$, where $p_0$ is a constant and $p$ is a real power. The~conservation law defined by Equation~\eqref{302d} becomes, with field expression inversion as $t\left(\phi\right)=\left(\frac{\phi}{p_0}\right)^{1/p}$:
\begin{align}
	0=	p\,p_0^{1/p}\,\left(\ln(A_2\,A_3^2)\right)'\,\phi^{1-1/p}+p(p-1)\,p_0^{2/p}\,\phi^{1-2/p}+\frac{dV}{d\phi} , \label{400}
\end{align}
where $\left(\ln(A_2\,A_3^2)\right)'$ is dependent on $\phi$ and the used ansatz for $A_2$ and $A_3$ components. There are a number of possible~ansatz:
\begin{enumerate}
	
	\item \textbf{{Power-law:} 
	} This is the most simple, used, and well-known ansatz. It is defined as~\cite{nonvacKSpaper}:
	\begin{align}\label{401}
		A_2=t^b \quad \text{and}\quad A_3=c_0\,t^c,
	\end{align}
	where $c_0$ is a constant. Then, we find that $\left(\ln(A_2\,A_3^2)\right)'=\frac{(b+2c)}{t\left(\phi\right)}=(b+2c)\,p_0^{1/p}\,\phi^{-1/p}$ and Equation~\eqref{400} becomes:
	\begin{align}
		0=& p\,p_0^{2/p}\,\left( b+2c+p-1\right)\,\phi^{1-2/p}+\frac{dV}{d\phi} ,
		\nonumber\\
		& \Rightarrow\; V\left(\phi\right)=\phi_0 -\frac{p^2\,p_0^{2/p}}{2(p-1)}\,\left( b+2c+p-1\right)\,\phi^{2-2/p} . \label{401a}
	\end{align}
	The $\alpha_Q$-index from Equation~\eqref{Quintessenceindex} is, by substituting Equation~\eqref{401a}:
	\begin{align}\label{401aq}
		\alpha_Q =& -1 +\frac{p^2\,p_0^{2/p}\,\phi^{2-2/p}}{\phi_0-\frac{p^2\,p_0^{2/p}}{2}\,\left(\frac{b+2c}{p-1}\right)\,\phi^{2-2/p}}.
	\end{align}
	
	There are special~cases:
	\begin{itemize}
		\item  {${\bf p\gg 1}$}: Equation~\eqref{401a} can be approximated as:
		\begin{align}
			V\left(\phi\right)\approx \phi_0 -\frac{p^2}{2}\,\phi^{2} .\label{401b}
		\end{align}
		Then, Equation~\eqref{401aq} will be, for this case:
		\begin{align}\label{401bq}
			\alpha_Q \approx &  -1 +\frac{p^2\,\phi^{2}}{\phi_0-\frac{p(b+2c)}{2}\,\phi^{2}} \approx -1-\frac{2p}{b+2c},
		\end{align}
		where we have the physical~situations:
		\begin{itemize}
			\item[$-$] {${\bf\frac{1}{3}< \frac{p}{b+2c} < 0}\;$}: dark energy quintessence process. If~$p>0$ , then $b+2c<0$.
			
			\item[$-$] {${\bf \frac{p}{b+2c}>0}$}: phantom energy process. If~$p>0$ , then $b+2c>0$.
		\end{itemize}
		
		If $b+2c$ is not large, Equation~\eqref{401bq} will rather be $\alpha_Q=-1+\frac{p^2}{\phi_0}\,\phi^{2}$, without any constraint on $b+2c$. This last result can be directly found by using Equation~\eqref{401b} and then the Equation~\eqref{Quintessenceindex} $\alpha_Q$ definition.

		\item {${\bf p=1}$}: Equation~\eqref{401a} is simplified and then becomes:
		\begin{align}
			0=& p_0^{2}\,\left( b+2c\right)\,\phi^{-1}+\frac{dV}{d\phi} ,
			\nonumber\\
			& \Rightarrow\; V\left(\phi\right)=\phi_0 -p_0^{2}\,\left( b+2c\right)\,\ln\,\phi . \label{401c}
		\end{align}
		Then, Equation~\eqref{Quintessenceindex} is, by substituting Equation~\eqref{401c}:
		\begin{align}\label{401cq}
			\alpha_Q =& -1+ \frac{p_0^2}{\frac{p_0^2}{2}+\phi_0 -p_0^{2}\,\left( b+2c\right)\,\ln\,\phi}.
		\end{align}
	\end{itemize}
	
	\item \textbf{{Exponential:}} This case is defined by~\cite{nonvacKSpaper}:
	\begin{align}\label{402}
		A_2=e^{bt}  \quad \text{and}\quad  A_3=c_0\,e^{ct} .
	\end{align}
	
	The Equation~\eqref{402} form leads to $\left(\ln(A_2\,A_3^2)\right)'=(b+2c)$, and Equation~\eqref{400} becomes, in this case:
	\begin{align}
		0=&	p\,p_0^{1/p}\,(b+2c)\,\phi^{1-1/p}+p(p-1)\,p_0^{2/p}\,\phi^{1-2/p}+\frac{dV}{d\phi} ,
		\nonumber\\
		& \Rightarrow\; V\left(\phi\right)=\phi_0-\frac{p^2\,(b+2c)}{(2p-1)}\,p_0^{1/p}\,\phi^{2-1/p}-\frac{p^2}{2}\,p_0^{2/p}\,\phi^{2-2/p} . \label{402a}
	\end{align}
	
	Then, Equation~\eqref{Quintessenceindex} will be:
	\begin{align}\label{402aq}
		\alpha_Q =& -1+\frac{p^2p_0^{2/p}\phi^{2-2/p}}{\phi_0-\frac{p^2\,(b+2c)}{(2p-1)}\,p_0^{1/p}\,\phi^{2-1/p}}.
	\end{align}

	The previous special cases~become:
	\begin{itemize}
		\item {${\bf p\gg 1}$}: Equations \eqref{402a} and \eqref{402aq} will be exactly Equations \eqref{401b} and \eqref{401bq}. 
		
		\item {${\bf p=1}$}: Equation~\eqref{402a} yields a linear potential:
		\begin{align}
			0=&	p_0\,(b+2c)+\frac{dV}{d\phi} ,
			\nonumber\\
			& \Rightarrow\; V\left(\phi\right)=\phi_0-p_0\,(b+2c)\,\phi , \label{402b}
		\end{align}
		and then Equation~\eqref{Quintessenceindex} becomes, with the Equation~\eqref{402b} potential:
		\begin{align}\label{402bq}
			\alpha_Q =& -1+\frac{p_0^2}{\frac{p_0^{2}}{2}+\phi_0-p_0\,(b+2c)\,\phi} .
		\end{align}
		
	\end{itemize}
\end{enumerate}

We also note the scalar potential $V(\phi)$ solutions to Equation~\eqref{400}, obtained for a power-law scalar field, confirm those found in one of the first studies on dark energy quintessence~\cite{quintessencecmbpeak}. This was just confirmed by the separate use of the Equations \eqref{401} and \eqref{402} ansatzes, which shows the rightness and relevance of the potential solutions compared to the first models studied in the past~literature.

\subsection{Power-Law Ansatz~Solutions}\label{sect31}

Equation~\eqref{301a}, in terms of Equation \eqref{401}, leads to the characteristic equation (see ref.~\cite{nonvacKSpaper}):
\begin{align}\label{403}
	0 =& 2c\left(c+2b\right)\,t^{-2} - \frac{2}{c_0^2}\,t^{-2c}-T,
\end{align}
where $c\neq 0$ ($c=0$ subcase leads to GR solutions). Then, Equation~\eqref{305} becomes, under the Equation~\eqref{401} ansatz and by replacing $V\left(\phi(T)\right)=\phi_0+V(T)$:
\begin{align}	
	{ \Lambda_0}+2\kappa\,V(T) =&  -F(T) +\left[{\frac{3}{2} T +\frac{(3b-2c)}{c_0^2\,(b-c)}t^{-2c}(T) }\right]\, F_T(T),
	\nonumber\\
	=&  -F(T) +A(T)\, F_T(T), \label{404}
\end{align}
where the function $A(T)$ is defined as
\begin{align}\label{404atdef}
	A(T)={\frac{3}{2} T +\frac{(3b-2c)}{c_0^2\,(b-c)}t^{-2c}(T) },
\end{align}
with $t(T)$ as the Equation~\eqref{403} solution { and $\Lambda_0=2\kappa\,\phi_0$ as the cosmological constant.} From this Equation~\eqref{403}, there are several possible solutions to Equation~\eqref{404} depending on the values of $c$, as in ref.~\cite{nonvacKSpaper}. Each value of $c$  will lead to specific $V(T)$ and $A(T)$ functions, and~then we will solve Equation~\eqref{404} with this form. In~general, the~Equation~\eqref{404} solution will be under the form:
\begin{align}\label{404solution}
	F(T) =& -{ \Lambda_0} + \exp\left[\int_{T}\,\frac{dT'}{A(T')}\right]\Bigg[F_0+2\kappa\,\int_{T}\,dT'\,\frac{V(T')}{A(T')}\,\exp\left[-\int_{T'}\,\frac{dT''}{A(T'')}\right]\Bigg] ,
\end{align} 
where $A(T)$ is Equation~\eqref{404atdef}. Equation~\eqref{404solution} is the general formula applicable for any subcases and will be used to find all new teleparallel $F(T)$ solutions in the current~paper. 

The subcases~are:
\begin{enumerate}
	\item {${\bf c=-2b}$} ($b \neq 0$): Equation~\eqref{403} is simplified as~\cite{nonvacKSpaper}:
	\begin{align}\label{405}
		t(T) =& \left(\frac{c_0^2}{2} (-T)\right)^{1/4b}.
	\end{align}
	
	Then, Equation~\eqref{404atdef} is simplified to $A(T)=\frac{T}{3}$. The~scalar field will be $\phi(T)=p_0\,\left(\frac{c_0^2}{2} (-T)\right)^{p/4b}$, and we will use this expression for the~following subcases:
	\begin{enumerate}
		\item \textbf{{General:}} Equation~\eqref{401a} for scalar field potential is:
		\begin{align}\label{406}
			V(T)= -\frac{V_p}{2} (-T)^{(p-1)/2b}, 
		\end{align}
		where $V_p=\frac{p^2\,p_0^{2}\,\left(p-1-3b\right)}{2^{(p-1)/2b}\,(p-1)}\,c_0^{(p-1)/b}$. We obtain that $V_p=0$ for $p=3b+1$ and $V(T)$ is undefined for $p=1$. Then, by substituting the simplifed $A(T)$ and Equations \eqref{405} and \eqref{406} into Equation~\eqref{404solution}, the~teleparallel $F(T)$ solution will be:
		\small
		\begin{align}\label{408}
			F(T) = -{ \Lambda_0}-\frac{6\kappa\,b\,V_p}{(p-1-6b)}\,(-T)^{(p-1)/2b} + F_0\,T^3 ,
		\end{align}
		\normalsize
		where $F_0$ is an integration constant and $p\neq 1$ and $p\neq 1+6b $.

		\item {${\bf p\gg 1}$ \textbf{case:}} By using the same scalar field $\phi(T)$, Equation~\eqref{401b} yields as potential:
		\begin{align}
			V(T)=\,-\frac{V_{\infty}}{2}\,(-T)^{p/2b} ,\label{409a}
		\end{align}
		where $V_{\infty}=\frac{p^2\,p_0^{2}\,c_0^{p/b}}{2^{p/2b}}$. Then, by using the same $A(T)$ and Equations \eqref{405} and \eqref{409a},~Equation~\eqref{404solution} is:
		\begin{align}\label{409c}
			F(T) \approx -{ \Lambda_0}-\frac{6\kappa\,b\,V_{\infty}}{p}\,(-T)^{p/2b}+ F_0\,T^3 \approx  -\frac{6\kappa\,b\,p\,p_0^{2}\,c_0^{p/b}}{2^{p/2b}}\,(-T)^{p/2b} ,
		\end{align}
		where $p\rightarrow \infty$ and $T \leq 0$ in some~situations.
		
		\item {${\bf p= 1}$ \textbf{case:}} By still using the same scalar field $\phi(T)$, Equation~\eqref{401c} becomes:
		\begin{align}
			V(T)=\tilde{\phi}_0-\phi_0+\frac{3p_0^{2}}{4}\,\ln(-T) , \label{409e}
		\end{align}
		where  $\tilde{\phi}_0=\phi_0 +\frac{3p_0^{2}}{4}\,\ln\,\left(\frac{p_0^{4b}\,c_0^{2}}{2}\right)$. Then, Equation~\eqref{404solution} becomes, by using the same $A(T)$ and Equations \eqref{405} and \eqref{409e}:
		\begin{align}\label{409g}
			F(T) = -{ \tilde{\Lambda}_0}-\frac{\kappa\,p_0^{2}}{2}-\frac{3\kappa\,p_0^{2}}{2}\,\ln(-T) + F_0\,T^3 ,
		\end{align}
		where $T \leq 0$ { and $\tilde{\Lambda}_0=2\kappa\,\tilde{\phi}_0$ is the modified cosmological constant}.
		
	\end{enumerate}
	
	By comparing Equations \eqref{408}, \eqref{409c}, and \eqref{409g}, we see that the homogeneous parts are very similar and their respective differences are only from the $V(T)$ parts (particular solution).

	\item {${\bf c=1}$}:
	Equation~\eqref{403} becomes~\cite{nonvacKSpaper}:
	\begin{align}\label{420}
		& 0=\left(2(1+2b) - \frac{2}{c_0^2}\right)\,t^{-2}-T,
		\nonumber\\
		&\Rightarrow\,t^{-2}(T)=\frac{T}{2\left(1+2b - \frac{1}{c_0^2}\right)}.
	\end{align}
	
	From Equation~\eqref{420}, we find, as Equation~\eqref{404atdef}:
	\begin{align}\label{420at}
		A(T)={\left[\frac{3}{2} +\frac{(3b-2)}{2c_0^2\,(b-1)\left(1+2b - c_0^{-2}\right)} \right]\,T}=C\,T ,
	\end{align}
	where $C=\frac{3}{2} +\frac{(3b-2)}{2c_0^2\,(b-1)\left(1+2b - c_0^{-2}\right)}$. The~scalar field for Equation~\eqref{420} is:
	\begin{align}\label{421}
		\phi(T)=\,p_0\,2^{p/2}\left(1+2b - c_0^{-2}\right)^{p/2}\,T^{-p/2} .
	\end{align}
	The potential $V(T)$ and $F(T)$ solutions are, for the~following subcases:
	\begin{enumerate}
		\item \textbf{{General:}} Equation~\eqref{401a} becomes:
		\begin{align}
			V(T)=-\frac{V_p}{2}\,T^{1-p} , \label{422}
		\end{align}
		where $p \neq 1$ and $V_p=\frac{p^2\,p_0^{2}\,2^{p-1}}{(p-1)}\,\left( b+p+1\right)\left(1+2b - c_0^{-2}\right)^{p-1}$. Equation~\eqref{422} will be constant for $b=-\frac{1}{2}+\frac{1}{2c_0^2}$ and $b=-p-1$. By~substituting Equations \eqref{420at} and \eqref{422} into Equation~\eqref{404solution}, we find, as a solution:
		\begin{align}\label{423a}
			F(T)=& -{ \Lambda_0} +\frac{\kappa\,V_p}{\left(1+C(p-1)\right)}\,T^{1-p} +F_0\,T^{1/C} ,
		\end{align}
		where $p \neq 1$.
		
		\item \textbf{{Equation}~\eqref{401b} {potential:}} This equation, in terms of Equation \eqref{421}, is ($p\gg 1$):
		\begin{align}
			V(T)= -\frac{V_{\infty}}{2} \,T^{-p} , \label{424}
		\end{align}
		where $V_{\infty}=p^2\,p_0^2\,2^{p}\left(1+2b - \frac{1}{c_0^2}\right)^{p}$. By~substituting Equations \eqref{420at} and \eqref{424} into Equation~\eqref{404solution}, we find as a solution:
		\begin{align}\label{426}
			F(T) =& -{ \Lambda_0}+\,F_0\,T^{1/C} +\frac{\kappa\,V_{\infty}}{\left[1+p\,C\right] }\,T^{-p} \approx -{ \Lambda_0}+\,F_0\,T^{1/C}.
		\end{align}
		
		\item\textbf{{Equation}~\eqref{401c} {potential:}} This equation, in terms of Equation \eqref{421}, is ($p=1$):
		\begin{align}
			V(T)=\tilde{\phi}_0-\phi_0 +\frac{p_0^{2}}{2}\,\left( b+2\right)\,\ln\,\left(T\right). \label{427}
		\end{align}
		where $b \neq -2$ and $\tilde{\phi}_0=\phi_0 -\frac{p_0^{2}}{2}\,\left( b+2\right)\,\ln\,\left[2p_0^2\,\left(1+2b - \frac{1}{c_0^2}\right)\right]$. By~substituting Equations \eqref{420at} and \eqref{427} into Equation~\eqref{404solution}, we find as a solution:
		\small
		\begin{align}\label{429}
			F(T) =& -{ \tilde{\Lambda}_0}+F_0\,T^{1/C}-\kappa\,p_0^{2}\,\left( b+2\right)\left[\ln\,\left(T\right)+C \right] .
		\end{align}
		\normalsize
	\end{enumerate}

	\item {${\bf c=-1}$}:
	Equation~\eqref{403} becomes~\cite{nonvacKSpaper}:
	\begin{align}\label{430}
		& 0=t^{4}+\frac{c_0^2\,T}{2}\,t^{2} - c_0^2(1-2b), \quad &
		\nonumber\\
		&\Rightarrow\,t^{2}(T)= \frac{c_0^2}{4}\left[-T+\delta_1\,\sqrt{T^2+16(1-2b)\,c_0^{-2}}\right] ,
	\end{align}
	where $\delta_1=\pm 1$. Equation~\eqref{404atdef} for Equation~\eqref{430} is:
	\begin{align}\label{430at}
		A(T)=& {\frac{3}{2} T +\frac{(3b+2)}{4\,(b+1)}\left[-T+\delta_1\,\sqrt{T^2+16(1-2b)\,c_0^{-2}}\right] }
		\nonumber\\
		=& \left(\frac{3}{2}-C_1\right)T+\delta_1\,C_1\sqrt{T^2+C_2},
	\end{align}
	where $C_1=\frac{(3b+2)}{4\,(b+1)}$ and $C_2=16(1-2b)\,c_0^{-2}$. The~scalar field for Equation~\eqref{430} is:
	\vspace{-12pt}
	
		\begin{align}\label{431}
			\phi(T)=&\,\frac{p_0\,c_0^p}{2^p}\left[-T+\delta_1\,\sqrt{T^2+16(1-2b)\,c_0^{-2}}\right]^{p/2}= \frac{p_0\,c_0^p}{2^p}\left[-T+\delta_1\,\sqrt{T^2+C_2}\right]^{p/2},
			\nonumber\\
			=& \,{ \frac{p_0\,c_0^p}{2^p}\,\left[u_{\pm}(T)\right]^{p/2},}
		\end{align}
	{ where $u_{\pm}(T)=-T+\delta_1\,\sqrt{T^2+C_2}$ ($+$ for $\delta_1=+1$ and $-$ for $\delta_1=-1$).} The potential $V(T)$ and $F(T)$ solutions are, for the~following subcases:
	\begin{enumerate}
		\item \textbf{{General:}} Equation~\eqref{401a} becomes:
		\begin{align}
			V(T)= -\frac{V_p}{2}\,\left[{ u_{\pm}(T)}\right]^{p-1} , \label{432}
		\end{align}
		where $p \neq 1$ and $V_p=\frac{p^2\,p_0^{2}\,c_0^{2p-2}}{2^{2p-2}\,(p-1)}\,\left( b+p-3\right)$. Equation~\eqref{432} will be constant for $b=3-p$. By~substituting Equations \eqref{430at} and \eqref{432} into Equation~\eqref{404solution}, we find as a solution:
		\small
		\begin{align}\label{433a}
			F(T) =& -{ \Lambda_0} +{\left[{ 3T+2C_1\,u_{\pm}(T)}\right]^{\frac{2(3-2C_1)}{3(3-4C_1)}}\,\left[{ u_{-}(T)}\right]^{-\frac{4C_1\delta_1}{3(3-4C_1)}}}\,\Bigg[F_0-2\kappa\,V_p\,
			\nonumber\\	
			&\,\times\,\int_{T}\,dT'\,\left[{ u_{\pm}(T')}\right]^{p-1}{\left[{ 3T'+2C_1\,u_{\pm}(T')}\right]^{-\frac{2(3-2C_1)}{3(3-4C_1)}-1}\,\left[{ u_{-}(T')}\right]^{\frac{4C_1\delta_1}{3(3-4C_1)}}}\Bigg].
		\end{align}
		\normalsize
		
		Equation~\eqref{433a} is difficult to solve under this current form, { but there are special case solutions. For~example, the~$C_2=0$ ($b=\frac{1}{2}$ and $C_1=\frac{7}{12}$) and $\delta_1=-1$ case where $u_{-}(T)=-2T$ (and $u_{+}(T)=0$ for $\delta_1=+1$) yields:}
		\begin{align}\label{433b}
			F(T) =& -{ \Lambda_0} +\,F_0\,T^{3}-\frac{3\kappa V_p(-2)^{p}}{(p-4)}\,T^{p-1}.
		\end{align}
		{ The other $C_2\neq 0$ possible cases are solved and presented in Appendix \ref{appena1}.}

		\item \textbf{{Equation}~\eqref{401b} {potential:}} This equation, in terms of Equation \eqref{431}, is ($p\gg 1$):
		\begin{align}
			V(T)= -\frac{V_{\infty}}{2}\,\left[{ u_{\pm}(T)}\right]^{p} ,\label{434}
		\end{align}
		where $V_{\infty}=\frac{p^2\,p_0^2\,c_0^{2p}}{2^{2p}}$. By~substituting Equations \eqref{430at} and \eqref{434} into Equation~\eqref{404solution}, we find as a solution:
		\small
		\begin{align}\label{436}
			F(T) =& -{ \Lambda_0} +{\left[{ 3T+2C_1\,u_{\pm}(T)}\right]^{\frac{2(3-2C_1)}{3(3-4C_1)}}\,\left[{ u_{-}(T)}\right]^{-\frac{4C_1\delta_1}{3(3-4C_1)}}}
			\Bigg[F_0-2\kappa\,V_{\infty}\,
			\nonumber\\	
			&\,\times\,\int_{T}\,dT'\,\left[{ u_{\pm}(T)}\right]^{p}\,{\left[{ 3T+2C_1\,u_{\pm}(T)}\right]^{-\frac{2(3-2C_1)}{3(3-4C_1)}-1}\,\left[{ u_{-}(T)}\right]^{\frac{4C_1\delta_1}{3(3-4C_1)}}}\Bigg].
		\end{align}
		\normalsize
		Once again, Equation~\eqref{436} does not yield a general solution, { but there are specific cases leading to analytical $F(T)$ solutions. For~example, the~$C_2=0$ ($b=\frac{1}{2}$ and $C_1=\frac{7}{12}$) and $\delta_1=-1$ case yields:}
		\begin{align}\label{436a}
			F(T) =& -{ \Lambda_0} +\,F_0\,T^3-\frac{3\kappa\,V_{\infty}(-2)^p}{p}\,{T^{p}}.
		\end{align}
		{ The other $C_2\neq 0$ possible cases are also solved and presented in Appendix \ref{appena1}.}

		\item \textbf{{Equation}~\eqref{401c} {potential:}} This equation, in terms of Equation \eqref{431}, is ($p=1$):
		\begin{align}
			V(T)=\tilde{\phi}_0-\phi_0-\frac{p_0^{2}}{2}\,\left( b-2\right)\,\ln\,\left[{ u_{\pm}(T)}\right] , \label{437}
		\end{align}
		where $b \neq 2$ and $\tilde{\phi}_0=\phi_0 -p_0^{2}\,\left( b-2\right)\,\ln\,\left[\frac{p_0\,c_0}{2}\right]$. By~substituting Equations \eqref{430at} and \eqref{437} into Equation~\eqref{404solution}, we find as a solution:
		\small
		\begin{align}\label{439}
			F(T) =& -{ \tilde{\Lambda}_0} +{\left[{ 3T+2C_1\,u_{\pm}(T)}\right]^{\frac{2(3-2C_1)}{3(3-4C_1)}}\,\left[{ u_{-}(T)}\right]^{-\frac{4C_1\delta_1}{3(3-4C_1)}}}
			\Bigg[F_0-2\kappa\,p_0^{2}\,\left( b-2\right)\,
			\nonumber\\	
			&\,\times\,\int_{T}\,dT'\,\ln\,\left[{ u_{\pm}(T)}\right]\,{\left[{ 3T+2C_1\,u_{\pm}(T)}\right]^{-\frac{2(3-2C_1)}{3(3-4C_1)}-1}\,\left[{ u_{-}(T)}\right]^{\frac{4C_1\delta_1}{3(3-4C_1)}}}\Bigg].
		\end{align}
		\normalsize
		{ There is no general solution for Equation~\eqref{439}, but~only specific solutions. For~example, the~$C_2=0$ ($b=\frac{1}{2}$ and $C_1=\frac{7}{12}$) and $\delta_1=-1$ case leads to:}
		\begin{align}\label{439a}
			F(T) =& -{ \tilde{\Lambda}_0} +F_0\,T^{\frac{2}{(3-4C_1)}}-\frac{\kappa\,p_0^{2}\,\left( b-2\right)}{2}\,\left(4C_1-3-2\ln\,(-2T)\right).
		\end{align}
		{ All $C_2\neq 0$ solvable cases are also presented in Appendix \ref{appena1}.}

	\end{enumerate}

	\item {${\bf c=2}$:} 
	Equation~\eqref{403} becomes~\cite{nonvacKSpaper}:
	\begin{align}\label{440}
		& 0=t^{-4}-4c_0^2\,(1+b)\,t^{-2} +\frac{c_0^2\,T}{2},
		\nonumber\\
		& \Rightarrow\,t^{-2}(T)=2c_0^2\left[(1+b)+\delta_1\,\sqrt{(1+b)^2-\frac{T}{8c_0^2}}\right],
	\end{align}
	where $\delta_1=\pm 1$.~Equation~\eqref{404atdef} for Equation~\eqref{440} is:
	\begin{align}\label{440at}
		A(T)=& {\frac{3}{2}T+ \frac{4(3b-4)(1+b)^2\,c_0^2}{(b-2)}\,\left[1+\delta_1\,\sqrt{1-\frac{T}{8c_0^2(1+b)^2}}\right]^2},
		\nonumber\\
		=& \frac{1}{(b-2)}\left[-T+C_1\left(1+\delta_1\,\sqrt{1-\frac{T}{C_2}}\right)\right],
		\nonumber\\
		=& { \frac{1}{(b-2)}\left[-T+C_1\,w_{\pm}(T)\right]} ,
	\end{align}
	where $C_1=(3b-4)C_2$, $C_2=8c_0^2(1+b)^2$, $b\neq 2$, and { $w_{\pm}(T)=1+\delta_1\,\sqrt{1-\frac{T}{C_2}}$ ($+$ for $\delta_1=1$ and $-$ for $\delta_1=-1$).} The scalar field for Equation~\eqref{440} is:
	\begin{align}\label{441}
		\phi(T)=& { \frac{p_0}{2^{p/2}c_0^{p}(1+b)^{p/2}}\,\left[w(T)\right]^{-p/2} .}
	\end{align}
	The potential $V(T)$ and $F(T)$ solutions are, for the following~subcases:
	\begin{enumerate}
		\item \textbf{{General:}} Equation~\eqref{401a} becomes:
		\begin{align}
			V(T)= -\frac{V_p}{2}\,\left[{ w_{\pm}(T)}\right]^{1-p} , \label{442}
		\end{align}
		where $p \neq 1$ and $V_p=\frac{p^2\,p_0^{2}}{2^{p-1}\,(p-1)\,c_0^{2p-2}}\,\left( b+p+3\right)(1+b)^{1-p}$. Equation~\eqref{442} will be constant for $b=-p-3$. By~substituting Equations \eqref{440at} and \eqref{442} into Equation~\eqref{404solution}, we find as a solution:
		
			\begin{align}\label{443a}
				F(T) =&  -{ \Lambda_0} +\left[-T+C_1\left({ w_{\pm}(T)}\right)\right]^{(2-b)}
				\,\left[{ y(T)}\right]^{\frac{(b-2)C_1}{2\left(C_1-2C_2\right)}}\,
				\nonumber\\
				&\,\times\,\left[\frac{T+\left(C_1-2C_2\right)\left({ w_{\pm}(T)}\right)}{T+\left(C_1-2C_2\right)\left({ 2-w_{\pm}(T)}\right)}\right]^{\frac{\delta_1\,(b-2)C_1}{2\left(C_1-2C_2\right)}}
				\,\Bigg[F_0\,C_2^{(2-b)/2}-\kappa\,V_p\,(b-2)
				\nonumber\\
				&\,\times\,\int_{T}\,dT'\,\left[{ w_{\pm}(T')}\right]^{1-p}\,\left[-T'+C_1\left({ w_{\pm}(T')}\right)\right]^{(b-3)}\,\left[{ y(T')}\right]^{-\frac{(b-2)C_1}{2\left(C_1-2C_2\right)}}\,
				\nonumber\\
				&\,\times\,\left[\frac{T'+\left(C_1-2C_2\right)\left({ w_{\pm}(T')}\right)}{T'+\left(C_1-2C_2\right)\left({ 2-w_{\pm}(T')}\right)}\right]^{-\frac{\delta_1\,(b-2)C_1}{2\left(C_1-2C_2\right)}}\Bigg] ,
			\end{align}
		{where $y(T)=\frac{-C_2\,T}{C_1\left(C_1-2C_2\right)+C_2\,T}$.} There is no general solution for Equation~\eqref{443a}, { but} for the specific~cases:
		\begin{itemize}
			\item $C_1=0$ and $\delta_1=1$:
			\begin{align}\label{443aa}
				F(T) =&  -{ \Lambda_0} +F_0\,(-T)^{(2-b)}
				\nonumber\\
				&\,-\kappa\,V_p\,(2)^{1-p}\,_3F_2\left(b-2,\frac{p}{2},\frac{p-1}{2};p,b-1;\frac{T}{C_2}\right) .
			\end{align}
			
			\item $C_1=0$ and $\delta_1=-1$:
			\begin{align}\label{443ab}
				F(T) =&  -{ \Lambda_0} +F_0\,(-T)^{(2-b)}-\kappa\,V_p\,\frac{(b-2)(-2C_2)^{p-1}}{(b-p-1)}\,(-T)^{1-p}
				\nonumber\\
				&\,\times\,\,_3F_2\left(b-p-1,\frac{2-p}{2},\frac{1-p}{2};b-p,2-p;\frac{T}{C_2}\right) .
			\end{align}
			
			\item { The $C_1=2C_2$ solutions are presented in Appendix \ref{appena2}.}
			
		\end{itemize}

		\item \textbf{{Equation}~\eqref{401b} {potential:}} This equation, in terms of Equation \eqref{441}, is ($p\gg 1$):
		\begin{align}
			V(T)= -\frac{V_{\infty}}{2}\,\left[{ w_{\pm}(T)}\right]^{-p} ,\label{444}
		\end{align}
		where $V_{\infty}=\frac{p^2\,p_0^2\,(1+b)^{-p}}{2^{p}c_0^{2p}}$. By~substituting Equations \eqref{440}, \eqref{440at}, and \eqref{444} into Equation~\eqref{404solution}, we find as a solution:
		\vspace{-12pt}
			\begin{align}\label{446}
				F(T) =&    -{ \Lambda_0} +\left[-T+C_1\left({ w_{\pm}(T)}\right)\right]^{(2-b)} \,\left[{ y(T)}\right]^{\frac{(b-2)C_1}{2\left(C_1-2C_2\right)}}\,
				\nonumber\\
				&\,\times\,\left[\frac{T+\left(C_1-2C_2\right)\left({ w_{\pm}(T)}\right)}{T+\left(C_1-2C_2\right)\left({ 2-w_{\pm}(T)}\right)}\right]^{\frac{\delta_1\,(b-2)C_1}{2\left(C_1-2C_2\right)}}
				\,\Bigg[F_0\,C_2^{(2-b)/2}-\kappa\,V_{\infty}\,(b-2)
				\nonumber\\
				&\,\times\,\int_{T}\,dT'\,\left[{ w_{\pm}(T')}\right]^{-p}\,\left[-T'+C_1\left({ w_{\pm}(T')}\right)\right]^{(b-3)}\,\left[{ y(T')}\right]^{-\frac{(b-2)C_1}{2\left(C_1-2C_2\right)}}\,
				\nonumber\\
				&\,\times\,\left[\frac{T'+\left(C_1-2C_2\right)\left({ w_{\pm}(T')}\right)}{T'+\left(C_1-2C_2\right)\left({ 2-w_{\pm}(T')}\right)}\right]^{-\frac{\delta_1\,(b-2)C_1}{2\left(C_1-2C_2\right)}}\Bigg] .
			\end{align}
		There is no general solution for Equation~\eqref{446}. There are solutions for the~ following cases:
		\newpage			
		\begin{itemize}
			\item $C_1=0$ and $\delta_1=1$:
				\begin{align}\label{446aa}
					F(T) \approx & -{ \Lambda_0} +\Bigg[F_0\,(-T)^{(2-b)}+\frac{\kappa\,V_{\infty}}{2^p}\,_3F_2\left(b-2,\frac{p}{2},\frac{p}{2};b-1,p;\frac{T}{C_2}\right)\Bigg] .
				\end{align}
			\item $C_1=0$ and $\delta_1=-1$:
				\begin{align}\label{446ab}
					F(T) \approx & -{ \Lambda_0} +\Bigg[F_0\,(-T)^{(2-b)}-\frac{\kappa\,V_{\infty}(-2C_2)^p}{p}\,(b-2)\,(-T)^{-p}
					\nonumber\\
					&\,\times\,_3F_2\left(-p,-\frac{p}{2},-\frac{p}{2};-p,p;\frac{T}{C_2}\right)\Bigg] .
				\end{align}

		\end{itemize}

		\item \textbf{{Equation}~\eqref{401c} {potential:}} This equation, in terms of Equation \eqref{441}, is ($p=1$):
		\begin{align}
			V(T)=\tilde{\phi}_0-\phi_0+\frac{p_0^{2}}{2}\,\left( b+4\right)\,\ln\,\left[{ w_{\pm}(T)}\right] , \label{447}
		\end{align}
		where $b \neq -1,\,-4$ and $\tilde{\phi}_0=\phi_0 -p_0^{2}\,\left( b+4\right)\,\ln\,\left[\frac{p_0}{2^{1/2}c_0\sqrt{1+b}}\right]$. By~substituting Equations \eqref{440}, \eqref{440at}, and \eqref{447} into Equation~\eqref{404solution}, we find as a solution:
		\vspace{-12pt}
		
			\begin{align}\label{449}
				F(T) =&   -{ \tilde{\Lambda}_0} +\left[-T+C_1\left({ w_{\pm}(T)}\right)\right]^{(2-b)}
				\,\left[{ y(T)}\right]^{\frac{(b-2)C_1}{2\left(C_1-2C_2\right)}}\,
				\nonumber\\
				&\,\times\,\left[\frac{T+\left(C_1-2C_2\right)\left({ w_{\pm}(T)}\right)}{T+\left(C_1-2C_2\right)\left({ 2-w_{\pm}(T)}\right)}\right]^{\frac{\delta_1\,(b-2)C_1}{2\left(C_1-2C_2\right)}}
				\,\Bigg[F_0\,C_2^{(2-b)/2}
				\nonumber\\
				&\,+\frac{\kappa\,p_0^{2}\,\left( b+4\right)(b-2)}{C_1}\,\int_{T}\,dT'\,\ln\,\left[{ w_{\pm}(T')}\right]\,\left[-T'+C_1\left({ w_{\pm}(T')}\right)\right]^{(b-3)}
				\nonumber\\
				&\,\times\,\left[{ y(T')}\right]^{-\frac{(b-2)C_1}{2\left(C_1-2C_2\right)}}\,\left[\frac{T'+\left(C_1-2C_2\right)\left({ w_{\pm}(T')}\right)}{T'+\left(C_1-2C_2\right)\left({ 2-w_{\pm}(T')}\right)}\right]^{-\frac{\delta_1\,(b-2)C_1}{2\left(C_1-2C_2\right)}}\Bigg] .
			\end{align}
		There is no general solution for Equation~\eqref{449}, { but only for $C_1 \neq 0$ cases, as presented in Appendix \ref{appena2}.}

	\end{enumerate}

	{
		\item \textbf{{Late Cosmology} $t\,\rightarrow\,\infty$ {limit:}} 
		
		In this case, there are some subcases according to Equation~\eqref{403}, such as:
		\begin{enumerate}
			\item ${\bf c>1}$: The Equation~\eqref{403} $t(T)$ relation will be, for the following cases:
			\begin{align}\label{470}
				0 \approx & 2c\left(c+2b\right)\,t^{-2} -T,
				\nonumber\\
				\Rightarrow\,& t^{-2}(T)\approx \frac{T}{2c\left(c+2b\right)}.
			\end{align}
			If $t(T)\,\rightarrow\,\infty$, we find from Equation~\eqref{470} that $T\,\rightarrow\,0$. Then, Equation~\eqref{404atdef} becomes:
			\begin{align}\label{470at}
				A(T)\approx &\,{\frac{3}{2} T +\frac{(3b-2c)}{c_0^2\,(b-c)\,\left(2c\left(c+2b\right)\right)^{c}}\,T^c },
				\nonumber\\
				\rightarrow &\,\frac{3}{2} T \quad \text{when}\quad T\,\rightarrow\,0.
			\end{align}
			For any $V(T)$ potential and by inserting Equations \eqref{470} and \eqref{470at} into Equation~\eqref{404solution}, we find, as the $F(T)$ solution:
			\begin{align}\label{470solution}
				F(T) =& -{ \Lambda_0} + T^{\frac{2}{3}}\,\Bigg[F_0+\frac{4\kappa}{3}\,\int_{T}\,dT'\,V(T')\, T'^{-\frac{5}{3}}\Bigg] .
			\end{align} 
			The Equation~\eqref{470solution} solutions for $\phi(T)=p_0\,\left(2c\left(c+2b\right)\right)^{p/2}\,T^{-p/2}$ are as follows:
			\begin{itemize}
				\item \textbf{{General}}: The $V(T)$ is exactly as Equation~\eqref{422}, with $V_p=p^2\,p_0^2\,\frac{(b+2c+p-1)}{(p-1)}$ $(2c(c+2b))^{p-1}$ as the proportionality constant. Equation~\eqref{470solution} will~be:
				\begin{itemize}
					\item[$-$] $p\neq \frac{1}{3}$:
					\begin{align}\label{470aa}
						F(T) =& -{ \Lambda_0} + F_0\,T^{\frac{2}{3}}-\frac{4\kappa}{\left(1-3p\right)}\,T^{1-p} .
					\end{align}
					
					\item[$-$] $p= \frac{1}{3}$:
					\begin{align}\label{470ab}
						F(T) =& -{ \Lambda_0} + T^{\frac{2}{3}}\,\Bigg[F_0-\frac{4\kappa}{3}\,\ln (T)\Bigg] .
					\end{align}
				\end{itemize}
				
				\item \textbf{{Equation}~\eqref{401b} {potential}}: The $V(T)$ is exactly as Equation~\eqref{424}, with $V_{\infty}=p^2\,p_0^2\,(2c(c+2b))^{p}$ as the proportionality constant. Equation~\eqref{470solution} will~be:
				\begin{itemize}
					\item[$-$] $p\neq -\frac{2}{3}$:
					\begin{align}\label{470ba}
						F(T) =& -{ \Lambda_0} + F_0\,T^{\frac{2}{3}}+\frac{2\kappa\,V_p}{\left(2+3p\right)}\,T^{-p} .
					\end{align}
					
					\item[$-$] $p= -\frac{2}{3}$:
					\begin{align}\label{470bb}
						F(T) =& -{ \Lambda_0} + T^{\frac{2}{3}}\,\Bigg[F_0-\frac{2\kappa\,V_p}{3}\,\ln (T)\Bigg] .
					\end{align}
					
				\end{itemize}
				
				\item \textbf{{Equation}~\eqref{401c} {potential}}: The $V(T)$ is exactly $V(T)=\tilde{\phi}_0-\phi_0+\frac{p_0^2}{2} (b+2c)\,\ln (T)$, a~very similar form to Equation~\eqref{427}. Equation~\eqref{470solution} will be ($p=1$):
				\begin{align}\label{470c}
					F(T) =& -{ \tilde{\Lambda}_0} + F_0\,T^{\frac{2}{3}}-\kappa\,p_0^2(b+2c)\,\left(\ln\,(T)+\frac{3}{2}\right) .
				\end{align}
				
			\end{itemize}
			
			\item {${\bf c=1}$}: The $t(T)$ solution will be described by Equation~\eqref{420}, $A(T)$ by Equation~\eqref{420at}, and $F(T)$ by Equations \eqref{423a}, \eqref{426}, and \eqref{429} under the $t\,\rightarrow\,\infty$ limit.
			
			\item {${\bf 0<c<1}$} ($c\neq 0$): Equation~\eqref{403} leads to:
			\begin{align}\label{471}
				0 \approx &  \frac{2}{c_0^2}\,t^{-2c}+T,
				\nonumber\\
				\Rightarrow\,& t^{-2c}(T)\approx \frac{c_0^2}{2}\,(-T).
			\end{align}
			If $t(T)\,\rightarrow\,\infty$, we find that $T\,\rightarrow\,0$ from Equation~\eqref{471}. Equation~\eqref{404atdef} becomes:
			\begin{align}\label{471at}
				A(T)\approx {\frac{c}{2\,(c-b)}\,T} \quad \text{when}\quad T\,\rightarrow\,0,
			\end{align}
			where $b\neq c$. By~inserting Equations \eqref{471} and \eqref{471at} into Equation~\eqref{404solution}, we find as the $F(T)$ solution for any potential:
			\begin{align}\label{471solution}
				F(T) =& -{ \Lambda_0} + T^{\frac{2\,(c-b)}{c}}\,\Bigg[F_0+\frac{4\kappa\,(c-b)}{c}\,\int_{T}\,dT'\,V(T')\,T'^{\frac{2b-3c}{c}}\Bigg] .
			\end{align} 
			The Equation~\eqref{471solution} solutions for $\phi(T)=p_0\,2^{p/2c}\,c_0^{-p/c}\,(-T)^{-p/2c}$ are as follows:
			\begin{itemize}
				\item \textbf{{General}}: Equation~\eqref{401a} becomes:
				\begin{align}\label{471apot}
					V(T) = -\frac{V_p}{2}\,(-T)^{\frac{(1-p)}{c}}  ,
				\end{align}
				where $V_p=\frac{p^2\,p_0^{2}}{(p-1)}\,\left( b+2c+p-1\right)2^{\frac{p-1}{c}}\,c_0^{\frac{2(1-p)}{c}}$, and Equation~\eqref{471solution} will~be:
				\begin{itemize}
					\item[$-$] {${\bf p\neq 1+2(b-c)}$}:
					\begin{align}\label{471aa}
						F(T) =& -{ \Lambda_0} + F_0\,T^{\frac{2\,(c-b)}{c}}-\frac{2\kappa\,(c-b)\,V_p}{(2b-2c-p+1)}\,(-T)^{\frac{(1-p)}{c}} .
					\end{align}
					
					\item[$-$] {${\bf p= 1+2(b-c)}$:}
					\begin{align}\label{471ab}
						F(T) =& -{ \Lambda_0} + T^{\frac{2\,(c-b)}{c}}\,\Bigg[F_0-\frac{2\kappa\,(c-b)\,V_p}{c}\,\ln(T)\Bigg] .
					\end{align}
				\end{itemize}

				\item \textbf{{Equation}~\eqref{401b} {potential}}: This equation, in terms of the $\phi(T)$, is:
				\begin{align}\label{471bpot}
					V(T) =  -\frac{V_{\infty}}{2}\,(-T)^{-p/c} ,
				\end{align}
				where $V_{\infty}=p^2\,p_0^2\,2^{p/c}\,c_0^{-2p/c}$, and Equation~\eqref{471solution} will~be:
				\begin{itemize}
					\item ${\bf p\neq 2(b-c)}$
					\begin{align}\label{471ba}
						F(T) =& -{ \Lambda_0} + F_0\,T^{\frac{2\,(c-b)}{c}}-\frac{2\kappa\,(c-b)\,V_{\infty}}{(2b-2c-p)}\,(-T)^{-\frac{p}{c}} .
					\end{align}
					
					\item ${\bf p=2(b-c)}$
					\begin{align}\label{471bb}
						F(T) =& -{ \Lambda_0} + T^{\frac{2\,(c-b)}{c}}\,\Bigg[F_0-\frac{2\kappa\,(c-b)\,V_{\infty}}{c}\,\ln(T)\Bigg] .
					\end{align}
				\end{itemize}

				\item \textbf{{Equation}~\eqref{401c} {potential}}: This equation, in terms of the $\phi(T)$, is ($p=1$):
				\begin{align}\label{471cpot}
					V(T) =\tilde{\phi}_0-\phi_0 +\frac{p_0^{2}}{2c}\,\left( b+2c\right)\,\ln\,(-T) ,
				\end{align}
				where $\tilde{\phi}_0=\phi_0-p_0^{2}\,\left( b+2c\right)\,\ln\,\left(p_0\,2^{1/2c}\,c_0^{-1/c}\right)$, and Equation~\eqref{471solution} will be:
				\begin{align}\label{471c}
					F(T) =& -{ \tilde{\Lambda}_0} + F_0\,T^{\frac{2\,(c-b)}{c}}-\frac{\kappa\,p_0^2}{c}\,\left( b+2c\right)\,\left(\ln(-T)-\frac{c}{2(b-c)} \right) .
				\end{align}
			\end{itemize}
			
			
			\item {${\bf c<0}$}: By setting $c=-|c|$ in Equations \eqref{403} and \eqref{404atdef}, we find Equations \eqref{471} and \eqref{471at} as exactly $t^{2|c|}(T)\approx \frac{c_0^2}{2}\,(-T)$ and $A(T)\approx {\frac{|c|}{2\,(|c|+b)}\,T}$. We also find that, for $T\,\rightarrow\,-\infty$ and then $A(T)\,\rightarrow\,-\infty$ under the $t\,\rightarrow\,\infty$ limit. In~this case, we will find that all $F(T)$ solutions are under the Equation~\eqref{471solution} form, and we will recover the Equations \eqref{471aa}--\eqref{471c} teleparallel $F(T)$ solutions by only performing the $c=-|c|$ transformation.

			\item {${\bf c=-2b}$}: We find that $t(T)$ is Equation~\eqref{405} and $A(T)=\frac{T}{3}$, leading to Equations \eqref{408}, \eqref{409c}, and \eqref{409g} as $F(T)$ solutions under the $t\,\rightarrow\,\infty$ limit. We can summarize the situation~as follows:
			\begin{itemize}
				\item $b>0$: $T\,\rightarrow\,-\infty$, $A(T)\,\rightarrow\,-\infty$ and $F(T)\,\rightarrow\,-\infty$,
				
				\item $b<0$: $T\,\rightarrow\,0y$, $A(T)\,\rightarrow\,0$ and $F(T)\,\rightarrow\,0$.
			\end{itemize}

		\end{enumerate}

		\item \textbf{{Early Cosmology} $t\,\rightarrow\,0$ {limit:}} For this limit,~Equation~\eqref{403} will lead~to:
		\begin{enumerate}
			\item {${\bf c>1}$:} $t^{-2c}(T)$ and $A(T)$ are, respectively, Equations \eqref{471} and \eqref{471at}, but~$t^{-2c}(T)\,\rightarrow\,\infty$ and $A(T)\,\rightarrow\,\infty$ when $T\,\rightarrow\,-\infty$. We will then recover the same $F(T)$ solution expressions as Equations \eqref{471solution}--\eqref{471c}, but~under the $T\,\rightarrow\,-\infty$ limit.

			\item {${\bf 0<c<1}$} and  {${\bf c<0}$}: $t^{-2}(T)$ and $A(T)$ are, respectively, Equations \eqref{470} and \eqref{470at}, but~$t^{-2}(T)\,\rightarrow\,\infty$ and $A(T)\,\rightarrow\,\infty$ when $T\,\rightarrow\,\infty$. We will then recover the same $F(T)$ solution expressions as Equations \eqref{470solution}--\eqref{470c}, but~under the $T\,\rightarrow\,\infty$ limit.
			
		\end{enumerate}

	}
	
\end{enumerate}

For other values of $b$, $c$, and/or subcases,~most Equation~\eqref{404solution} integral cases will not lead to a closed and analytic $F(T)$ solutions, as in ref.~\cite{nonvacKSpaper}. { All the previous new teleparallel $F(T)$ solutions are new results, and some of those are comparable to several $F(T)$ solutions found in ref.~\cite{nonvacKSpaper} for some perfect fluid sources. Some $c=1$ and $t\,\rightarrow\,\infty$ $F(T)$ solutions can also be compared to some TRW $F(T)$ solutions found in refs.~\cite{coleylandrygholami,scalarfieldTRW,FTBcosmogholamilandry} for perfect fluid and also scalar field sources, a~typical case of isotropic cosmological spacetime. This last limit is possible because of additional symmetries (six KVs for TRW spacetime instead of four in a KS spacetime) from the linear isotropy group, as mentioned in Section~\ref{sect1}.}

\subsection{Exponential Ansatz~Solutions}\label{sect32}

Equation~\eqref{301a}, in terms of Equation \eqref{402}, leads to the characteristic equation:
\begin{align}\label{451}
	e^{-2ct(T)} =& \frac{c_0^2}{2}\left(T_0 -T\right),
\end{align}
where $c\neq 0$ and $T_0=2c\left(c+2b\right)$ ($c=0$ subcase leads to GR solutions). By~substituting Equation~\eqref{451} into Equation~\eqref{305}, we find the Equation~\eqref{404} DE form with
\begin{align}\label{451atdef}
	A(T) = {\frac{c}{2(b-c)}\left[\left(\frac{3b}{2}-c\right)\,T_0 -T \right]}=\frac{c}{2(b-c)}(T_1-T),
\end{align}
where $T_1=c\,\left(3b-2c\right)\left(c+2b\right)$, and the solution is also described by Equation~\eqref{404solution}. The~subcases~are:
\begin{enumerate}
	\item {${\bf c=-2b}$}: $T_0=0$, and Equation~\eqref{451} is simplified to:
	\begin{align}\label{453}
		e^{4bt(T)} =& \frac{c_0^2}{2}\left(-T\right),
	\end{align}
	where $T\leq 0$. Then, Equation~\eqref{451atdef} is simplified to $A(T)=\frac{T}{3}$; the~scalar field will be $\phi(T)=\frac{p_0}{(4b)^p}\,\left(\ln\left(\frac{c_0^2}{2}\,(-T)\right)\right)^p$. The~potential $V(T)$ and $F(T)$ solutions are, for the following~subcases:
	\begin{enumerate}
		
		\item \textbf{{General}:} Equation~\eqref{402a} becomes:
		\begin{align}
			V(T)=\frac{V_{1p}}{2}\, \left(\ln\left(\frac{c_0^2}{2}(-T)\right)\right)^{2p-1}-\frac{V_{2p}}{2}\,\left(\ln\left(\frac{c_0^2}{2}\,(-T)\right)\right)^{2p-2} , \label{454}
		\end{align}
		where $V_{1p}=\frac{3 p^2\,p_0^{2}}{2(2p-1)(4b)^{2p-2}}$ and $V_{2p}=\frac{p^2\,p_0^{2}}{(4b)^{2p-2}}$. By~substituting $A(T)$ and  Equation~\eqref{454} into Equation~\eqref{404solution}, we find as a solution:
		
			\begin{align}\label{456}
				F(T)=& -{ \Lambda_0}+F_0\,T^3+3\kappa\,T^3\left[ V_{1p}\,\mathcal{N}_{2p-1}\left(-\frac{c_0^2}{2},T\right)-V_{2p} \,\mathcal{N}_{2p-2}\left(-\frac{c_0^2}{2},T\right) \right],
			\end{align}
		where $F_0$ is an integration constant, $p\neq \frac{1}{2}$, and $\mathcal{N}_{k}\left(-\frac{c_0^2}{2},T \right)$ is a new special function class defined as:
		\begin{equation}\label{457}
			\mathcal{N}_{k}\left(B_1,x\right)=\int\,dx\,\frac{\left(\ln\left(B_1\,x\right)\right)^{k}}{x^4}.
		\end{equation}
		{ Some simple values of Equation~\eqref{457} are shown in the Table \ref{table1}.}
			\begin{table}
				\setlength{\tabcolsep}{5.5mm}
				\caption{{Some values} 
					of Equation~\eqref{457} $\mathcal{N}_{k}\left(B_1,x\right)$ special~functions.}
				\begin{tabular}{|c|c|}
					\hline
					\boldmath{$k$} & \boldmath{$\mathcal{N}_{k}\left(B_1,x\right)$} \\	
					\hline
					$0$ &  $-\frac{1}{3x^3}$\\
					\hline
					$1$ & $-\frac{1}{3x^3}\,\left[\ln(A\,x)+\frac{1}{3}\right]$ \\
					\hline
					$2$ & $-\frac{1}{3x^3}\,\left[\ln^2(A\,x)+\frac{2}{3}\ln(A\,x)+\frac{2}{9}\right]$ \\
					\hline
					$3$ & $-\frac{1}{3x^3}\,\left[\ln^3(A\,x)+\ln^2(A\,x)+\frac{2}{3}\ln(A\,x)+\frac{2}{9}\right]$ \\
					\hline
					$4$ & $-\frac{\ln^4(A\,x)}{3x^3}-\frac{4}{9x^3}\,\left[\ln^3(A\,x)+\ln^2(A\,x)+\frac{2}{3}\ln(A\,x)+\frac{2}{9}\right]$ \\
					\hline
					$5$ & $-\frac{1}{3x^3}\,\left[\ln^5(A\,x)+\frac{5}{3}\ln^4(A\,x)\right]-\frac{20}{27x^3}\,\left[\ln^3(A\,x)+\ln^2(A\,x)+\frac{2}{3}\ln(A\,x)+\frac{2}{9}\right]$ \\
					\hline	
					$6$ & $-\frac{1}{3x^3}\,\left[\ln^6(A\,x)+2\ln^5(A\,x)+\frac{10}{3}\ln^4(A\,x)\right]$\\
					& $-\frac{40}{27x^3}\,\left[\ln^3(A\,x)+\ln^2(A\,x)+\frac{2}{3}\ln(A\,x)+\frac{2}{9}\right]$ \\
					\hline	
				\end{tabular}
				\label{table1}
			\end{table}
		
		\item \textbf{Equation~\eqref{401b} potential:} This equation, in terms of the Equation \eqref{453} scalar field, is ($p\gg 1$):
		\begin{align}
			V(T)= -\frac{V_{\infty}}{2}\,\left(\ln\left(\frac{c_0^2}{2}\,(-T)\right)\right)^{2p} ,\label{458}
		\end{align}
		where $V_{\infty}=\frac{p^2\,p_0^2}{(4b)^{2p}}$. By~substituting $A(T)=\frac{T}{3}$ and \eqref{458} into Equation~\eqref{404solution}, we find as a solution:
		\begin{align}\label{459b}
			F(T) =&   -{ \Lambda_0}+F_0\,T^3 -3\kappa\,V_{\infty}\,T^3\,\mathcal{N}_{2p}\left(-\frac{c_0^2}{2},T\right),
		\end{align}
		where $\mathcal{N}_{2p}\left(-\frac{c_0^2}{2},T\right)$ is defined by Equation~\eqref{457}.

		\item \textbf{Equation~\eqref{402b} potential:} This equation, in terms of the Equation \eqref{453} scalar field, is ($p=1$):
		\begin{align}
			V(T)=\tilde{\phi}_0-\phi_0 +\frac{3p_0^2}{4}\,\ln(-T) , \label{458a}
		\end{align}
		where $\tilde{\phi}_0=\phi_0+\frac{3p_0^2}{4}\,\ln\left(\frac{c_0^2}{2}\right)$. By~substituting $A(T)=\frac{T}{3}$ and \eqref{458a} into Equation~\eqref{404solution}, we find as a solution:
		\begin{align}\label{459c}
			F(T) =&   -{ \tilde{\Lambda}_0} +F_0\,T^3+\frac{9\kappa\,p_0^2}{2}\,T^3\,\mathcal{N}_{1}\left(-1,T\right) ,
		\end{align} 
		where $\mathcal{N}_{1}\left(-1,T\right)$ is defined by Equation~\eqref{457}.

	\end{enumerate}

	\item ${\bf c\neq -2b}$ (General case): From Equation~\eqref{451}, we find that the scalar field is:
	\begin{equation}\label{460}
		\phi(T)=\frac{p_0}{(-2c)^p}\,\left[\ln\left(\frac{c_0^2}{2}\left(T_0 -T\right)\right) \right]^p.
	\end{equation}
	For potential $V(T)$ and $F(T)$ solutions, we find for the following~subcases:
	\begin{enumerate}
		\item \textbf{General:} Equation~\eqref{402a} becomes:
		\small
		\begin{align}
			V(T)=&-\frac{V_{1p}}{2}\,\left[\ln\left(-\frac{c_0^2}{2}\left(T-T_0\right)\right) \right]^{2p-1}
			-\frac{V_{2p}}{2}\,\left[\ln\left(-\frac{c_0^2}{2}\left(T-T_0\right)\right) \right]^{2p-2} , \label{461}
		\end{align}
		\normalsize
		where $V_{1p}=\frac{2p^2\,p_0^{2}(b+2c)}{(2p-1)(-2c)^{2p-1}}$ and $V_{2p}=\frac{p^2\,p_0^{2}}{(-2c)^{2p-2}}$. By~substituting  Equations \eqref{451atdef} and \eqref{461} into Equation~\eqref{404solution}, we find as a solution:
		\small
		\begin{align}\label{463}
			F(T)=& -{ \Lambda_0} +F_0\,\left(T-T_1\right)^{2(c-b)/c}-\frac{2\kappa\,(c-b)}{c}\,\left(T-T_1\right)^{2(c-b)/c}
			\nonumber\\
			&\;\times\,\Bigg[V_{1p}\,\mathcal{Q}_{2p-1}\left(-\frac{c_0^2}{2},T_0,T_1,T\right)+V_{2p}\,\mathcal{Q}_{2p-2}\left(-\frac{c_0^2}{2},T_0,T_1,T\right) \Bigg],
		\end{align}
		\normalsize
		where $\mathcal{Q}_{2p}\left(B_1,B_2,B_3,x\right)$ is defined as:
		\begin{equation}\label{457gen}
			\mathcal{Q}_{k}\left(B_1,B_2,B_3,x\right)=\int\,dx\,\left(\ln\left(B_1\,(x-B_2)\right)\right)^{k}\left(x-B_3\right)^{2b/c-3}.
		\end{equation}
		Equation~\eqref{457gen} is a generalization of the Equation~\eqref{457} special function class. { Some values of Equation~\eqref{457gen} are shown in the Table \ref{table2}.}
			\begin{table}
				\setlength{\tabcolsep}{2.9mm}
				\caption{{Some values} 
					of Equation~\eqref{457gen} for $\mathcal{Q}_{k}\left(B_1,B_2,B_3,x\right)$ special~functions.}
				\begin{tabular}{|c|c|c|}
					\hline
					\boldmath{$k$} & \boldmath{$b/c$} &  \boldmath{$\mathcal{Q}_{k}\left(B_1,B_2,B_3,x\right)$} \\
					\hline
					$0$ & all & $\frac{c}{2(b-c)}\left(x-B_3\right)^{2(b-c)/c}$ \\
					\hline
					$1$ & $0$ & $\frac{2}{\left(B_2-B_3\right)^2}\left[\frac{\left(B_2-B_3\right)}{\left(x-B_3\right)}-\ln\left(B_1\,(x-B_3)\right)+\frac{(x-B_2)(B_2-2B_3+x)}{\left(x-B_3\right)^2}\ln\left(B_1\,(x-B_2)\right)\right] $ \\
					\hline
					$1$ & $1$ & $ dilog \left(\frac{x-B_{3}}{B_{2}-B_{3}}\right)+\ln\! \left(B_{1} \left(x-B_{2}\right)\right) \ln\! \left(\frac{x-B_{3}}{B_{2}-B_{3}}\right)$ \\
					\hline
					$1$ & $\frac{3}{2}$ & $\left(\ln\! \left(B_{1} \left(x-B_{2}\right)\right)-1\right) \left(x-B_{2}\right)$ \\
					\hline
					$1$ & $2$ & $\frac{\left(x-B_{2}\right)}{2} \left[\left(B_{2}-2 B_{3}+x\right) \ln\! \left(B_{1} \left(x-B_{2}\right)\right)-\frac{x}{2}-\frac{3 B_{2}}{2}+2 B_{3}\right]$\\
					\hline
					$2$ & $1$ & $\ln\! \left(B_{1} \left(x-B_{2}\right)\right)^{2} \ln\! \left(\frac{x-B_{3}}{B_{2}-B_{3}}\right)+2 \ln\! \left(B_{1} \left(x-B_{2}\right)\right) \mathrm{polylog}\! \left(2,\frac{-x+B_{2}}{B_{2}-B_{3}}\right)$\\
					& & $-2 \mathrm{polylog}\! \left(3,\frac{-x+B_{2}}{B_{2}-B_{3}}\right)$ \\
					\hline
					$2$ & $\frac{3}{2}$ & $\left(\ln\! \left(B_{1} \left(x-B_{2}\right)\right)^{2}-2 \ln\! \left(B_{1} \left(x-B_{2}\right)\right)+2\right) \left(x-B_{2}\right)$ \\
					\hline
					$2$ & $2$ & $\frac{\left(x-B_{2}\right)}{2} \left[\left(B_{2}-2 B_{3}+x\right) \ln\! \left(B_{1} \left(x-B_{2}\right)\right)^{2}+\left(-x-3 B_{2}+4 B_{3}\right) \ln\! \left(B_{1} \left(x-B_{2}\right)\right)\right]$\\
					& & $+\frac{\left(x-B_{2}\right)}{2}\left[\frac{x}{2}+\frac{7 B_{2}}{2}-4 B_{3}\right]$ \\
					\hline
				\end{tabular}
				\label{table2}
			\end{table}
		
		\item \textbf{Equation~\eqref{401b} potential:} This equation, in terms of the Equation \eqref{460} scalar field, is ($p\gg 1$):
		\begin{align}
			V(T)= -\frac{V_{\infty}}{2}\,\left[\ln\left(-\frac{c_0^2}{2}\left(T-T_0\right)\right) \right]^{2p} ,\label{464}
		\end{align}
		where $V_{\infty}=\frac{p_0^2\,p^2}{(-2c)^{2p}}$. By~substituting Equations \eqref{451atdef} and \eqref{464} into Equation~\eqref{404solution}, we find as a solution:
		\vspace{-12pt}
		
			\begin{align}\label{466}
				F(T) =&  -{ \Lambda_0} + \left(T-T_1\right)^{2(c-b)/c}\Bigg[F_0-\frac{2\kappa\,(c-b)\,V_{\infty}}{c}\,\mathcal{Q}_{2p}\left(-\frac{c_0^2}{2},T_0,T_1,T\right)\Bigg],
			\end{align}
		where $\mathcal{Q}_{2p}\left(-\frac{c_0^2}{2},T_0,T_1,T\right)$ is defined by Equation~\eqref{457gen}.
		
		\item \textbf{Equation~\eqref{402b} potential:} This equation, in terms of the Equation \eqref{460} scalar field, is ($p=1$):
		\begin{align}
			V(T)=\tilde{\phi}_0-\phi_0+\frac{p_0^2\,(b+2c)}{2c}\,\ln\left(T_0 -T\right) , \label{467}
		\end{align}
		where $\tilde{\phi}_0=\phi_0+\frac{p_0^2\,(b+2c)}{2c}\,\ln\left(\frac{c_0^2}{2}\right)$. By~substituting Equation \eqref{467} into Equation~\eqref{404solution}, we find as a solution:
		\vspace{-12pt}
			\begin{align}\label{469}
				F(T) =&  - { \tilde{\Lambda}_0}+\left(T-T_1\right)^{2(c-b)/c}\Bigg[F_0+\frac{2\kappa\,p_0^2\,(c-b)(b+2c)}{c^2}\,\mathcal{Q}_{1}\left(-1,T_0,T_1,T\right)\Bigg],
			\end{align}
		where $\mathcal{Q}_{1}\left(-1,T_0,T_1,T\right)$ is defined by Equation~\eqref{457gen}.

	\end{enumerate}
	
	{
		\item \textbf{Late Cosmology $t\,\rightarrow\,\infty$ limit:} 
		\begin{enumerate}
			\item {${\bf c>0}$}: Equation~\eqref{451} will lead to $T=T_0=$ constant under this limit, a~GR solution similar to a Teleparallel de Sitter (TdS) spacetime ($c\neq -2b$ general case) \cite{TdSpaper}. For~$c=-2b$, we will find that $T=0$, a~null torsion scalar spacetime~\cite{landryvandenhoogen1}. In~the previous situations, $A(T)=$ constant according to Equation~\eqref{451atdef}, and we will only find GR~solutions.

			\item {${\bf c<0}$}: We will find that $\exp \left(2|c|t(T)\right)\,\rightarrow\,\infty$ and then $T\,\rightarrow\,-\infty$ under the same limit with Equation~\eqref{451}:
			\begin{align}\label{480}
				e^{2|c|t(T)} =& \frac{c_0^2}{2}\left(T_0 -T\right)\approx \frac{c_0^2}{2} (-T)\,\rightarrow\,\infty,
			\end{align}
			where $T_0=2|c|\left(|c|-2b\right)\ll |T|$. Equation~\eqref{451atdef} becomes:
			\begin{align}\label{480atdef}
				A(T) \approx \frac{|c|}{2(b+|c|)}\,T\,\rightarrow\,-\infty,
			\end{align}
			where $T_1=|c|\,\left(3b+2|c|\right)\left(|c|-2b\right)\ll |T|$. For~any $V(T)$ potential,~Equation~\eqref{404solution} will be:
			\begin{align}\label{480solution}
				F(T) =& -{ \Lambda_0} + T^{\frac{2(b+|c|)}{|c|}}\,\Bigg[F_0+\frac{4\kappa\,(b+|c|)}{|c|}\,\int_{T}\,dT'\,V(T')\,T'^{-\frac{2b+3|c|}{|c|}}\Bigg] .
			\end{align}
			Equation~\eqref{480solution} is for $\phi(T)=\frac{p_0}{(2|c|)^p}\,\left(\ln\left(\frac{c_0^2}{2}(-T)\right)\right)^p$:
			\begin{itemize}
				\item \textbf{General}: Equation~\eqref{402a} becomes Equation~\eqref{461}, with $T_1=T_0=0$, $V_{1p}=\frac{2p^2\,p_0^{2}\,(b-2|c|)}{(2p-1)\,(2|c|)^{2p-1}}$ and $V_{2p}=\frac{p^2\,p_0^{2}}{(2|c|)^{2p-2}}$. Then, Equation~\eqref{480solution} is:
				\begin{align}\label{480a}
					F(T) =& -{ \Lambda_0} + T^{\frac{2(b+|c|)}{|c|}}\,\Bigg[F_0-\frac{2\kappa\,(b+|c|)}{|c|}\,\Bigg[V_{1p}\,\mathcal{Q}_{2p-1}\left(-\frac{c_0^2}{2},0,0,T\right)
					\nonumber\\
					&\,+V_{2p}\,\mathcal{Q}_{2p-2}\left(-\frac{c_0^2}{2},0,0,T\right) \Bigg]\Bigg] ,
				\end{align}
				where $\mathcal{Q}_{2p}\left(B_1,0,0,x\right)$ is Equation~\eqref{457gen}.~Equation~\eqref{480a} is also a simplified form of Equation~\eqref{463}.
				
				\item \textbf{Equation~\eqref{401b} potential}: This equation becomes Equation~\eqref{464}, with $T_1=T_0=0$ and $V_{\infty}=\frac{p^2\,p_0^2}{(2|c|)^{2p}}$. Equation~\eqref{480solution} is:
				\begin{align}\label{480b}
					F(T) =& -{ \Lambda_0} + T^{\frac{2(b+|c|)}{|c|}}\,\Bigg[F_0-\frac{2\kappa\,(b+|c|)\,V_{\infty}}{|c|}\,\mathcal{Q}_{2p}\left(-\frac{c_0^2}{2},0,0,T\right)\Bigg] ,
				\end{align}
				where $\mathcal{Q}_{2p}\left(B_1,0,0,x\right)$ is defined by Equation~\eqref{457gen}.~Equation~\eqref{480b} is also a simplified form of Equation~\eqref{466}.
				
				\item \textbf{Equation~\eqref{402b} potential}: The $V(T)$ for $c=-|c|$ will exactly lead to Equation~\eqref{471cpot}, with  $\tilde{\phi}_0=\phi_0-\frac{p_0^2\,(b-2|c|)}{(2|c|)}\ln\left(\frac{c_0^2}{2}\right)$, and Equation~\eqref{480solution} will be Equation~\eqref{471c} for $c=-|c|$. This case proves that some $F(T)$ solutions will be invariant for any coframe ansatz. In~this case, the~power-law and exponential ansatzes both yield to the same $F(T)$ solution.
				
			\end{itemize}
			
		\end{enumerate}
		
		\item \textbf{Early Cosmology $t\,\rightarrow\,0$ limit:}
		For any non-zero value of $c$,~Equation~\eqref{451} leads to $T\,\rightarrow\,T_0-\frac{2}{c_0^2}=$ constant and then $A(T)=$ constant, a~GR solution. This is a TdS-like case model because of the constant torsion scalar~\cite{TdSpaper}.
		
	}
	
\end{enumerate}

All the teleparallel $F(T)$ solutions in this section are { new and comparable to those found in ref.~\cite{nonvacKSpaper}}. However, there are, in principle, other additional subcases leading to new $F(T)$ solutions.


\section{Exponential Scalar Field~Solutions}\label{sect4}

In this section, we will set an exponential scalar field $\phi(t)=p_0\,\exp\left(p\,t\right)$, where $p_0$ is a constant and $p$ is the exponential coefficient. The~conservation law defined by Equation~\eqref{302d} becomes:
\begin{align}
	0=	p\,\left(\ln(A_2\,A_3^2)\right)'\,\phi+p^2\,\phi+\frac{dV}{d\phi} , \label{500}
\end{align}
where $\left(\ln(A_2\,A_3^2)\right)'$ is still dependent on $\phi$ and the used ansatz for $A_2$ and $A_3$ components, as for Equation~\eqref{400}. Equation \eqref{500} is, at first glance, significantly simpler than Equation~\eqref{400}. The~possible ansatz~are as follows:
\begin{enumerate}
	\item \textbf{Power-law:} By substituting Equation~\eqref{401} into Equation~\eqref{500}, we find that:
	\begin{align}\label{501}
		0=& p\,(b+2c)\,\frac{\phi}{\ln \left(\phi/p_0\right)}+p^2\,\phi+\frac{dV}{d\phi} ,
		\nonumber\\
		& \Rightarrow\; V\left(\phi\right)=\phi_0 -\frac{p^2}{2}\,\phi^2+ p\,(b+2c)\,p_0^2\,Ei\left(-2\ln\left(\frac{\phi}{p_0}\right)\right).
	\end{align}
	Equation~\eqref{501}, by the last integral term, is a different kind of potential, and this last can be considered as an additional interaction term. Equation~\eqref{Quintessenceindex} for $\alpha_Q$ is, by substituting Equation~\eqref{501}:
	\begin{align}\label{501q}
		\alpha_Q =& -1+ \frac{p^2\phi^2}{\phi_0 + p\,(b+2c)\,p_0^2\,Ei\left(-2\ln\left(\frac{\phi}{p_0}\right)\right)}.
	\end{align}

	${\bf p\gg 1}$ \textbf{:} Equations \eqref{501} and \eqref{501q} become exactly Equations \eqref{401b} and \eqref{401bq}, where the $p\,(b+2c)\,p_0^2\,Ei\left(-2\ln\left(\frac{\phi}{p_0}\right)\right)$ term is negligible in this~case.
	
	\item \textbf{Exponential:} This is the most simple and naturally adapted ansatz for a pure exponential scalar field. By~substituting Equation~\eqref{402} into Equation~\eqref{500}, the~conservation law becomes:
	\begin{align}
		0=&	p\,(b+2c+p)\,\phi+\frac{dV}{d\phi} , 
		\nonumber\\	
		& \Rightarrow\; V\left(\phi\right)=\phi_0-\frac{p}{2}\,(b+2c+p)\,\phi^2 ,	\label{502}
	\end{align}
	and then Equation~\eqref{Quintessenceindex} will be:
	\begin{align}\label{502q}
		\alpha_Q =& -1+\frac{p^2\phi^2}{\phi_0-\frac{p}{2}\,(b+2c)\,\phi^2}.
	\end{align} 
	
	\begin{itemize}
		
		\item ${\bf p\gg 1}$ \textbf{:} Equations \eqref{502} and \eqref{502q} become Equations \eqref{401b} and \eqref{401bq}, with the same physical process~scenarios. 
		
		\item ${\bf p\,(b+2c+p)<0}$ \textbf{:} Equation~\eqref{502} is a simple harmonic oscillator (SHO) potential of angular frequency $\omega_{\phi}^2 = -p\,(b+2c+p)$. In such a case, the~scalar field will be:
		\begin{align}\label{503}
			\phi(t) = C_1\,\cos\left(\omega_{\phi}\,t\right) +C_2\,\sin\left(\omega_{\phi}\,t\right).
		\end{align}
		Equation~\eqref{503} is an oscillating scalar field, but~it is not really relevant as a scalar field source for usual cosmological solutions. Beyond~that, there are two situations leading to this~case:
		\begin{itemize}
			\item[$-$] $p<0$ and $p > -b-2c$.
			
			\item[$-$] $p>0$ and $p < -b-2c$.
		\end{itemize}
		This quadratic $V\left(\phi\right)$ potential case can be easily used for dark energy quintom oscillating models~\cite{quintom1,quintom2,quintom3}. The~Equation~\eqref{503} scalar field definition can only be relevant for this type of physical~model.

		\item \textbf{${\bf p=0}$ and/or ${\bf p=-b-2c}$:} Equation~\eqref{502} leads to a constant scalar field $V\left(\phi\right)=\phi_0$.

	\end{itemize}
\end{enumerate}

As for the power-law potentials obtained in Section \ref{sect3}, the~Equation~\eqref{500} solutions are going in the same direction as the first quintessence process studies~\cite{quintessencecmbpeak}. This situation is also confirmed separately by the Equations \eqref{401} and \eqref{402} ansatz approaches in this~section.



\subsection{Power-Law Ansatz~Solutions}\label{sect41}

By using Equations \eqref{403} and \eqref{404} with Equation~\eqref{401} ansatz, we  once again obtain the Equation~\eqref{404} DE form, with $A(T)$ defined by Equation~\eqref{404atdef}. By~using Equation~\eqref{501} and setting $V\left(\phi(T)\right)=\phi_0+V(T)$ in the Equation~\eqref{404} potential, the~general solution will be described again by Equation~\eqref{404solution}. We will find new $F(T)$ solutions by computing Equation~\eqref{404solution} for each $V(T)$ potential~case:
\begin{enumerate}
	\item ${\bf c=-2b}$: By using $A(T)=\frac{T}{3}$ and substituting Equation~\eqref{405} for the $t(T)$ solution, the~scalar field is $\phi(T)=p_0\,\exp\left(p\,\left(\frac{c_0^2}{2} (-T)\right)^{1/4b}\right)$, and Equation~\eqref{501} for $V(T)$ potential will be:
	\begin{align}
		V(T) = -\frac{p^2\,p_0^2}{2}\,\exp\left(2p\,\left(\frac{c_0^2}{2} (-T)\right)^{1/4b}\right)-3 p\,b\,p_0^2\,Ei\left(-2p\,\left(\frac{c_0^2}{2} (-T)\right)^{1/4b}\right). \label{511} 
	\end{align}
	By substituting $A(T)$ and Equation~\eqref{511} into Equation~\eqref{404solution}, we find as a solution:
	
	\vspace{-12pt}
	
		\begin{align}\label{512}
			F(T) =&  -{ \Lambda_0} + F_0\,T^3-\frac{\kappa\,p^2\,p_0^2}{\left(12 b-1\right)}\,\Bigg[{\left(2p \left(-\frac{c_{0}^{2}}{2} T\right)^{\frac{1}{4 b}}\right)}^{6 b} \exp{\left(-p \left(-\frac{c_{0}^{2}}{2} T\right)^{\frac{1}{4 b}} \right)} 
			\nonumber\\
			&\,\times\,{WhittakerM}\! \left(-6 b,-6 b+\frac{1}{2},2p \left(-\frac{c_{0}^{2}}{2} T\right)^{\frac{1}{4 b}}\right)+\left(1-12 b\right) \exp{\left(-2p \left(-\frac{c_{0}^{2}}{2} T\right)^{\frac{1}{4 b}}\right)}\Bigg]
			\nonumber\\
			&\,-\frac{6\kappa\,p\,p_0^2}{\left(b-\frac{1}{12}\right)}\,\Bigg[2 p\,b^2 \left(-\frac{c_{0}^{2}}{2} T\right)^{\frac{1}{4 b}} \,_3F_3\left(1,1,1-12 b;2,2,-12 b+2;-2p \left(-\frac{c_{0}^{2}}{2} T\right)^{\frac{1}{4 b}}\right) 
			\nonumber\\
			&\,-\left(b-\frac{1}{12}\right) \left( b \ln\! \left(2p \left(-\frac{c_{0}^{2}}{2} \right)^{\frac{1}{4 b}} \right)+\gamma  b+\frac{\ln\left(T\right)}{4}+\frac{1}{12}\right)\Bigg]
			, 
		\end{align}
	
	where ${WhittakerM}\!\left(B_1,B_2,x\right)=\exp\left(-\frac{x}{2}\right)\,x^{\frac{1}{2}+B_2}\,_1F_1\left(\frac{1}{2}+B_2-B_1;1+2B_2;x\right)$ is a special~function.
	
	${\bf p\gg 1}$ \textbf{:} By using $A(T)=\frac{T}{3}$ and substituting Equation~\eqref{405} into the Equation~\eqref{401b} potential, $V(T)$ becomes:
	\begin{align}
		V(T)= -\frac{p^2\,p_0^2}{2}\,\exp\left(2p\,\left(\frac{c_0^2}{2} (-T)\right)^{1/4b}\right). \label{513} 
	\end{align}
	By substituting Equation~\eqref{513} into Equation~\eqref{404solution}, we find as a solution:
	\begin{align}\label{514}
		F(T) \approx &  -{ \Lambda_0} + F_0\,T^3-\frac{\kappa\,p^2\,p_0^2}{\left(12 b-1\right) }\,\Bigg[{\left(2p \left(-\frac{c_{0}^{2}}{2} T\right)^{\frac{1}{4 b}} \right)}^{6 b} \exp{\left(-p \left(-\frac{c_{0}^{2}}{2} T\right)^{\frac{1}{4 b}}\right) }
		\nonumber\\
		&\,\times\, {WhittakerM}\! \left(-6 b,-6 b+\frac{1}{2},2p \left(-\frac{c_{0}^{2}}{2} T\right)^{\frac{1}{4 b}}\right)\Bigg].
	\end{align}

	\item ${\bf c=1}$: By using Equation~\eqref{420at} and substituting Equation~\eqref{420}, we find that:
	\begin{align}\label{530}
		\phi(T)=p_0\,\exp\left(\frac{\sqrt{2}p\left(1+2b - \frac{1}{c_0^2}\right)^{1/2}}{\sqrt{T}}\right) ,
	\end{align}
	and the Equation~\eqref{501} potential becomes:
	\begin{align}
		V(T)=&  -\frac{p^2\,p_0^2}{2}\,\exp\left(\frac{T_2}{\sqrt{T}}\right)
		+p\,(b+2)\,p_0^2\,Ei\left(-\frac{T_2}{\sqrt{T}}\right)  , \label{531}
	\end{align}
	where $T_2=2\sqrt{2}p\left(1+2b - \frac{1}{c_0^2}\right)^{1/2}$. By~substituting Equation~\eqref{531} into Equation~\eqref{404solution}, we find that:
	\begin{align}\label{532}
		F(T) =&   -{ \Lambda_0} + F_0\,T^{1/C}-\frac{\kappa\,p_0^2}{C}\Bigg[p^2\,T^{1/C}\,\int_{T}\,dT'\,T'^{-1/C-1}\,\exp\left(\frac{T_2}{\sqrt{T'}}\right)
		\nonumber\\
		&\,-2p\,(b+2)\,T^{1/C}\,\int_{T}\,dT'\,T'^{-1/C-1}\,Ei\left(-\frac{T_2}{\sqrt{T'}}\right)\Bigg].
	\end{align}
	Equation~\eqref{532} solutions are possible only for { non-zero specific values of $C$, such as, for example, $C=1$:}
	\begin{align}\label{532a}
		F(T) =&   -{ \Lambda_0} + F_0\,T-\kappa\,p_0^2\Bigg[\frac{2p^2}{T_{2}^{2}}\,\sqrt{T}\left({\sqrt{T}}-T_{2}\right)\,\exp{\left(\frac{T_{2}}{\sqrt{T}}\right)} 
		\nonumber\\
		&\,+\frac{2p\,(b+2)}{ T_{2}^{2}}\left[\left(\sqrt{T} T_{2}+T\right) \exp{\left(-\frac{T_{2}}{\sqrt{T}}\right)}+{Ei}\! \left(-\frac{T_{2}}{\sqrt{T}}\right) T_{2}^{2}\right] \Bigg].
	\end{align}
	There are several possible values of $C$ yielding to analytical $F(T)$ solutions, { as presented in Appendix \ref{appenb1}}.

	${\bf p\gg 1}$ \textbf{:} By using Equation~\eqref{420at} and substituting Equation~\eqref{420}, we find as the Equation~\eqref{401b} potential:
	\begin{align}
		V(T)=&  -\frac{p^2\,p_0^2}{2}\,\exp\left(\frac{T_2}{\sqrt{T}}\right). \label{533}
	\end{align}
	By substituting Equation~\eqref{533} into Equation~\eqref{404solution}, we find that:
	\begin{align}\label{534}
		F(T) =&   -{ \Lambda_0} + T^{1/C}\Bigg[F_0-\frac{\kappa\,p^2\,p_0^2}{C}\,\int_{T}\,dT'\,T'^{-1/C-1}\,\exp\left(\frac{T_2}{\sqrt{T'}}\right)\Bigg].
	\end{align}
	As for Equation~\eqref{532}, there is no general solution, but~we can solve some cases for $T_2\gg 1$ {, such as the $C=1$ case:}
	\begin{align}\label{534a}
		F(T) =&   -{ \Lambda_0} + F_0\,T-2\kappa\,p^2\,p_0^2\,\frac{T}{T_2^2}\left(1-\frac{T_2}{\sqrt{T}}\right)\,\exp\left(\frac{T_2}{\sqrt{T}}\right) .
	\end{align}
	There are several other possible $F(T)$ solutions arising from Equation~\eqref{534} {, as presented in Appendix \ref{appenb1}}.

	\item ${\bf c=-1}$: By using Equation~\eqref{430at} and substituting Equation~\eqref{430}, we find that:
	\begin{align}\label{540}
		\phi(T)=p_0\,\exp\left(\frac{c_0\,p}{2}\left[{ u_{\pm}(T)}\right]^{1/2}\right),
	\end{align}
	and the Equation~\eqref{501} potential becomes:
	\begin{align}
		V(T)=& -\frac{p^2\,p_0^2}{2}\,\exp\left(c_0\,p\left[{ u_{\pm}(T)}\right]^{1/2}\right)+p\,(b-2)\,p_0^2\,Ei\left(-c_0\,p\left[{ u_{\pm}(T)}\right]^{1/2}\right). \label{541}
	\end{align}
	By substituting Equation~\eqref{541} into Equation~\eqref{404solution}, we find that:
	\vspace{-12pt}
	
		\begin{align}\label{542}
			F(T) =&    -{ \Lambda_0} +{\left[{ 3T+2C_1\,u_{\pm}(T)}\right]^{\frac{2(3-2C_1)}{3(3-4C_1)}}\,\left[{ u_{-}(T)}\right]^{-\frac{4C_1\delta_1}{3(3-4C_1)}}}
			\Bigg[F_0-2\kappa\,p_0^2
			\nonumber\\
			&\,\times\,\int_{T}\,dT'\,\Bigg[p^2\,\exp\left(c_0\,p\left[{ u_{\pm}(T')}\right]^{1/2}\right)-2p\,(b-2)\,Ei\left(-c_0\,p\left[{ u_{\pm}(T')}\right]^{1/2}\right)\Bigg]
			\nonumber\\
			&\,\times\,{\left[{ 3T'+2C_1\,u_{\pm}(T')}\right]^{-\frac{2(3-2C_1)}{3(3-4C_1)}-1}\,\left[{ u_{-}(T')}\right]^{\frac{4C_1\delta_1}{3(3-4C_1)}}}\Bigg].
		\end{align}
	As in Section \ref{sect31}, there is no general solution for Equation~\eqref{542}. However, there are specific cases yielding to analytical~solutions:
	\begin{itemize}
		\item $C_2=0$ and $\delta_1=-1$:
		\vspace{-12pt}
			\begin{align}\label{542a}
				F(T) =&    -{ \Lambda_0} +F_0\,T^{\frac{2}{(3-4C_1)}}
				\nonumber\\
				&\, +2\kappa\,p_0^2\Bigg[\frac{p^2}{(4C_1-3)}\,T^{\frac{2}{(3-4C_1)}}\,\mathcal{R}\left(p,C_1,-2c_0^2\,T\right)+p\,(b-2)
				\nonumber\\
				&\,\times\,\Bigg(\frac{4c_{0} p}{\left(4 C_{1}+1\right)} \sqrt{-2T}\,_3F_3\left(1,1,\frac{4 C_{1}+1}{4 C_{1}-3};2,2,\frac{8 C_{1}-2}{4 C_{1}-3};-c_{0} p \sqrt{-2T}\right)
				\nonumber\\
				&-\left(\Psi \! \left(\frac{4}{4 C_{1}-3}\right)+\gamma -\Psi \! \left(\frac{1+4 C_{1}}{4 C_{1}-3}\right)+\ln\! \left(c_{0} p \sqrt{-2T}\right)\right) \Bigg) \Bigg],
			\end{align}
		where $\Psi(x)$ is the Digamma function and $\mathcal{R}\left(p,C_1,x\right)$ is defined by:
		\begin{align}\label{542aa}
			\mathcal{R}\left(p,C_1,x\right)=\int_0^x\,dx'\,x'^{-\frac{2}{(3-4C_1)}-1}\,\exp\left(p\,x'^{1/2}\right),
		\end{align}
		where $p\neq 0$ and $C_1 \neq \frac{3}{4}$. { Some simple values of Equation~\eqref{542aa} are shown in the Table \ref{table3}.}
		\begin{table}
			\setlength{\tabcolsep}{15.5mm}
			\caption{{Some values} 
				of Equation~\eqref{542aa} for the $\mathcal{R}\left(p,C_1,x\right)$ special~function.}
			\begin{tabular}{|c|c|}
				\hline
				\boldmath{$C_1$} & \boldmath{$\mathcal{R}\left(p,C_1,x\right)$} \\
				\hline
				$0$ & $\frac{9 \left({\mathrm e}^{p \mathrm{\sqrt{x}}} \left(-p \mathrm{\sqrt{x}}-\frac{1}{3}\right) \left(-p \mathrm{\sqrt{x}}\right)^{\frac{2}{3}}+p^{2} x \left(\Gamma \! \left(\frac{2}{3}\right)-\Gamma \! \left(\frac{2}{3},-p \mathrm{\sqrt{x}}\right)\right)\right)}{2 x^{\frac{2}{3}} \left(-p \mathrm{\sqrt{x}}\right)^{\frac{2}{3}}}$\\
				\hline
				$1$ & $\frac{2 {\mathrm e}^{p \mathrm{\sqrt{x}}} \left(x^{\frac{3}{2}} p^{3}-3 p^{2} x+6 p \mathrm{\sqrt{x}}-6\right)}{p^{4}}$ \\
				\hline
				$-1$ & $\frac{7 \left(-{\mathrm e}^{p \mathrm{\sqrt{x}}} \left(-p \mathrm{\sqrt{x}}\right)^{\frac{3}{7}}+\mathrm{\sqrt{x}}\, p \left(\Gamma \! \left(\frac{3}{7}\right)-\Gamma \! \left(\frac{3}{7},-p \mathrm{\sqrt{x}}\right)\right)\right)}{2 x^{\frac{2}{7}} \left(-p \mathrm{\sqrt{x}}\right)^{\frac{3}{7}}}$ \\
				\hline
				$\frac{5}{4}$  & $\frac{2 {\mathrm e}^{p \mathrm{\sqrt{x}}} \left(p \mathrm{\sqrt{x}}-1\right)}{p^{2}}$ \\
				\hline
				$2$   &  $\frac{2 x^{\frac{2}{5}} \left(\Gamma \! \left(\frac{4}{5}\right)-\Gamma \! \left(\frac{4}{5},-p \mathrm{\sqrt{x}}\right)\right)}{\left(-p \mathrm{\sqrt{x}}\right)^{\frac{4}{5}}}$\\
				\hline
				$-2$ & $\frac{11 \left(-{\mathrm e}^{p \mathrm{\sqrt{x}}} \left(-p \mathrm{\sqrt{x}}\right)^{\frac{7}{11}}+p \mathrm{\sqrt{x}}\, \left(\Gamma \! \left(\frac{7}{11}\right)-\Gamma \! \left(\frac{7}{11},-p \mathrm{\sqrt{x}}\right)\right)\right)}{2 x^{\frac{2}{11}} \left(-p \mathrm{\sqrt{x}}\right)^{\frac{7}{11}}}$ \\
				\hline
			\end{tabular}
			\label{table3}
		\end{table}

		\item Other values of $C_1$ and/or $C_2$: No analytical and/or closed form of the $F(T)$ solution.
	\end{itemize}

	${\bf p\gg 1}$ \textbf{:} By using Equation~\eqref{430at} and substituting Equation~\eqref{430}, we find as the Equation~\eqref{401b} potential:
	\begin{align}
		V(T)=&  -\frac{p^2\,p_0^2}{2}\,\exp\left(c_0\,p\left[{ u_{\pm}(T)}\right]^{1/2}\right). \label{543}
	\end{align}
	By substituting Equation~\eqref{543} into Equation~\eqref{404solution}, we find that:
	\small
	\begin{align}\label{544}
		F(T) =&   -{ \Lambda_0} +{\left[{ 3T+2C_1\,u_{\pm}(T)}\right]^{\frac{2(3-2C_1)}{3(3-4C_1)}}\,\left[{ u_{-}(T)}\right]^{-\frac{4C_1\delta_1}{3(3-4C_1)}}}
		\Bigg[F_0-2\kappa\,p^2\,p_0^2
		\nonumber\\
		& \,\times\,\int_{T}\,dT'\,\exp\left(c_0\,p\left[{ u_{\pm}(T')}\right]^{1/2}\right)\,{\left[{ 3T'+2C_1\,u_{\pm}(T')}\right]^{-\frac{2(3-2C_1)}{3(3-4C_1)}-1}\left[{ u_{-}(T')}\right]^{\frac{4C_1\delta_1}{3(3-4C_1)}}}\Bigg].
	\end{align}
	\normalsize
	Once again, there is no general solution for Equation~\eqref{544}, except~for $C_2=0$ and $\delta_1=-1$:
	\begin{align}\label{544a}
		F(T) \approx &   -{ \Lambda_0} +{T^{\frac{2}{(3-4C_1)}}}\,\Bigg[F_0+\frac{2\kappa\,p^2\,p_0^2}{(4C_1-3)}\,\mathcal{R}\left(p,C_1,-2c_0^2\,T\right)\Bigg],
	\end{align}
	where $\mathcal{R}\left(p,C_1,-2c_0^2\,T\right)$ is described by Equation~\eqref{542aa}.

	\item ${\bf c=2}$: By using Equation~\eqref{440at} and substituting Equation~\eqref{440}, we find that:
	\begin{align}\label{550}
		\phi(T)=p_0\,\exp\left(\frac{p(1+b)}{\sqrt{2}c_0\,\left[{ w_{\pm}(T)}\right]^{1/2}}\right) ,
	\end{align}
	and the Equation~\eqref{501} potential becomes:
	\vspace{-12pt}
	
		\begin{align}
			V(T)=&  -\frac{p^2\,p_0^2}{2}\,\exp\left(\frac{2p(1+b)}{\sqrt{2}c_0\,\left[{ w_{\pm}(T)}\right]^{1/2}}\right)+p\,(b+4)\,p_0^2\,Ei\left(-\frac{2p(1+b)}{\sqrt{2}c_0\,\left[{ w_{\pm}(T)}\right]^{1/2}}\right)  . \label{551}
		\end{align}
	By substituting Equation~\eqref{551} into Equation~\eqref{404solution}, we find that:

		\begin{align}\label{552}
			F(T) =&    -{ \Lambda_0} +\left[-T+C_1\left({ w_{\pm}(T)}\right)\right]^{(2-b)}\left[{ y(T)}\right]^{\frac{(b-2)C_1}{2\left(C_1-2C_2\right)}}\,
			\left[\frac{T+\left(C_1-2C_2\right)\left({ w_{\pm}(T)}\right)}{T+\left(C_1-2C_2\right)\left({ 2-w_{\pm}(T)}\right)}\right]^{\frac{\delta_1\,(b-2)C_1}{2\left(C_1-2C_2\right)}}
			\nonumber\\
			&\times\,\Bigg[F_0+\kappa\,(b-2)\,p_0^2\,\int_{T}\,dT'\,\Bigg[-p^2\,\exp\left(\frac{\sqrt{2}p(1+b)}{c_0\,\left[{ w_{\pm}(T')}\right]^{1/2}}\right)+2p\,(b+4)
			\nonumber\\
			&\times\,Ei\left(-\frac{\sqrt{2}p(1+b)}{c_0\,\left[{ w_{\pm}(T')}\right]^{1/2}}\right) \Bigg]\,
			\left[-T'+C_1\left({ w_{\pm}(T')}\right)\right]^{(b-3)}\,\left[{ y(T')}\right]^{-\frac{(b-2)C_1}{2\left(C_1-2C_2\right)}}
			\nonumber\\
			&\times\,\left[\frac{T'+\left(C_1-2C_2\right)\left({ w_{\pm}(T')}\right)}{T'+\left(C_1-2C_2\right)\left({ 2-w_{\pm}(T')}\right)}\right]^{-\frac{\delta_1\,(b-2)C_1}{2\left(C_1-2C_2\right)}}\Bigg].
		\end{align}
	
	There is no general $F(T)$ solution for Equation~\eqref{552}. However, there are analytical $F(T)$ solutions for { several subcases, as presented in Appendix \ref{appenb2}.}

	${\bf p\gg 1}$ \textbf{:} By using Equation~\eqref{440at} and substituting Equation~\eqref{440}, we find as the Equation~\eqref{401b} potential:
	\begin{align}
		V(T)=& -\frac{p^2\,p_0^2}{2}\,\exp\left(\frac{\sqrt{2}p(1+b)}{c_0\,\left[{ w_{\pm}(T)}\right]^{1/2}}\right). \label{553}
	\end{align}
	By substituting Equation~\eqref{553} into Equation~\eqref{404solution}, we find that:
	
	\vspace{-12pt}
		\begin{align}\label{554}
			F(T) =&    -{ \Lambda_0} +\left[-T+C_1\left({ w_{\pm}(T)}\right)\right]^{(2-b)}\,\left[{ y(T)}\right]^{\frac{(b-2)C_1}{2\left(C_1-2C_2\right)}}\left[\frac{T+\left(C_1-2C_2\right)\left({ w_{\pm}(T)}\right)}{T+\left(C_1-2C_2\right)\left({ 2-w_{\pm}(T)}\right)}\right]^{\frac{\delta_1\,(b-2)C_1}{2\left(C_1-2C_2\right)}}
			\nonumber\\
			&\times\,\Bigg[F_0-\kappa\,p^2\,p_0^2\,\int_{T}\,dT'\exp\left(\frac{\sqrt{2}p(1+b)}{c_0\,\left[{ w_{\pm}(T')}\right]^{1/2}}\right)
			\left[-T'+C_1\left({ w_{\pm}(T')}\right)\right]^{(b-3)}\left[{ y(T')}\right]^{-\frac{(b-2)C_1}{2\left(C_1-2C_2\right)}}
			\nonumber\\
			&\times\,\left[\frac{T'+\left(C_1-2C_2\right)\left({ w_{\pm}(T')}\right)}{T'+\left(C_1-2C_2\right)\left({ 2-w_{\pm}(T')}\right)}\right]^{-\frac{\delta_1\,(b-2)C_1}{2\left(C_1-2C_2\right)}}\Bigg].
		\end{align}
	
	There is no general $F(T)$ solution for Equation~\eqref{554}. However, there are analytical $F(T)$ solution for { several subcases, as presented in Appendix \ref{appenb2}.}

	{
		\item \textbf{Late Cosmology $t\,\rightarrow\,\infty$ limit:} We have the same type of scenarios as in Section~\ref{sect31}, and we can find some $F(T)$ solutions:
		\begin{enumerate}
			\item ${\bf c>1}$: By using Equations \eqref{470}--\eqref{470solution}, we find for the following potentials with $\phi(T)=p_0\,\exp\left(p\,\left({2c\left(c+2b\right)}\right)^{1/2}\,T^{-1/2}\right)$:
			\begin{itemize}
				\item Equation~\eqref{501} potential:
				\begin{align}\label{590pota}
					V(T)=& -\frac{p^2\,p_0^2}{2}\,\Bigg[\exp\left(2p\,\left({2c\left(c+2b\right)}\right)^{1/2}\,T^{-1/2}\right)
					\nonumber\\
					&\,- \frac{2(b+2c)}{p}\,Ei\left(-2p\,\left({2c\left(c+2b\right)}\right)^{1/2}\,T^{-1/2}\right)\Bigg],
				\end{align}
				and then Equation~\eqref{470solution}:
				
					\begin{align}\label{590a}
						F(T) =& -{ \Lambda_0} + T^{\frac{2}{3}}\,\Bigg[F_0-\frac{2\kappa\,p^2\,p_0^2}{3}\,\Bigg[\frac{\sqrt{2}}{2 \left(-p \sqrt{c \left(c+2 b\right)}\right)^{\frac{4}{3}}}
						\nonumber\\
						&\,\times\,\Bigg[{e}^{\frac{2 p \sqrt{2}\, \sqrt{c \left(c+2 b\right)}}{\sqrt{T}}} \left(-\frac{p \sqrt{c \left(c+2 b\right)}}{\sqrt{T}}\right)^{\frac{1}{3}} +\frac{\Gamma \left(\frac{1}{3},-\frac{2 p \sqrt{2}\, \sqrt{c \left(c+2 b\right)}}{\sqrt{T}}\right)}{3\sqrt{2}}\Bigg]
						\nonumber\\
						&\,+ \frac{(b+2c)}{4p \left(p \sqrt{c \left(c +2 b \right)}\right)^{\frac{4}{3}}} \Bigg[\Gamma \left(\frac{1}{3},\frac{2 p \sqrt{2}\, \sqrt{c \left(c +2 b \right)}}{\sqrt{T}}\right)
						\nonumber\\
						&\,-12\,  {Ei}_{1}\! \left(\frac{2 p \sqrt{2}\, \sqrt{c \left(c +2 b \right)}}{\sqrt{T}}\right) \left(p \sqrt{c \left(c +2 b \right)}\right)^{\frac{4}{3}}\,T^{-2/3}
						\nonumber\\
						&\,+3\sqrt{2}\,  {e}^{-\frac{2 p \sqrt{2}\, \sqrt{c \left(c +2 b \right)}}{\sqrt{T}}} \, \left(\frac{p \sqrt{c \left(c +2 b \right)}}{\sqrt{T}}\right)^{\frac{1}{3}}\Bigg] \Bigg]\Bigg] .
					\end{align} 

				\item Equation~\eqref{401b} potential:
				\begin{align}\label{590potb}
					V(T)=-\frac{p^2\,p_0^2}{2}\,\exp\left(2p\,\left({2c\left(c+2b\right)}\right)^{1/2}\,T^{-1/2}\right) ,
				\end{align}
				and then Equation~\eqref{470solution}:
				
					\begin{align}\label{590b}
						F(T) =& -{ \Lambda_0} + T^{\frac{2}{3}}\,\Bigg[F_0-\frac{\sqrt{2}\kappa\,p^2\,p_0^2}{3 \left(-p \sqrt{c \left(c+2 b\right)}\right)^{\frac{4}{3}}}
						\nonumber\\
						&\,\times\,\Bigg[{e}^{\frac{2 p \sqrt{2}\, \sqrt{c \left(c+2 b\right)}}{\sqrt{T}}} \left(-\frac{p \sqrt{c \left(c+2 b\right)}}{\sqrt{T}}\right)^{\frac{1}{3}} +\frac{\Gamma \left(\frac{1}{3},-\frac{2 p \sqrt{2}\, \sqrt{c \left(c+2 b\right)}}{\sqrt{T}}\right) }{3\sqrt{2}}\Bigg]\Bigg] .
					\end{align} 
				
			\end{itemize}

			\item ${\bf 0<c<1}$ and {${\bf c<0}$}: By using Equations \eqref{471}--\eqref{471solution}, we find for the following potentials with $\phi(T)=p_0\,\exp\left(p\,\left(\frac{2}{c_0^2}\right)^{1/2c}\,(-T)^{-1/2c}\right)$ the $F(T)$ solutions:
			\begin{itemize}
				\item Equation~\eqref{501} potential:
				\begin{align}\label{591pota}
					V(T)=& -\frac{p^2\,p_0^2}{2}\,\Bigg[\exp\left(2p\,\left(\frac{2}{c_0^2}\right)^{1/2c}\,(-T)^{-1/2c}\right)
					\nonumber\\
					&\,- \frac{2(b+2c)}{p}\,Ei\left(-2p\,\left(\frac{2}{c_0^2}\right)^{1/2c}\,(-T)^{-1/2c}\right)\Bigg],
				\end{align}
				and then Equation~\eqref{471solution}:

					\begin{align}\label{591a}
						F(T) =& -{ \Lambda_0} + T^{\frac{2\,(c-b)}{c}}\,\Bigg[F_0-2\kappa\,p^2\,p_0^2\,(c-b)\,\Bigg[2 \,\left(p^{2c} \frac{2^{2 c+1}}{c_{0}^{2}}\right)^{2 (b-c)/c} 
						\nonumber\\
						&\,\times\,\Bigg(-\Gamma \! \left(4 (c-b)\right)+\Gamma \! \left(4 (c-b),-2p\left(\frac{2}{c_{0}^{2}}\right)^{\frac{1}{2 c}} \left(-T\right)^{-\frac{1}{2 c}}\right)\Bigg)
						\nonumber\\
						&\,- \frac{2(b+2c)}{p} \,T^{\frac{2 \left(b-c\right)}{c}} \Bigg[\frac{1}{8 \left(b-c\right)^{2}}+\frac{\gamma}{2 \left(b-c\right)}-\frac{\ln\! \left(T\right)}{4 c \left(b-c\right)}+\frac{\ln\! \left(2\right)}{2 \left(b-c\right)}
						\nonumber\\
						&\,+\frac{\ln\! \left(p \left(\frac{2}{c_{0}^{2}}\right)^{\frac{1}{2 c}} \left(-1\right)^{-\frac{1}{2 c}}\right)}{2 \left(b-c\right)}
						+\frac{4 p}{1+4 (c-b)}  \left(-T\right)^{-\frac{1}{2 c}} \left(\frac{2}{c_{0}^{2}}\right)^{\frac{1}{2 c}} 
						\nonumber\\
						&\,\times\,_3F_3 \left(1,1,1+4 (c-b);2,2,2+4 (c-b);-2p \left(\frac{2}{c_{0}^{2}}\right)^{\frac{1}{2 c}} \left(-T\right)^{-\frac{1}{2 c}}\right)\Bigg]\Bigg]\Bigg] .
					\end{align}

				\item Equation~\eqref{401b} potential:
				\begin{align}\label{591potb}
					V(T)=-\frac{p^2\,p_0^2}{2}\,\exp\left(2p\,\left(\frac{2}{c_0^2}\right)^{1/2c}\,(-T)^{-1/2c}\right) ,
				\end{align}
				and then Equation~\eqref{471solution}:
				
					\begin{align}\label{591b}
						F(T) =& -{ \Lambda_0} + T^{\frac{2\,(c-b)}{c}}\,\Bigg[F_0-4\kappa\,p^2\,p_0^2\,(c-b)\,\left(p^{2c} \frac{2^{2 c+1}}{c_{0}^{2}}\right)^{2 (b-c)/c} 
						\nonumber\\
						&\,\times\,\Bigg(-\Gamma \! \left(4 (c-b)\right)+\Gamma \! \left(4 (c-b),-2p\left(\frac{2}{c_{0}^{2}}\right)^{\frac{1}{2 c}} \left(-T\right)^{-\frac{1}{2 c}}\right)\Bigg)\Bigg] .
					\end{align}

			\end{itemize}
		\end{enumerate}

		\item \textbf{Early Cosmology $t\,\rightarrow\,0$ limit:}
		\begin{enumerate}
			\item ${\bf c>1}$: $t^{-2c}(T)$ and $A(T)$ are, respectively, Equations \eqref{471} and \eqref{471at}, but~satisfying the $T\,\rightarrow\,-\infty$ limit in the Section~\ref{sect31} case. We will then recover Equations \eqref{591a} and \eqref{591b} under the same~limit.

			\item ${\bf 0<c<1}$ and  {${\bf c<0}$}: $t^{-2}(T)$ and $A(T)$ are, respectively, Equations \eqref{470} and \eqref{470at}, but satisfying the $T\,\rightarrow\,\infty$ limit in the Section~\ref{sect31} case. We will recover  Equations \eqref{590a} and \eqref{590b} under the same~limit.
			
		\end{enumerate}
	}
	
\end{enumerate}

There are several other cases yielding to new analytical $F(T)$ solutions. All the previous teleparallel $F(T)$ solutions are new { and comparable to those found in ref~\cite{nonvacKSpaper}.}

\subsection{Exponential Ansatz~Solutions}\label{sect42}

By using Equations \eqref{451} and \eqref{451atdef} with the Equation~\eqref{402} ansatz, we substitute Equation~\eqref{502} into Equation~\eqref{404} for again finding the $F(T)$ solution defined by the Equation~\eqref{404solution} formula. We will compute the Equation~\eqref{404solution} results in the following potential~cases:
\begin{enumerate}
	\item ${\bf c=-2b}$: By using $A(T)=\frac{T}{3}$ and Equation~\eqref{453}, we find that $\phi(T)=p_0\,\left(\frac{c_0^2}{2}\right)^{p/4b}$ $\left(-T\right)^{p/4b}$, and the Equation~\eqref{502} $V(T)$ potential is:
	\begin{align}
		V(T)=-\frac{ p\,(p-3b)\,p_0^2}{2}\,\left(\frac{c_0^2}{2}\right)^{p/2b}\left(-T\right)^{p/2b} , \label{560}
	\end{align}
	where $p\neq 3b$. By~substituting Equation~\eqref{560} into Equation~\eqref{404solution}, the~$F(T)$ solution is:
	\begin{align}\label{561}
		F(T) = &  -{ \Lambda_0} +F_0\,T^3-\frac{6\kappa\,p\,b\,(p-3b)\,p_0^2}{(p-6b)}\,\left(\frac{c_0^2}{2}\right)^{p/2b}\,(-T)^{p/2b}.
	\end{align}
	${\bf p\gg 1}$ \textbf{:} Under this limit,~Equation~\eqref{401b} becomes:
	\begin{align}
		V(T)=-\frac{p^2\,p_0^2}{2}\,\left(\frac{c_0^2}{2}\right)^{p/2b}\left(-T\right)^{p/2b} . \label{562}
	\end{align}
	By substituting Equation~\eqref{562} into Equation~\eqref{404solution}, we find as a solution:
	\begin{align}\label{563}
		F(T) \approx &  -6\kappa\,p\,b\,p_0^2\,\left(\frac{c_0^2}{2}\right)^{p/2b}\,(-T)^{p/2b}.
	\end{align}

	\item ${\bf c\neq -2b}$ (General case): From Equations \eqref{451} and \eqref{451atdef}, we find that:
	\begin{align}
		\phi(T)=p_0\,\left(\frac{c_0^2}{2}\right)^{-p/2c}\left(T_0 -T\right)^{-p/2c}	,
	\end{align}
	and the Equation~\eqref{502} potential becomes:
	\small
	\begin{align}
		V(T)=-\frac{p\,(b+2c+p)\,p_0^2}{2}\,\left(\frac{2}{c_0^2}\right)^{p/c}\left(T_0 -T\right)^{-p/c} , \label{570}
	\end{align}
	\normalsize
	where $p\neq -b-2c$. By~substituting Equation~\eqref{570} into Equation~\eqref{404solution}, the~$F(T)$ solution is:
	\vspace{-12pt}
		\begin{align}\label{571}
			F(T) =&  -{ \Lambda_0} +(T-T_1)^{2(c-b)/c}\Bigg[F_0+\frac{2\kappa\,p\,p_0^2\,(b-c)(b+2c+p)}{c}\,\left(-\frac{2}{c_0^2}\right)^{p/c}
			\nonumber\\
			&\,\times\,\int_{T}\,dT'\,\left(T'-T_0\right)^{-p/c}\,(T'-T_1)^{(2b-3c)/c}\Bigg].
		\end{align}
	There is no general solution to Equation~\eqref{571}, but~there are $F(T)$ solutions for the following~cases:
	\begin{itemize}
		\item $T_1=T_0$:
		\vspace{-12pt}
			\begin{align}\label{571a}
				F(T) =&  -{ \Lambda_0} +F_0\,(T-T_0)^{2-2b/c}+2\kappa\,p\,p_0^2\,\frac{(b-c)(b+2c+p)}{\left(2b-p-2c\right)}\left(\frac{-2}{c_0^2\left(T-T_0\right)}\right)^{p/c}.
			\end{align}
		
		\item { $T_1\neq T_0$: There are several possible $F(T)$ solutions only when $b \geq \frac{3c}{2}$, as presented in Appendix \ref{appenb3}.}
	\end{itemize}

	${\bf p\gg 1}$ \textbf{:} Under this limit,~Equation~\eqref{401b} becomes:
	\begin{align}
		V(T)= -\frac{p^2\,p_0^2}{2}\,\left(\frac{2}{c_0^2}\right)^{p/c}\left(T_0 -T\right)^{-p/c}. \label{572}
	\end{align}
	By substituting Equation~\eqref{570} into Equation~\eqref{404solution}, we find as a solution:
	\begin{align}\label{573}
		F(T) =&  -{ \Lambda_0} + (T-T_1)^{2(c-b)/c}\Bigg[F_0-\frac{2\kappa\,p^2\,p_0^2 (c-b)}{c}\,\left(-\frac{2}{c_0^2}\right)^{p/c} 
		\nonumber\\
		&\,\times\,\int_{T}\,dT'\,\left(T'-T_0\right)^{-p/c}\,(T'-T_1)^{(2b-3c)/c}\Bigg]. 
	\end{align}
	There is no general solution to Equation~\eqref{573}, but~there are $F(T)$ solutions for the following~cases:
	\begin{itemize}
		\item $T_1=T_0$:
		\vspace{-12pt}
			\begin{align}\label{573a}
				F(T) \approx &  -{ \Lambda_0} +F_0\,(T-T_0)^{2-2b/c}-2\kappa\,p\,p_0^2\,(b-c)\,\left(-\frac{2}{c_0^2}\right)^{p/c}\,\left(T-T_0\right)^{-p/c}.
			\end{align}
		For $c>0$ and $p\rightarrow \infty$, Equation~\eqref{573a} becomes:
		\begin{align}\label{573aa}
			F(T)\,\rightarrow\,-{ \Lambda_0} +F_0\,(T-T_0)^{2-2b/c}.
		\end{align}
		For $c<0$ and $p\rightarrow \infty$, Equation~\eqref{573a} becomes:
		\begin{align}\label{573ab}
			F(T)\,\rightarrow\,-2\kappa\,p\,p_0^2\,(b+|c|)\,\left(-\frac{c_0^2}{2}\right)^{p/|c|}\,\left(T-T_0\right)^{p/|c|}\,\rightarrow\,\infty.
		\end{align}
		
		\item { $T_1\neq T_0$ and $b \geq \frac{3c}{2}$: There are several possible $F(T)$ solutions, as presented in Appendix \ref{appenb3}.}
		
		\item {  $T_1\neq T_0$,   $c>0$ and $p\rightarrow \infty$ limit: We} find that all cases go to the Equation~\eqref{573aa} limit. For~the $c<0$ and $p\rightarrow \infty$ limit, we obtain that:
		\begin{align}\label{573ac}
			F(T) \sim \,\left(T-T_{0}\right)^{\frac{p}{|c|}} \,\left(T-T_{1}\right)^{-1} .
		\end{align}
	\end{itemize}
	
	{
		\item \textbf{Late Cosmology $t\,\rightarrow\,\infty$ limit:} 
		\begin{enumerate}
			\item {${\bf c>0}$}: Equation~\eqref{451} leads to $T=T_0=$ constant when $c\neq -2b$ (TdS-like spacetime), $T=0$ (null torsion scalar spacetime) when $c=-2b$, and then $A(T)=$ constant under this limit, GR solutions~\cite{TdSpaper}, as in Section~\ref{sect32}.

			\item {${\bf c<0}$}: $t(T)$ and $A(T)$ satisfy Equations \eqref{480} and \eqref{480atdef}, and the $F(T)$ solutions are described by Equation~\eqref{480solution} because this is the same situation as in Section~\ref{sect32}. By~using $\phi(T)=p_0\,\left(\frac{c_0^2}{2}\right)^{\frac{p}{2|c|}}\,(-T)^{\frac{p}{2|c|}}$, where $c=-|c|$, we find the $F(T)$ solutions for the~following cases:
			\begin{itemize}
				\item \textbf{General:} The Equation~\eqref{502} potential is:
				\begin{align}\label{595pota}
					V(T)=& -\frac{V_p}{2}\,(-T)^{\frac{p}{|c|}},
				\end{align}
				where $V_p =p\,p_0^2\,(b+2c+p)\,\left(\frac{c_0^2}{2}\right)^{\frac{p}{|c|}}$, and then Equation~\eqref{480solution} will~be:
				\begin{itemize}
					\item[$-$] $p\neq 2\left(b+|c|\right)$
					\begin{align}\label{595a}
						F(T) =& -{ \Lambda_0} + F_0\,T^{\frac{2(b+|c|)}{|c|}}+\frac{2\kappa\,(b+|c|)\,V_p}{(2b+2|c|-p)}\,(-T)^{\frac{p}{|c|}}\Bigg] .
					\end{align}
					
					\item[$-$] $p= 2\left(b+|c|\right)$
					\begin{align}\label{595b}
						F(T) =& -{ \Lambda_0} + T^{\frac{2(b+|c|)}{|c|}}\,\Bigg[F_0-\frac{2\kappa\,(b+|c|)\,V_p}{|c|}\,\ln(T)\Bigg] .
					\end{align}

				\end{itemize}
				
				\item \textbf{Equation~\eqref{401b} potential:} The $V(T)$ expression is under the same form as Equation~\eqref{595pota}, but~we replace $V_p$ with  $V_{\infty}=p^2\,p_0^2\,\left(\frac{c_0^2}{2}\right)^{\frac{p}{|c|}}$. Equation~\eqref{480solution} will be under the same forms as Equations \eqref{595a} and \eqref{595b}, but~we replace $V_p$ with $V_{\infty}$ inside these equations. For~$p\,\rightarrow\,\infty$, Equation~\eqref{595a} will become $F(T) \sim (-T)^{\frac{p}{|c|}}$, a~pure teleparallel cosmological~solution.
				
			\end{itemize}

		\end{enumerate}

		\item \textbf{Early Cosmology $t\,\rightarrow\,0$ limit:} 	
		For any non-zero value of $c$,~Equation~\eqref{451} also leads to $T\,\rightarrow\,T_0-\frac{2}{c_0^2}=$ constant and $A(T)=$ constant, a~GR solution (TdS-like spacetime), as in Section~\ref{sect32}  \cite{TdSpaper,landryvandenhoogen1}.
		
	}

\end{enumerate}

All the previous teleparallel $F(T)$ solutions are new, { and most of those are comparable with the solutions found in refs.~\cite{nonvacKSpaper,coleylandrygholami,scalarfieldTRW,FTBcosmogholamilandry}, including the late/early cosmological limit cases.} However, there are other possible subcases leading to additional teleparallel $F(T)$ solutions.

{
	\section{Other Scalar Field Source~Solutions}\label{sect5}
	\unskip

	\subsection{General Methods of Teleparallel Field Equations~Solving}\label{sect50}

	In Sections~\ref{sect3} and \ref{sect4}, we used the power-law and exponential scalar field $\phi(t)$ sources for finding the main classes of teleparallel $F(T)$ solutions. These two types of scalar field sources are the most simple and usual in the literature and allow a large number of new analytical $F(T)$ solutions~\cite{leon1,leon2,leon3,FTBcosmogholamilandry,scalarfieldTRW}. But~there are, in principle, several other possible scalar field source cases, such as logarithmic, oscillating (see Equation~\eqref{503}), polynomial (superposition of power-law terms), and several others. As~previously, we use an appropriate coframe ansatz; we solve for $t(T)$ by using the torsion scalar expression defined by Equation~\eqref{301a} and then for $A(T)$. Then, we will find the scalar field $\phi(T)$, the~$V(\phi)$ potential from Equation~\eqref{302d}, and the $V(T)$ expression from the coframe ansatz. Finally, after being sure that we have an Equation~\eqref{404} DE form, we will compute the $F(T)$ solutions by using and substituting the previous elements into Equation~\eqref{404solution}. This is the same process used in Sections~\ref{sect3} and \ref{sect4} for computing the huge number of new teleparallel $F(T)$ solutions. The~teleparallel $F(T)$ solution finding approach (and/or the extensions) is not a wall-to-wall and/or an absolute method, and there are some other approaches in the literature. The main advantage of the current approach is that we can control the used scalar field sources and then find the relevant and exact $F(T)$ solutions for a specific type of source, including the homogeneous part of solutions (fundamental FE $F(T)$ solutions). This is also why we used this approach in previous and current works~\cite{TdSpaper,SSpaper,nonvacSSpaper,nonvacKSpaper,coleylandrygholami,FTBcosmogholamilandry,roberthudsonSSpaper,scalarfieldTRW}.
	
	
	Therefore, we may also proceed by another manner to solve the same problem, as briefly mentionned in Section~\ref{sect25}. We can look for the exact scalar field $\phi(t)$ and $V(\phi)$ instead of looking for a teleparallel $F(T)$ solution. We will set the coframe ansatz components and the~characteristic equation defined from Equation~\eqref{301a}, as done in the current development, and~we will study some relevant and simple $F(T)$ function ansatz. By~inserting the coframe ansatz components and the $F(T)$ function, we will compute the unified FE defined by Equation~\eqref{305} and find the associated potential $V(\phi)$. By~using the conservation law defined in Equation~\eqref{302d}, we will find the exact expression for the scalar field $\phi(t)$, as desired in such a case. We can easily expect to recover the same $\phi(T)$ solutions by setting similar $F(T)$ functions to those found in the current paper. However, if we set more complex and/or different $F(T)$ functions, we will probably find that the $\phi(T)$ expressions might be slightly more complex and/or different compared to those used in the current paper. However,~the scalar field finding approach has the disadvantage of losing control of the type of scalar field source used and to potentially miss the fundamental FEs $F(T)$ solutions without the source (homogeneous parts of solution). These are the reasons for using the $F(T)$ solution finding approach, as done in the current~paper.

	\subsection{Logarithmic Scalar Field~Source}\label{sect51}

	The logarithmic scalar field source $\phi(t)=p_0\,\ln \left(p\,t \right)$ case (or $t(\phi)=p^{-1}\,\exp \left(\frac{\phi}{p_0}\right)$) is a good example, leading to simple teleparallel $F(T)$ solutions. By~using the conservation law defined by Equation~\eqref{302d}, we find as potential $V(\phi)$ for the~ansatz:
	\begin{enumerate}
		\item \textbf{Power-law:} By using the Equations \eqref{401} ansatz, we find that:
		\begin{align}\label{601}
			\frac{dV}{d\phi} =& p_0\,p^2\,\left(1-b-2c\right)\exp\left(-\frac{2\phi}{p_0}\right),
			\nonumber\\
			\Rightarrow\,& V(\phi)= \phi_0+ \frac{p_0^2\,p^2}{2}\,(b+2c-1)\,\exp \left(-\frac{2\phi}{p_0}\right). 
		\end{align}
		Equation~\eqref{Quintessenceindex} for $\alpha_Q$ is:
		\begin{align}\label{602}
			\alpha_Q =& -1+ \left[\left(\frac{\phi_0}{p^2\,p_0^2}\right)\,\exp \left(\frac{2\phi}{p_0}\right)+\frac{(b+2c)}{2}\right]^{-1}.
		\end{align}
		There are some examples leading to simple $F(T)$ solutions:
		\begin{enumerate}
			\item ${\bf c=1}$: We will use Equations \eqref{420} and \eqref{420at} to find, as scalar field:
			\begin{align}\label{603a}
				\phi(T)=p_0\,\ln \left(\sqrt{2}p\,\left(1+2b - \frac{1}{c_0^2}\right)^{1/2}\right)-\frac{p_0}{2}\,\ln(T) ,
			\end{align}
			and then:
			\begin{align}\label{603b}
				V(T)=& \frac{p_0^2\,(b+2c-1)}{4\left(1+2b - \frac{1}{c_0^2}\right)}\,T .
			\end{align}
			Equation~\eqref{404solution} becomes, for the Equation~\eqref{601} potential:
			\begin{align}\label{603}
				F(T) =& -{ \Lambda_0} + F_0\,T^{1/C}+\frac{2\kappa\,p_0^2\,(b+2c-1)}{4\left(1+2b - \frac{1}{c_0^2}\right)\left(C-1\right)}\,T .
			\end{align}
			Equation~\eqref{603b} shows that a logarithmic scalar field can lead to a simple $V(T)$ potential expression and then to a TEGR-like $F(T)$ source term, as obtained in Equation~\eqref{603}. This term is similar to a case of Teleparallel Robertson--Walker (TRW) spacetime $F(T)$ solutions for flat cosmological cases, as found in   {refs.~\cite{coleylandrygholami,scalarfieldTRW,FTBcosmogholamilandry}}.
			
			\newpage			
			\item \textbf{Late Cosmology $t\,\rightarrow\,\infty$ limit:} 
			\begin{itemize}
				\item ${\bf c>1}$: From Equation~\eqref{470}, we find that $\phi(T)=p_0\,\ln \left(p\,\left(\frac{T}{2c\left(c+2b\right)}\right)^{-1/2} \right)$ and the $V(T)$ from Equation~\eqref{601} becomes:
				\begin{align}\label{604}
					V(T)=& \frac{p_0^2\,(b+2c-1)}{4c\left(c+2b\right)}\,T .
				\end{align}
				By using Equations \eqref{470}, \eqref{470at}, and \eqref{604}, we will find that Equation~\eqref{470solution} becomes:
				\begin{align}\label{605}
					F(T) =& -{ \Lambda_0} + F_0\,T^{\frac{2}{3}}+\frac{\kappa\,p_0^2\,(b+2c-1)}{c\left(c+2b\right)}\,T .
				\end{align} 
				Here again, the~Equation~\eqref{605} source term is a TEGR-like contribution as for Equation~\eqref{603}, as in refs.~\cite{coleylandrygholami,scalarfieldTRW,FTBcosmogholamilandry}. 
				
				\item ${\bf c<1}$: From Equation~\eqref{471}, we find that $\phi(T)=p_0\,\ln \left(p\,\left(\frac{c_0^2}{2}\,(-T)\right)^{-1/2c} \right)$  and the $V(T)$ from Equation~\eqref{601} becomes:
				\begin{align}\label{606}
					V(T)=& \frac{p_0^2}{2}\,(b+2c-1)\,\left(\frac{c_0^2}{2}\right)^{1/c}\,(-T)^{1/c}.
				\end{align}
				By using Equations \eqref{471}, \eqref{471at}, and \eqref{606}, we will find that Equation~\eqref{471solution} becomes:
				\begin{align}\label{607}
					F(T) =& -{ \Lambda_0} + F_0\,T^{\frac{2\,(c-b)}{c}}+\frac{2\kappa\,p_0^2\,(c-b)\,(b+2c-1)}{(2b-2c+1)}\,\left(-\frac{c_0^2}{2}\right)^{1/c}\,T^{\frac{1}{c}} .
				\end{align}
				Equation~\eqref{607} is a sum of power-law terms as for the TRW $F(T)$ spacetimes, as found in refs.~\cite{coleylandrygholami,scalarfieldTRW,FTBcosmogholamilandry}.
				
				
			\end{itemize}
			
			\item \textbf{Early Cosmology $t\,\rightarrow\,0$ limit:} For this limit,~Equation~\eqref{403} will lead to the cases~satisfying:
			\begin{enumerate}
				\item ${\bf c>1}$: Equations \eqref{471} and \eqref{471at} for the $T\,\rightarrow\,-\infty$ limit, and we recover Equation~\eqref{607} under the same~limit.

				\item ${\bf c<1}$: Equations \eqref{470} and \eqref{470at} for the $T\,\rightarrow\,\infty$ limit, and we recover Equation~\eqref{605} under the same~limit.
				
			\end{enumerate}
			
		\end{enumerate}

		\item \textbf{Exponential:} By using the Equations \eqref{402} ansatz, we find that:
		\begin{align}\label{611}
			\frac{dV}{d\phi} =& p_0\,p^2\,\exp\left(-\frac{2\phi}{p_0}\right)-p_0\,p\,(b+2c)\,\exp\left(-\frac{\phi}{p_0}\right),
			\nonumber\\
			\Rightarrow\,& V(\phi)= \phi_0- \frac{p_0^2\,p^2}{2}\,\exp \left(-\frac{2\phi}{p_0}\right)+p_0^2\,p\,(b+2c)\,\exp \left(-\frac{\phi}{p_0}\right). 
		\end{align}
		Equation~\eqref{Quintessenceindex} for $\alpha_Q$ is:
		\begin{align}\label{612}
			\alpha_Q =& -1+ \left[\left(\frac{\phi_0}{p_0^2\,p^2}\right)\,\exp\left(\frac{2\phi}{p_0}\right)+\frac{(b+2c)}{p}\,\exp\left(\frac{\phi}{p_0}\right)\right]^{-1}.
		\end{align}
		There are some examples leading to simple $F(T)$ solutions:
		\begin{enumerate}
			\item {${\bf c\neq -2b}$: By using Equation~\eqref{451}, we find that $\phi(T)=p_0\,\ln \left(-\frac{p}{2c}\,\ln\left(\frac{c_0^2}{2}\left(T_0 -T\right)\right) \right)$}, and then $V(T)$ becomes, for the Equation~\eqref{611} potential:
			\vspace{-12pt}
			
				\begin{align}\label{613}
					V(T)=& - 2p_0^2\,c^2\,\left[\ln\left(\frac{c_0^2}{2}\left(T_0 -T\right)\right)\right]^{-2}-2p_0^2\,c\,(b+2c)\,\left[\ln\left(\frac{c_0^2}{2}\left(T_0 -T\right)\right)\right]^{-1} .
				\end{align}
			Then, by substituting Equations \eqref{451}, \eqref{451atdef}, and \eqref{613} into Equation~\eqref{404solution}, we find as the $F(T)$ solution:
			\begin{align}\label{614}
				F(T) =& -{ \Lambda_0} +(T-T_1)^{\frac{2(c-b)}{c}} \Bigg[F_0+8\kappa\,p_0^2\,c\,(b-c)\,
				\nonumber\\
				&\,\times\,\Bigg[\int_{T}\,dT'\,\left[\ln\left(\frac{c_0^2}{2}\left(T_0 -T'\right)\right)\right]^{-2}\,(T'-T_1)^{\frac{2b-3c}{c}}
				\nonumber\\
				&\,+\frac{(b+2c)}{c}\,\int_{T}\,dT'\,\left[\ln\left(\frac{c_0^2}{2}\left(T_0 -T'\right)\right)\right]^{-1}\,(T'-T_1)^{\frac{2b-3c}{c}}\Bigg]\Bigg] .
			\end{align}
			There is no general solution for Equation~\eqref{614}. By~setting $b=\frac{3c}{2}$, we can find, as a simple $F(T)$ solution:
			\vspace{-12pt}
			
				\begin{align}\label{614a}
					F(T) =& -{ \Lambda_0} +(T-T_1)^{-1} \Bigg[F_0+\frac{4\kappa\,p_0^2\,c^2}{c_0^2}\,\Bigg[\frac{c_{0}^{2} \left(T_{0}-T\right)}{-\ln\left(2\right)+\ln\left(-c_{0}^{2} \left(T-T_{0}\right)\right)}
					\nonumber\\
					&\,+9 \,\mathrm{Ei}_{1}\! \left(\ln\! \left(2\right)-\ln\! \left(-c_{0}^{2} \left(T-T_{0}\right)\right)\right)\Bigg]\Bigg] .
				\end{align}
			There are several possible examples for $b>\frac{3c}{2}$ and/or $T_1\neq T_0$.
			
			\item \textbf{Late Cosmology $t\,\rightarrow\,\infty$ limit:} 
			\begin{itemize}
				\item {${\bf c>0}$}: As in Sections~\ref{sect32} and \ref{sect42},~Equation~\eqref{451} leads to $T=T_0=$ constant when $c\neq -2b$ (TdS-like spacetime) and $T=0$ (null torsion scalar spacetime) when $c=-2b$, GR solutions as for TdS spacetimes~\cite{TdSpaper}.
				
				\item {${\bf c<0}$}: By using Equation \eqref{480}, we find that $\phi(T)=p_0\,\ln \left(\frac{p}{2|c|}\,\ln\left(\frac{c_0^2}{2} (-T)\right) \right)$, and $V(T)$ becomes, for Equation~\eqref{611}:
				\vspace{-12pt}
					\begin{align}\label{615}
						V(T)=- 2c^2\,p_0^2\,\left[\ln\left(\frac{c_0^2}{2} (-T)\right)\right]^{-2}+2p_0^2\,|c|\,(b-2|c|)\,\left[\ln\left(\frac{c_0^2}{2} (-T)\right)\right]^{-1} .
					\end{align}
				Then, by substituting Equations \eqref{480} and \eqref{615} into Equation~\eqref{480solution}, we find as the $F(T)$ solution:
				\begin{align}\label{616}
					F(T) =& -{ \Lambda_0} + T^{\frac{2(b+|c|)}{|c|}}\,\Bigg[F_0-8\kappa\,p_0^2\,(b+|c|)\,|c|
					\nonumber\\
					&\,\times\,\Bigg[\int_{T}\,dT'\,\left[\ln\left(\frac{c_0^2}{2} (-T')\right)\right]^{-2}\,T'^{-\frac{2b+3|c|}{|c|}}
					\nonumber\\
					&\,+\frac{(2|c|-b)}{|c|}\,\int_{T}\,dT'\,\left[\ln\left(\frac{c_0^2}{2} (-T')\right)\right]^{-1}\,T'^{-\frac{2b+3|c|}{|c|}}\Bigg]\Bigg] .
				\end{align}
				There is no general solution for Equation~\eqref{616}. By~setting $b=-\frac{3|c|}{2}$, we can find, as a simple $F(T)$ solution:
				\begin{align}\label{616a}
					F(T) =& -{ \Lambda_0} + T^{-1}\,\Bigg[F_0+\frac{4\kappa\,p_0^2\,|c|^2}{c_{0}^{2}}\,\Bigg[\frac{T}{\ln\! \left(2\right)-\ln\! \left(-c_{0}^{2} T\right)}
					\nonumber\\
					&\,+9 \,{Ei}_{1}\! \left(\ln\! \left(2\right)-\ln\! \left(-c_{0}^{2} T\right)\right)\Bigg]\Bigg] .
				\end{align}
				
				{ The new teleparallel $F(T)$ solutions found from the Equation~\eqref{402} ansatz are more complex than in refs.~\cite{coleylandrygholami,FTBcosmogholamilandry,scalarfieldTRW}, but~they are comparable to the solution classes in ref.~\cite{nonvacKSpaper}.}

			\end{itemize}
			
			\item \textbf{Early Cosmology $t\,\rightarrow\,0$ limit:}
			As in Sections~\ref{sect32} and \ref{sect42},~Equation~\eqref{451} leads to $T\,\rightarrow\,T_0-\frac{2}{c_0^2}=$ constant, a~GR solution (a TdS-like spacetime) \cite{TdSpaper,landryvandenhoogen1}. 
			
		\end{enumerate}

	\end{enumerate}
	
	All the previous teleparallel $F(T)$ solutions are new { and comparable to those found in refs.~\cite{coleylandrygholami,SSpaper,nonvacKSpaper,scalarfieldTRW,FTBcosmogholamilandry}, especially for the late and/or early cosmological limit scenarios.}

}
\section{Discussion and~Conclusions}

The primary aim of this study was to find the teleparallel $F(T)$ solutions coming from the FEs precisely described by Equations \eqref{301a}--\eqref{301d} with two usual types of scalar field $\phi(t)$: power-law and exponential. { In addition and to complete the study, we briefly solved the same FEs with the same ansatzes for a logarithmic scalar field source to find more simple $F(T)$ solutions comparable to the TRW spacetime solutions found in refs.~\cite{coleylandrygholami,scalarfieldTRW,FTBcosmogholamilandry}.} In { each} cases and depending on the values of $p$ in the scalar field, we solved the laws of conservation and the FEs for various types of potential $V(T)$. We first found that the FEs transform by various substitutions into the single Equation~\eqref{404}, for any type of scalar field and ansatz. From~this last equation, the~possible teleparallel $F(T)$ solutions all boil down to Equation~\eqref{404solution} using the function $A(T)$ described by Equation~\eqref{404atdef}. All of the new $F(T)$ solutions obtained in Sections~\ref{sect31} through to the end of Section~\ref{sect51} were found by applying Equation~\eqref{404solution} for various types of scalar potential. For~the $A_2$ and $A_3$ coframe components, the~power-law ansatz has been used in Sections~\ref{sect31} and \ref{sect41}; the application has been restricted to the values of $c=-2b$, $1$, $-1$, and $2$. Most of other values of $c$ do not result in closed analytic functions and/or are not analytically solvable. { In addition, we studied and found that the teleparallel $F(T)$ solutions under the late/early cosmological limit (i.e. $t\,\rightarrow\,\infty$) for the simple cases are close to the TRW solutions, as found in refs.~\cite{coleylandrygholami,scalarfieldTRW,FTBcosmogholamilandry}.} In Sections~\ref{sect32} and \ref{sect42}, the~exponential coframe ansatz has been used to study a well-known case of infinite sum of power terms. At~the same time, this ansatz is much simpler to treat and only brings out two types of case instead of an infinity: general and $c=-2b$. { In Section~\ref{sect51}, we used the same coframe ansatzes, but~we restricted the study to $c=1$ (power-law only), $c\neq -2b$ (exponential only), and the late/early cosmological limit cases.} For the rest, each type of case treated here leads to its own class of $F(T)$ solutions, with its~own particularities.

For the power-law $\phi(t)$ treated in Section~\ref{sect3}, we first studied in Section~\ref{sect31} via the power-law coframe ansatz the cases of potentials described by Equations \eqref{401a}, \eqref{401b}, and \eqref{401c}, i.e.,~the general case, $p\gg 1$, and $p=1$. This was intended to study the effects of different scalar potential types associated with gravitational sources. The~$p\gg 1$ potential cases can be useful for phantom energy model studies involving fast accelerating universe expansion. Often, physical phenomena described by a scalar field are characterized by its associated potential, and teleparallel $F(T)$-gravity cosmology is no exception to the rule. In~Section~\ref{sect31}, we find general and exact $F(T)$ solutions for the $c=-2b$ and $c=1$ cases. For~the $c=-1$ and $c=2$ cases, even if there is no real general and analytic large coverage solution, there are analytical $F(T)$ solutions for specific cases depending on the parameters $C_1$ and $C_2$, specially defined to simply solve these cases. Therefore, larger values of $c$ subcases can be useful for studies on teleparallel phantom energy and quintom models. { For late and early cosmological limit cases, an~important number of new $F(T)$ solutions are comparable to those obtained in refs.~\cite{nonvacKSpaper,scalarfieldTRW,FTBcosmogholamilandry}, especially with the isotropic TRW solutions.} In Section~\ref{sect32}, via the exponential ansatz coframe and for the potentials described by Equations \eqref{402a} and \eqref{402b}, we obtain purely analytic $F(T) $solutions, but~this required the introduction and definition of new special functions depending on the parameters of the scalar field and the ansatz used to achieve them. Even in cosmology and teleparallel gravity, special functions are a necessary evil to achieve analytic solutions. Other types of approach, such as quantum gravity theories or fundamental particle interactions with gravity, regularly require special functions to express certain new solutions (see, for example, refs.~\cite{Landry:2019otf,Hammad:2019fgu}). It would not be surprising if the special functions were to be used even more in some future work in teleparallel gravity, more specifically for solutions involving more complex and sophisticated source terms to be~solved.

By considering exponential $\phi(t)$ in Section~\ref{sect4}, we first have, in Section~\ref{sect41}, studied the power-law coframe ansatz and the potentials described in Equations \eqref{401b} and \eqref{501}. We obtain teleparallel $F(T)$ solutions that are slightly more complex compared to the results obtained in Section~\ref{sect31}, but~with some clearly visible similarities. Again, the~$c=-2b$ and $c=1$ cases lead to analytical $F(T)$ solutions. However, we only obtain $F(T)$ solutions for very specific subcases in the $c=-1$ and $c=2$ situations. This is not surprising compared to the results of Section~\ref{sect31}. For~Section~\ref{sect42}, with the exponential ansatz coframe and the potentials described by Equations \eqref{401b} and \eqref{502}, we essentially obtain purely analytical and easily usable $F(T)$ solutions. These latter solutions confirm, at the same time, some new simple solutions obtained in Section~\ref{sect3}. Let us not forget that the exponential ansatz and scalar field are cases fundamentally expressed by an infinite sum of terms in powers {, as stated in ref.~\cite{nonvacKSpaper}}. This last fact also explains some common points in the newly obtained $F(T)$ solutions. In~comparison, the~new simplest $F(T)$ solutions obtained in this paper { and including  most of the late cosmological limit $F(T)$ solutions} are also similar to the teleparallel cosmology solutions obtained in the recent literature (see, for example, refs.~\cite{nonvacKSpaper,coleylandrygholami,scalarfieldTRW,attractor,Kofinas}). This finding is also an additional argument in favor of the new solutions' rightness obtained in the present~paper.

Beyond the new solutions, this approach aimed to obtain purely analytical teleparallel $F(T)$ solutions that could be used in the future to study complex cosmological models in more detail using teleparallel $F(T)$-gravity tools. One need only think in particular of the quintessence dark energy, the~cosmological phantom energy (negative energy), or even cosmological quintom models. In~these latter cases, one will be able, in the near future, to study models involving perfect fluids and scalar fields at the same time, in~order to represent reality more faithfully. In~such a case, we will use the common (or very similar) $F(T)$ solutions of the present paper and of some recent papers concerning teleparallel $F(T)$ solutions in perfect fluids~\cite{nonvacKSpaper,roberthudsonSSpaper,coleylandrygholami}. This will ultimately lead to complete and realistic cosmological models that can fully describe and explain the dark energy quintessence process ($-1<\alpha_Q<-\frac{1}{3}$), a~realistic and most probable scenario according to the recent literature~\cite{steinhardt1,steinhardt2,steinhardt3,steinhardt2024}. After~that, one could just as well study with a similar approach some physical models of the cosmological phantom energy type ($\alpha_Q<-1$), leading to the extreme scenario of the acceleration of uncontrolled universe expansion leading to the Big Rip~\cite{caldwell1,farnes,baumframpton}. We can also add the quintom dark energy models, because~this is, by definition, a mix of the quintessence and phantom models~\cite{quintom1,quintom2,quintom3,quintom4,quintomcoleytot,quintomteleparallel1}. { As a safe and right comparison basis, we can also use some scalar field TRW $F(T)$ solutions found in ref.~\cite{scalarfieldTRW} as a comparison basis for the late/early cosmological limit solutions.} Like the fundamental scalar field associated with the dark energy quintessence process, one could obtain, for the different models involving phantom energy, such fundamental scalar fields. Perhaps the cosmological process of uncontrolled universe acceleration leading to the Big Rip would in fact be explainable by a fundamental scalar field. The~same questions will also be relevant and useful for future quintom model studies in teleparallel gravity. The~questions will then be as follows: What type(s) of $V(\phi(T))$ exactly? What are the most appropriate teleparallel $F(T)$ solutions? What are the coframes, spin-connections, and other conditions? All these questions and hypotheses deserve to be answered seriously and tactfully in future works by using the teleparallel gravity framework and~toolkit.


\vspace{6pt}

\section*{Acknowledgements}

{A.L. is supported}	by an Atlantic Association of Research in Mathematical Sciences (AARMS) fellowship. Thanks to A.A. Coley for useful and constructive comments. Thanks also to R.J. van den Hoogen for relevant discussions on the~topic.







\section*{Abbreviations}

{
	The following abbreviations are used in {this manuscript:} 
	\\
	
	\noindent 
	\begin{tabular}{@{}ll}
		AL & Alexandre Landry\\
		CK & Cartan--Karlhede\\
		DE & Differential Equation\\
		EoS & Equation of State \\
		FE & Field Equation\\
		GR & General Relativity \\
		KV & Killing Vector \\
		NGR & New General Relativity \\
		SHO & Simple Harmonic Oscillator \\
		TEGR & Teleparallel Equivalent of General Relativity \\
		{ TRW } & { Teleparallel Robertson--Walker } \\
		{ TdS } & { Teleparallel de Sitter } \\
	\end{tabular}
}

\appendix
{
	\section{Additional Power-Law Scalar Field \boldmath{$F(T)$} Solutions}\label{appena}
	\unskip
	
	\subsection{Power-Law Ansatz $c=-1$ Solutions}\label{appena1}
	
	Equation~\eqref{433a} $C_2\neq 0$ teleparallel $F(T)$ solutions:
	\begin{enumerate}
		\item $C_1=\frac{3}{2}$ ($b=-\frac{4}{3}$ and $C_2=\frac{176}{3c_0^2}$): There are three possible~solutions:
		\begin{itemize}
			\item $p\neq \frac{1}{3}$:
			\begin{align}\label{433c}
				F(T) =& -{ \Lambda_0} +F_0{\left[{ -u_{-}(T)}\right]^{\frac{2\delta_1}{3}}}+\frac{2\kappa\,V_p}{(3p-1)}\,\delta_1^{p-1}\left[u_{\pm}(T)\right]^{p-1}.
			\end{align}
			
			\item $p=\frac{1}{3}$ and $\delta_1=+1$:
			\begin{align}\label{433ca}
				F(T) =& -{ \Lambda_0} +{\left[{ -u_{-}(T)}\right]^{\frac{2}{3}}}\Bigg[F_0-\frac{2\kappa\,V_p}{3\,C_2^{\frac{2}{3}}}\,\ln\left[{ -u_{-}(T)}\right]\Bigg].
			\end{align}
			
			\item $p=\frac{1}{3}$ and $\delta_1=-1$:
			\begin{align}\label{433cb}
				F(T) =& -{ \Lambda_0}+{\left[{ -u_{-}(T)}\right]^{\frac{2}{3}}}\Bigg[F_0+\frac{2\kappa\,V_p}{3\,(-1)^{\frac{2}{3}}}\,\ln\left[{ -u_{-}(T)}\right]\Bigg].
			\end{align}
		\end{itemize}

		\item  $C_1\,\rightarrow\,\frac{3}{4}$: By setting $C_1=\frac{3}{4} +\epsilon$ where $\epsilon \ll 1$, we find:
		\begin{align}\label{433cc}
			F(T) =& -{ \Lambda_0} +F_0+\frac{4\kappa\,V_p}{3C_2\,(p+1)(p-1)}\,\left[{ u_{\pm}(T)}\right]^{p}\left[T+\delta_1 p\sqrt{T^2+C_2}\right].
		\end{align}
	\end{enumerate}
	
	Equation~\eqref{436} $C_2\neq 0$ teleparallel $F(T)$ solutions ($p\gg 1$ case):
	\begin{enumerate}
		\item $C_1=\frac{3}{2}$ ($b=-\frac{4}{3}$ and $C_2=\frac{176}{3c_0^2}$):
		\begin{align}\label{436b}
			F(T) =& -{ \Lambda_0} +F_0{\left[{ -u_{-}(T)}\right]^{\frac{2\delta_1}{3}}}+\frac{2\kappa\,V_{\infty}}{3p}\,\left[{ u_{\pm}(T)}\right]^{p}.
		\end{align}

		\item  $C_1\,\rightarrow\,\frac{3}{4}$: By setting $C_1=\frac{3}{4} +\epsilon$, where $\epsilon \ll 1$, we find:
		\begin{align}\label{436d}
			F(T) =& -{ \Lambda_0} +F_0+\frac{4\kappa\,V_{\infty}}{3C_2\,p}\,\left[{ u_{\pm}(T)}\right]^{p}\left(C_2-T\left[{ u_{\pm}(T)}\right] \right).
		\end{align}
		
	\end{enumerate}
	
	Equation~\eqref{439} $C_2\neq 0$ teleparallel $F(T)$ solutions ($p=1$ case):
	\begin{enumerate}
		\item $C_1=\frac{3}{2}$ ($b=-\frac{4}{3}$ and $C_2=\frac{176}{3c_0^2}$):
		\small
		\begin{align}\label{439b}
			F(T) =& -{ \tilde{\Lambda}_0} +F_0{\left[{ -u_{-}(T)}\right]^{\frac{2\delta_1}{3}}}+\kappa\,p_0^{2}\,\left( b-2\right)\,\left(\ln\,\left[{ u_{\pm}(T)}\right]-\frac{3}{2}\right).
		\end{align}
		\normalsize
		
		\item  $C_1\,\rightarrow\,\frac{3}{4}$: By setting $C_1=\frac{3}{4} +\epsilon$, where $\epsilon \ll 1$, we find:
		\small
		\begin{align}\label{439d}
			F(T) =& -{ \tilde{\Lambda}_0} +F_0
			\nonumber\\
			&\,-\frac{\kappa\,p_0^{2}}{3}\,\left( b-2\right)\,\Bigg[\delta_1\ln\,\left[{ -u_{-}(T)}\right]-\ln^2\,\left[{ u_{\pm}(T)}\right]
			+\frac{2T}{C_2}\left[{ u_{\pm}(T)}\right]\left(\ln\,\left[{ u_{\pm}(T)}\right]-\frac{1}{2}\right)\Bigg].
		\end{align}
		\normalsize
	\end{enumerate}

	\subsection{Power-Law Ansatz $c=2$ Solutions}\label{appena2}
	
	The general Equation~\eqref{443a} teleparallel $F(T)$ solution for $C_1=2C_2$ is:
	\begin{align}\label{443ac}
		F(T) =&  -{ \Lambda_0} +\left[-\frac{T}{C_2}+2\left({ w_{\pm}(T)}\right)\right]^{(2-b)}\,\Bigg[F_0-\frac{\kappa\,V_p\,(b-2)}{C_2}
		\nonumber\\
		&\,\times\,\int_{T}\,dT'\,\left[{ w_{\pm}(T')}\right]^{1-p}\left[-\frac{T'}{C_2}+2\left({ w_{\pm}(T')}\right)\right]^{(b-3)}\Bigg] .
	\end{align}
	\normalsize
	Equation~\eqref{443ac} solutions are possible only for the~following subcases:
	\begin{enumerate}
		\item $b=3$ and $\delta_1=1$:
		
		\vspace{-12pt}
		
			\begin{align}\label{443ad}
				F(T) =&  -{ \Lambda_0} +\left[-\frac{T}{C_2}+2\left(w_{+}(T)\right)\right]^{-1}\Bigg[F_0-\frac{\kappa\,V_p}{C_2}\,(2)^{1-p}\,
				T\,_3F_2\left(1,\frac{p}{2},\frac{p-1}{2};2,p;\frac{T}{C_2}\right)\Bigg] .
			\end{align}
		
		\item $b=3$ and $\delta_1=-1$:
			\begin{align}\label{443ae}
				F(T) =&  -{ \Lambda_0} +\left[-\frac{T}{C_2}+2\left(w_{-}(T)\right)\right]^{-1}\Bigg[F_0+\frac{2^{p-1}\kappa\,V_p}{C_2^{2-p}(p-2)}\,
				T^{2-p}\,_2F_1\left(\frac{2-p}{2},\frac{1-p}{2};3-p;\frac{T}{C_2}\right)\Bigg].
			\end{align}

		\item $b=4$ and $\delta_1=1$:
		
		\vspace{-12pt}
			\begin{align}\label{443af}
				F(T) =&  -{ \Lambda_0} +\left[-\frac{T}{C_2}+2\left(w_{+}(T)\right)\right]^{-2}\,\Bigg[F_0-\frac{\kappa\,V_p}{2^{p-1}}\,\Bigg[-\left(\frac{T}{C_2}\right)^2
				\,_3F_2\left(2,\frac{p}{2},\frac{p-1}{2};3,p;\frac{T}{C_2}\right)
				\nonumber\\
				&\,+8\left(\frac{T}{C_2}\right)\,_3F_2\left(1,\frac{p-2}{2},\frac{p-1}{2};2,p-1;\frac{T}{C_2}\right)\Bigg]\Bigg] .
			\end{align}

		\item $b=4$ and $\delta_1=-1$:
		\vspace{-12pt}
		
			\begin{align}\label{443ag}
				F(T) =&  -{ \Lambda_0} +\left[-\frac{T}{C_2}+2\left(w_{-}(T)\right)\right]^{-2}\,\Bigg[F_0-\frac{8\kappa\,V_p}{(p-3)}\,\left(\frac{T}{2C_2}\right)^{(3-p)}
				\nonumber\\
				&\,\times\,\Bigg[\,_3F_2\left(3-p,\frac{2-p}{2},\frac{1-p}{2};4-p,2-p;\frac{T}{C_2}\right) +\,_2F_1\left(\frac{2-p}{2},\frac{3-p}{2};4-p;\frac{T}{C_2}\right)\Bigg]\Bigg] .
			\end{align}

	\end{enumerate}

	\noindent  The general Equation~\eqref{446} teleparallel $F(T)$ solution for $C_1=2C_2$ is ($p\gg 1$ case):
	\small
	\begin{align}\label{446b}
		F(T) =&    -{ \Lambda_0} +\left[-T+2C_2\left({ w_{\pm}(T)}\right)\right]^{(2-b)}\,\Bigg[F_0-\kappa\,V_{\infty}\,(b-2)\,
		\nonumber\\
		&\,\times\,\int_{T}\,dT'\,\left[{ w_{\pm}(T')}\right]^{-p}\,\left[-T'+2C_2\left({ w_{\pm}(T')}\right)\right]^{(b-3)}\Bigg] .
	\end{align}
	\normalsize
	Equation~\eqref{446b} solutions are possible only for the~following subcases:
	\begin{enumerate}
		\item $b=3$ and $\delta_1=1$:
		\small
		\begin{align}\label{446ba}
			F(T) \approx &   -{ \Lambda_0} +\left[-\frac{T}{C_2}+2\left(w_{+}(T)\right)\right]^{-1}
			\,\Bigg[F_0-\frac{\kappa\,V_p}{C_2\,2^p}\,T\,_3F_2\left(1,\frac{p}{2},\frac{p}{2};2,p;\frac{T}{C_2}\right)\Bigg] .
		\end{align}
		\normalsize
		
		\item $b=3$ and $\delta_1=-1$:
		\small
		\begin{align}\label{446bb}
			F(T) \approx &  -{ \Lambda_0} -\left[\frac{T}{C_2}-2\left(w_{-}(T)\right)\right]^{-1}\,
			\Bigg[F_0+\frac{\left(2C_2\right)^{p}\kappa\,V_p}{p}\,T^{-p}\,_2F_1\left(-\frac{p}{2},-\frac{p}{2};-p;\frac{T}{C_2}\right)\Bigg].
		\end{align}
		\normalsize
		
		\item $b=4$ and $\delta_1=1$:
		\small
		\begin{align}\label{446bc}
			F(T) \approx &  -{ \Lambda_0} +\left[-\frac{T}{C_2}+2\left(w_{+}(T)\right)\right]^{-2}\,\Bigg[F_0-\frac{\kappa\,V_p}{2^{p}}\,\Bigg[-\left(\frac{T}{C_2}\right)^2
			\,_3F_2\left(2,\frac{p}{2},\frac{p}{2};3,p;\frac{T}{C_2}\right)
			\nonumber\\
			&\,+8\left(\frac{T}{C_2}\right)\,_3F_2\left(1,\frac{p}{2},\frac{p}{2};2,p;\frac{T}{C_2}\right)\Bigg]\Bigg] .
		\end{align}
		\normalsize
		
		\item $b=4$ and $\delta_1=-1$:
		\small
		\begin{align}\label{446bd}
			F(T) \approx &   -{ \Lambda_0} +\left[\frac{T}{C_2}-2\left(w_{-}(T)\right)\right]^{-2}\,\Bigg[F_0-\frac{2\kappa\,V_p}{p}\,\left(\frac{T}{2C_2}\right)^{-p}
			\,_2F_1\left(-\frac{p}{2},-\frac{p}{2};-p;\frac{T}{C_2}\right)\Bigg] .
		\end{align}
		\normalsize
	\end{enumerate}
	
	The general Equation~\eqref{449} teleparallel $F(T)$ solution for $C_1=2C_2$ is ($p=1$ case):
	\begin{align}\label{449b}
		F(T) =&   -{ \tilde{\Lambda}_0} +\left[-T+2C_2\left({ w_{\pm}(T)}\right)\right]^{(2-b)}\,\Bigg[F_0+\frac{\kappa\,p_0^{2}\,\left( b+4\right)(b-2)}{2C_2}
		\nonumber\\
		&\,\times\,\int_{T}\,dT'\,\ln\,\left[{ w_{\pm}(T')}\right]\left[-T'+2C_2\left({ w_{\pm}(T')}\right)\right]^{(b-3)}\Bigg] .
	\end{align}
	
	The simplest solutions of Equation~\eqref{449} are:
	\begin{enumerate}
		\item $b=3$:
		
		\vspace{-12pt}
			\begin{align}\label{449ba}
				F(T) =&   -{ \tilde{\Lambda}_0} +\left[-T+2C_2\left(w_{+}(T)\right)\right]^{-1}\Bigg[F_0-\frac{7\kappa\,p_0^{2}}{2C_2}\,
				\Bigg[\frac{T}{2}+C_2\,w_{\pm}(T)-T\,\ln\,\left[{ w_{\pm}(T)}\right]\Bigg]\Bigg] .
			\end{align}

		\item $b=4$:
		
		\vspace{-12pt}
			\begin{align}\label{449bc}
				F(T) =&   -{ \tilde{\Lambda}_0} +\left[-T+2C_2\left(w_{+}(T)\right)\right]^{-2}\Bigg[F_0+\frac{8\kappa\,p_0^{2}}{C_2}
				\,\Bigg[\Bigg[-\frac{4\delta_1\,C_2^2}{3}\left(1-\frac{T}{C_2}\right)^{3/2}-3\,T^2+12C_2\,T
				\nonumber\\
				&\,-8C_2^2\Bigg]\,\ln\,\left[{ w_{\pm}(T)}\right]+\frac{\delta_1\,C_2}{18}\left(2C_2-5\,T\right)\sqrt{1-\frac{T}{C_2}}+\frac{T^2}{8}-\frac{C_2}{3}\,T+\frac{C_2^2}{9}\Bigg]\Bigg] .
			\end{align}

	\end{enumerate}
	
	\section{Additional Exponential Scalar Field \boldmath{$F(T)$} Solutions}\label{appenb}
	\unskip

	\subsection{Power-Law Ansatz $c=1$ Solutions}\label{appenb1}
	
	Equation~\eqref{532} $C_1\neq 1$ teleparallel $F(T)$ solutions:
	\begin{enumerate}
		\item $C=-1$:
		\vspace{-12pt}
		
			\begin{align}\label{532b}
				F(T) =&   -{ \Lambda_0} + F_0\,T^{-1}+\kappa\,p_0^2\Bigg[p^2\,T^{-1}\,\left({Ei}_{1}\! \left(-\frac{T_{2}}{\sqrt{T}}\right) T_{2}^{2}+\exp{\left(\frac{T_{2}}{\sqrt{T}}\right)} \left(T_{2} {\sqrt{T}}+T\right)\right)
				\nonumber\\
				&\,-p\,(b+2)\,\left(\frac{\left(-T_{2} {\sqrt{T}}+T\right) \exp{\left(-\frac{T_{2}}{\sqrt{T}}\right)}}{T}+\frac{\left(T_{2}^{2}-2 T\right) {Ei}_{1}\! \left(\frac{T_{2}}{\sqrt{T}}\right)}{T}\right)\Bigg].
			\end{align}

		\item $C=2$:
		
		\vspace{-12pt}
			\begin{align}\label{532c}
				F(T) =&   -{ \Lambda_0} + F_0\,T^{1/2}
				+\kappa\,p_0^2\Bigg[\left(p^2-2p\,(b+2)\right)\,\frac{\sqrt{T}}{T_{2}} \exp\left({\frac{T_{2}}{\sqrt{T}}}\right)-2p\,(b+2)\,{Ei}\! \left(-\frac{T_{2}}{\sqrt{T}}\right)\Bigg].
			\end{align}

		\item $C=-2$:
		\vspace{-12pt}
		
			\begin{align}\label{532d}
				F(T) =&   -{ \Lambda_0} + F_0\,T^{-1/2}+\kappa\,p_0^2\,\Bigg[p^2\,\left( \exp{\left(\frac{T_{2}}{\sqrt{T}}\right)}+\frac{T_{2}}{\sqrt{T}}\,{Ei}_{1}\! \left(-\frac{T_{2}}{\sqrt{T}}\right) \right)
				\nonumber\\
				&\,-2p\,(b+2)\,\Bigg( \exp{\left(-\frac{T_{2}}{\sqrt{T}}\right)} -\left(\frac{T_{2}}{\sqrt{T}}+ 1\right) {Ei}_{1}\! \left(\frac{T_{2}}{\sqrt{T}}\right)\Bigg)\Bigg].
			\end{align}
		
	\end{enumerate}
	
	\noindent Equation~\eqref{534} $C_1\neq 1$ teleparallel $F(T)$ solutions ($p\gg 1$ case):
	\begin{enumerate}
		\item $C=-1$:
		\small
		\begin{align}\label{534b}
			F(T) =&   -{ \Lambda_0} + F_0\,T^{-1}+\kappa\,p^2\,p_0^2\,\left[\frac{T_2^2}{T}\,Ei_1\left(-\frac{T_2}{\sqrt{T}}\right)+\left(\frac{T_2}{\sqrt{T}}+1\right)\right] .
		\end{align}
		\normalsize
		
		\item $C=2$:
		\small
		\begin{align}\label{534c}
			F(T) =&   -{ \Lambda_0} + F_0\, T^{1/2}+\kappa\,p^2\,p_0^2\, \frac{T^{1/2}}{T_2}\,\exp\left(\frac{T_2}{\sqrt{T}}\right) .
		\end{align}
		\normalsize
		
		\item $C=-2$:
		\small
		\begin{align}\label{534d}
			F(T) =&   -{ \Lambda_0} + F_0\,T^{-1/2}+\kappa\,p^2\,p_0^2\,\left[\frac{T_2}{\sqrt{T}}\,Ei_1\left(-\frac{T_2}{\sqrt{T}}\right)+\exp\left(\frac{T_2}{\sqrt{T}}\right)\right].
		\end{align}
		\normalsize
	\end{enumerate}

	\subsection{Power-Law Ansatz $c=2$ Solutions}\label{appenb2}

	Equation~\eqref{552} teleparallel $F(T)$ solutions~are:
	\begin{enumerate}
		\item $C_1=0$:
		
		\vspace{-12pt}
			\begin{align}\label{552a}
				F(T) =&    -{ \Lambda_0} +(-T)^{(2-b)}\,\Bigg[F_0-\kappa\,(b-2)\,p_0^2\,
				\Bigg[p^2\,\int_{T}\,dT'\,(-T')^{(b-3)}\,\exp\left(\frac{\sqrt{2}p(1+b)}{c_0\,\left[{ w_{\pm}(T')}\right]^{1/2}}\right)
				\nonumber\\
				&-2p\,(b+4)\,\int_{T}\,dT'\,(-T')^{(b-3)}\,Ei\left(-\frac{\sqrt{2}p(1+b)}{c_0\,\left[{ w_{\pm}(T')}\right]^{1/2}}\right) \Bigg]\Bigg].
			\end{align}
		
		There are solutions for specific values of $b$, such~as:
		\begin{itemize}
			\item $b=1$:
			\vspace{-12pt}
			
				\scriptsize					
				\begin{align}\label{552aa}
					F(T) =&    -{ \Lambda_0} +(-T)\,\Bigg[F_0-\kappa\,p_0^2\,\Bigg[-\frac{1}{4 \sqrt{{ w_{\pm}(T)}}\, c_{0} C_{2} { \left(w_{\pm}(T)-2\right)}}
					\nonumber\\
					&\,\times\,\Bigg[- \sqrt{{ w_{\pm}(T)}}\, p^{3} { \left(w_{\pm}(T)-2\right)} \,\Bigg({\mathrm e}^{\frac{2 p}{c_{0}}} \mathrm{Ei}_{1}\! \left(-\frac{2 p \left(\sqrt{2}-\sqrt{{ w_{\pm}(T)}}\right)}{c_{0} \sqrt{{ w_{\pm}(T)}}}\right)
					\nonumber\\
					&\,- {\mathrm e}^{-\frac{2 p}{c_{0}}}\mathrm{Ei}_{1}\! \left(-\frac{2 p \left(\sqrt{2}+\sqrt{{ w_{\pm}(T)}}\right)}{c_{0} \sqrt{{ w_{\pm}(T)}}}\right)\Bigg)+\Bigg(\Bigg(\left(p^{2}+\frac{c_{0}^{2}}{2}\right) { \left(w_{\pm}(T)-1\right)}
					\nonumber\\
					&\,+\left(p^{2}-\frac{c_{0}^{2}}{2}\right)\Bigg) \sqrt{{ w_{\pm}(T)}}-c_{0} \sqrt{2}\, p { \left(w_{\pm}(T)-2\right)}\Bigg)
					\,\exp{\left(\frac{2 \sqrt{2}\, p}{c_{0} \sqrt{{ w_{\pm}(T)}}}\right)} c_{0}\Bigg]
					\nonumber\\
					&\,+\frac{5}{2 \left({ w_{\pm}(T)}\right)^{\frac{3}{2}} C_{2}^{2} p  { \left(w_{\pm}(T)-2\right)}}
					\Bigg[\Bigg(\frac{T}{2}\mathrm{Ei}_{1}\left(\frac{2 p \left(\sqrt{2}-\sqrt{{ w_{\pm}(T)}}\right)}{c_{0} \sqrt{{ w_{\pm}(T)}}}\right) {\mathrm e}^{-\frac{2 p}{c_{0}}} 
					\nonumber\\
					&\,+\frac{T}{2}\mathrm{Ei}_{1}\left(\frac{2 p \left(\sqrt{2}+\sqrt{{ w_{\pm}(T)}}\right)}{c_{0} \sqrt{{ w_{\pm}(T)}}}\right) {\mathrm e}^{\frac{2 p}{c_{0}}} 
					-4C_2 \,\mathrm{Ei}_{1}\! \left(\frac{2 \sqrt{2}\, p}{c_{0} \sqrt{{ w_{\pm}(T)}}}\right) \Bigg) \sqrt{{ w_{\pm}(T)}}\, p^{2}
					\nonumber\\
					&\,+\frac{T c_{0}}{4} \left(2 \sqrt{2}\, p +c_{0} \sqrt{{ w_{\pm}(T)}}\right) \exp\left({-\frac{2 \sqrt{2}\, p}{c_{0} \sqrt{{ w_{\pm}(T)}}}}\right)\Bigg] \Bigg]\Bigg].
				\end{align}

			
			\item $b=3$:

				\scriptsize				
				\begin{align}\label{552ab}
					F(T) =&    -{ \Lambda_0} -T^{-1}\,\Bigg[F_0+\kappa\,p_0^2\,\Bigg[-\frac{p^2}{3 c_{0}^{4}}\Bigg[-512 \left(p^{2}-\frac{3 c_{0}^{2}}{8}\right) p^{2} C_{2}
					\, {Ei}_{1}\! \left(-\frac{4 \sqrt{2}\, p}{c_{0} \sqrt{{ w_{\pm}(T)}}}\right)
					\nonumber\\
					&\,+ \Bigg(-64 pC_{2} \sqrt{2} \left(\frac{c_{0}^{2}}{16} { \left(w_{\pm}(T)-1\right)}+p^{2}-\frac{5 c_{0}^{2}}{16}\right) \sqrt{{ w_{\pm}(T)}}
					\nonumber\\
					&\,+c_{0} \Bigg(-16 p^{2} C_{2} \left({ w_{\pm}(T)}\right)+3T c_{0}^{2}\Bigg)\Bigg) c_{0} \exp\left({\frac{4 \sqrt{2}\, p}{c_{0} \sqrt{{ w_{\pm}(T)}}}}\right)\Bigg]
					\nonumber\\
					&\,+\,\frac{7p}{3 c_{0}^{4}}\Bigg[2 \left(96 C_{2} p^{2} c_{0}^{2}-128 p^{4} C_{2}-3 c_{0}^{4} T \right) \,{Ei}_{1}\! \left(\frac{4 \sqrt{2}\, p}{c_{0} \sqrt{{ w_{\pm}(T)}}}\right)
					\nonumber\\
					&\, + c_{0} \Bigg(32p C_{2} \sqrt{2} \left(\frac{c_{0}^{2}}{16} { \left(w_{\pm}(T)-1\right)}+p^{2}-\frac{11 c_{0}^{2}}{16}\right)\sqrt{{ w_{\pm}(T)}}
					\nonumber\\
					&\,+\frac{3c_{0}}{2} \left(2 C_{2} \left(-\frac{8 p^{2}}{3}+c_{0}^{2}\right) { \left(w_{\pm}(T)-1\right)}+\left(T +2 C_{2}\right) c_{0}^{2}-\frac{16 p^{2} C_{2}}{3}\right)\Bigg)
					\nonumber\\
					&\,\times\, \exp\left({-\frac{4 \sqrt{2}\, p}{c_{0} \sqrt{{ w_{\pm}(T)}}}}\right) \Bigg] \Bigg]\Bigg].
				\end{align}

			\item $b=4$:
			\vspace{-12pt}
			
				\scriptsize					
				\begin{align}\label{552ac}
					F(T) =&    -{ \Lambda_0} +(-T)^{-2}\,\Bigg[F_0+2\kappa\,p_0^2\,\Bigg[-\frac{p^2}{126 c_{0}^{8}}\Bigg[-78125 p^{4} C_{2}^{2} \left(p^{4}-\frac{84}{25} p^{2} c_{0}^{2}+\frac{168}{125} c_{0}^{4}\right)
					\nonumber\\
					&\,\times\, \mathrm{Ei}_{1}\! \left(-\frac{5 \sqrt{2}\, p}{c_{0} \sqrt{{ w_{\pm}(T)}}}\right)-63 \Bigg(-\frac{1}{21}\Bigg(25 \Bigg(\frac{3 c_{0}^{2}}{5} \Bigg(\left(T -\frac{32 C_{2}}{15}\right) c_{0}^{4} +20 C_{2} p^{2} c_{0}^{2}
					\nonumber\\
					&\,-\frac{125 p^{4} C_{2}}{18}\Bigg){ \left(w_{\pm}(T)-1\right)}-\frac{c_{0}^{6}}{25}\left(39 T +32 C_{2}\right) +p^{2} \left(T -128 C_{2}\right) c_{0}^{4}
					+\frac{2075 p^{4} C_{2} c_{0}^{2}}{6}
					\nonumber\\
					&\,-\frac{625 p^{6} C_{2}}{6}\Bigg)\,p C_{2} \sqrt{2}\, \sqrt{{ w_{\pm}(T)}}\Bigg)+c_{0} \Bigg(-\frac{25}{21} \Bigg(\left(T -\frac{76 C_{2}}{5}\right) c_{0}^{4}+65 C_{2} p^{2} c_{0}^{2}
					\nonumber\\
					&\,-\frac{125 p^{4} C_{2}}{6}\Bigg)  \,p^{2} C_{2} { \left(w_{\pm}(T)-1\right)}+T^{2} c_{0}^{6}+\frac{45 p^{2} C_{2} \left(T +\frac{76 C_{2}}{27}\right) c_{0}^{4}}{7}
					\nonumber\\
					&\,-\frac{125 p^{4} C_{2} \left(T +26 C_{2}\right) c_{0}^{2}}{42}
					+\frac{3125 p^{6} C_{2}^{2}}{126}\Bigg)\Bigg) \,c_{0} \exp\left({\frac{5 \sqrt{2}\, p}{c_{0} \sqrt{{ w_{\pm}(T)}}}}\right)\Bigg]
					\nonumber\\
					&\,  +\,\frac{p}{63 c_{0}^{8}}\Bigg[\left(-210000 p^{4} c_{0}^{4} C_{2}^{2}+350000 c_{0}^{2} p^{6} C_{2}^{2}-78125 C_{2}^{2} p^{8}+504 T^{2} c_{0}^{8}\right) 
					\nonumber\\
					&\,\times\,\mathrm{Ei}_{1}\! \left(\frac{5 \sqrt{2}\, p}{c_{0} \sqrt{{ w_{\pm}(T)}}}\right)+252 \Bigg(\frac{1}{4}\Bigg(-\frac{25 p C_{2}}{21}\sqrt{2}\, \Bigg(\frac{3}{5}\,c_{0}^{2} \Bigg(\left(T -\frac{116 C_{2}}{15}\right) c_{0}^{4}
					\nonumber\\
					&\,+\frac{250 C_{2} p^{2} c_{0}^{2}}{9}-\frac{125 p^{4} C_{2}}{18}\Bigg) { \left(w_{\pm}(T)-1\right)}+\frac{\left(-67 T -116 C_{2}\right) c_{0}^{6}}{25}
					\nonumber\\
					&\,+p^{2} \left(T -\frac{790 C_{2}}{3}\right) c_{0}^{4}
					+\frac{925 p^{4} C_{2} c_{0}^{2}}{2}-\frac{625 p^{6} C_{2}}{6}\Bigg) \sqrt{{ w_{\pm}(T)}}-c_{0} \Bigg(-\frac{25}{21} 
					\nonumber\\
					&\,\times\,\Bigg(-\frac{28 \left(T +2 C_{2}\right) c_{0}^{6}}{25}+p^{2} \left(T -\frac{188 C_{2}}{5}\right) c_{0}^{4}+\frac{265 p^{4} C_{2} c_{0}^{2}}{3}-\frac{125 p^{6} C_{2}}{6}\Bigg) C_{2}
					\nonumber\\
					&\,\times\, { \left(w_{\pm}(T)-1\right)}+\left(T^{2}+\frac{8 C_{2}^{2}}{3}\right) c_{0}^{6}+\frac{205 p^{2} C_{2} \left(T +\frac{188 C_{2}}{41}\right) c_{0}^{4}}{21}
					\nonumber\\
					&\,-\frac{125 p^{4} C_{2} \left(T +\frac{106 C_{2}}{3}\right) c_{0}^{2}}{42}+\frac{3125 p^{6} C_{2}^{2}}{126}\Bigg)\Bigg)
					\, \exp\left({-\frac{5 \sqrt{2}\, p}{c_{0} \sqrt{{ w_{\pm}(T)}}}}\right)\Bigg) c_{0}\Bigg] \Bigg]\Bigg].
				\end{align}
				

		\end{itemize}
		There are other values of $b$ yielding to analytical $F(T)$ solutions.

		\item $C_1=2C_2$:
		
		\vspace{-12pt}
			\begin{align}\label{552c}
				F(T) =&    -{ \Lambda_0} +\left[-T+2C_2\left({ w_{\pm}(T)}\right)\right]^{(2-b)}\,\Bigg[F_0-\kappa\,(b-2)\,p_0^2\,
				\int_{T}\,dT'\,\Bigg[p^2\,\exp\left(\frac{\sqrt{2}p(1+b)}{c_0\,\left[{ w_{\pm}(T')}\right]^{1/2}}\right)
				\nonumber\\
				&-2p\,(b+4)\,Ei\left(-\frac{\sqrt{2}p(1+b)}{c_0\,\left[{ w_{\pm}(T')}\right]^{1/2}}\right) \Bigg]\,
				\left[-T'+2C_2\left({ w_{\pm}(T')}\right)\right]^{(b-3)}\Bigg].
			\end{align}
		
		There are solutions for specific values of $\delta_1$ and $b$:
		\begin{itemize}
			\item $b=3$:
			\vspace{-12pt}
			

				\scriptsize																			
				\begin{align}\label{552ca}
					F(T) =&    -{ \Lambda_0} +\left[-T+2C_2\left({ w_{\pm}(T)}\right)\right]^{-1}\,\Bigg[F_0+\kappa\,p_0^2\,
					\nonumber\\
					&\,\times\,\Bigg[-\frac{p^2}{3 c_{0}^{4}}\Bigg[-512 p^{2} \left(p^{2}-\frac{3 c_{0}^{2}}{8}\right) C_{2} \mathrm{Ei}_{1}\! \left(-\frac{4 \sqrt{2}\, p}{c_{0} \sqrt{{ w_{\pm}(T)}}}\right)
					\nonumber\\
					&\,+ \exp\left({\frac{4 \sqrt{2}\, p}{c_{0} \sqrt{{ w_{\pm}(T)}}}} \right) \Bigg(-64 p \sqrt{2}\, C_{2} \left( \frac{c_{0}^{2}}{16} { \left(w_{\pm}(T)-1\right)}+p^{2}-\frac{5 c_{0}^{2}}{16} \right) 
					\nonumber\\
					&\,\times\,\sqrt{{ w_{\pm}(T)}}+c_{0} \left(-16 p^{2} C_{2} \left({ w_{\pm}(T)}\right)+3T c_{0}^{2}\right)\Bigg) c_{0}\Bigg]
					\nonumber\\
					&\,+\frac{7p}{3 c_{0}^{4}}\Bigg[2 \left(96 C_{2} c_{0}^{2} p^{2}-128 C_{2} p^{4}-3 T c_{0}^{4}\right) \,\mathrm{Ei}_{1}\! \left(\frac{4 \sqrt{2}\, p}{c_{0} \sqrt{{ w_{\pm}(T)}}}\right)
					\nonumber\\
					&\,+\Bigg(\Bigg(32 p \sqrt{2}C_{2} \, \left(\frac{c_{0}^{2}}{16} { \left(w_{\pm}(T)-1\right)}+p^{2}-\frac{11 c_{0}^{2}}{16}\right) \sqrt{{ w_{\pm}(T)}}
					\nonumber\\
					&\,+\frac{3c_{0}}{2} \left(2 C_{2} \left(-\frac{8 p^{2}}{3}+c_{0}^{2}\right) { \left(w_{\pm}(T)-1\right)}+\left(T +2 C_{2}\right) c_{0}^{2}-\frac{16 p^{2} C_{2}}{3}\right)\Bigg)
					\nonumber\\
					&\,\times\,\exp\left({-\frac{4 \sqrt{2}\, p}{c_{0} \sqrt{{ w_{\pm}(T)}}}}\right)\Bigg) c_{0}\Bigg] \Bigg]\Bigg].
				\end{align}

			\item $b=4$:

				\scriptsize				
				\begin{align}\label{552cb}
					F(T) =&    -{ \Lambda_0} +\left[-T+2C_2\left({ w_{\pm}(T)}\right)\right]^{-2}\,\Bigg[F_0+2\kappa\,p_0^2\,
					\nonumber\\
					&\,\times\,\Bigg[-\frac{p^2}{126 c_{0}^{8}}\Bigg[-78125 p^{6} C_{2}^{2} \left(p^{2}-\frac{28 c_{0}^{2}}{25}\right) \mathrm{Ei}_{1}\! \left(-\frac{5 \sqrt{2}\, p}{c_{0} \sqrt{{ w_{\pm}(T)}}}\right)
					\nonumber\\
					&\,-63 \,\exp\left({\frac{5 \sqrt{2}\, p}{c_{0} \sqrt{{ w_{\pm}(T)}}}}\right) c_{0} \Bigg(-\frac{25}{21} p \sqrt{2}\, C_{2}\Bigg(\frac{3 c_{0}^{2}}{5} \Bigg(\left(T -\frac{4 C_{2}}{15}\right) c_{0}^{4}
					\nonumber\\
					&\, +\frac{40 c_{0}^{2} p^{2} C_{2}}{9}-\frac{125 p^{4} C_{2}}{18}\Bigg) { \left(w_{\pm}(T)-1\right)}+\frac{\left(17 T -4 C_{2}\right) c_{0}^{6}}{25}+p^{2} \left(T +\frac{8 C_{2}}{3}\right) c_{0}^{4}  
					\nonumber\\
					&\,+\frac{225 p^{4} C_{2} c_{0}^{2}}{2}-\frac{625 p^{6} C_{2}}{6}\Bigg)\sqrt{{ w_{\pm}(T)}}+\Bigg(-\frac{25}{21} C_{2}\Bigg(\frac{56 \left(T -C_{2}\right) c_{0}^{6}}{25}
					\nonumber\\
					&\,+p^{2} \left(T +\frac{8 C_{2}}{5}\right) c_{0}^{4} +\frac{55 p^{4} C_{2} c_{0}^{2}}{3}-\frac{125 p^{6} C_{2}}{6}\Bigg) { \left(w_{\pm}(T)-1\right)}+\left(T^{2}-4 T C_{2}+\frac{8}{3} C_{2}^{2}\right) 
					\nonumber\\
					&\,\times\,c_0^6-\frac{5 p^{2} C_{2} \left(8 C_{2}+T \right) c_{0}^{4}}{21}-\frac{125 p^{4} \left(T +\frac{22 C_{2}}{3}\right) C_{2} c_{0}^{2}}{42} +\frac{3125 p^{6} C_{2}^{2}}{126}\Bigg) c_{0}\Bigg)\Bigg]
					\nonumber\\
					&\,+\frac{p}{378 c_{0}^{8}}\Bigg[\Bigg(8064 C_{2}^2 c_{0}^{8}{ \left(w_{\pm}(T)-1\right)^3} +\Bigg(8064 C_{2}^{2}-12096 T C_{2}+3024 T^{2}\Bigg) c_{0}^{8}
					\nonumber\\
					&\,+700000 C_{2}^{2} p^{6} c_{0}^{2}
					-468750 p^{8} C_{2}^{2}\Bigg) \mathrm{Ei}_{1}\! \left(\frac{5 \sqrt{2}\, p}{c_{0} \sqrt{{ w_{\pm}(T)}}}\right)+1512 c_{0}
					\nonumber\\
					&\,\times\, \Bigg(\Bigg(-\frac{25 p}{84} \Bigg(\frac{3}{5} \left(\left(T +\frac{44 C_{2}}{45}\right) c_{0}^{4}+\frac{190 c_{0}^{2} p^{2} C_{2}}{27}-\frac{125 p^{4} C_{2}}{18}\right) c_{0}^{2}{ \left(w_{\pm}(T)-1\right)}
					\nonumber\\
					&\, +\left(\frac{23 T}{75}+\frac{44 C_{2}}{75}\right) c_{0}^{6}+p^{2} \left(T +\frac{38 C_{2}}{9}\right) c_{0}^{4}+\frac{2725 p^{4} C_{2} c_{0}^{2}}{18}-\frac{625 p^{6} C_{2}}{6}\Bigg)
					\nonumber\\
					&\,\times\, \sqrt{2}\, C_{2} \sqrt{{ w_{\pm}(T)}}-\frac{c_{0}}{4} \Bigg(-\frac{25}{21} C_{2} \Bigg(\left(\frac{28 T}{15}-\frac{56 C_{2}}{75}\right) c_{0}^{6}
					\nonumber\\
					&\,+p^{2} \left(T +\frac{52 C_{2}}{15}\right) c_{0}^{4}+\frac{235 p^{4} C_{2} c_{0}^{2}}{9}-\frac{125 p^{6} C_{2}}{6}\Bigg) { \left(w_{\pm}(T)-1\right)}
					\nonumber\\
					&\,+\left(T^{2}-\frac{8}{3} T C_{2}+\frac{8}{9} C_{2}^{2}\right) c_{0}^{6}+\frac{55 p^{2} \left(T -\frac{52 C_{2}}{11}\right) C_{2} c_{0}^{4}}{63}-\frac{125 p^{4} \left(T +\frac{94 C_{2}}{9}\right) C_{2} c_{0}^{2}}{42}
					\nonumber\\
					&\,+\frac{3125 p^{6} C_{2}^{2}}{126}\Bigg)\Bigg) \,\exp\left({-\frac{5 \sqrt{2}\, p}{c_{0} \sqrt{{ w_{\pm}(T)}}}}\right)\Bigg)\Bigg]\Bigg]\Bigg].
				\end{align}

		\end{itemize}
	\end{enumerate} 
	There are other values of $b$ yielding to analytical $F(T)$ solutions.

	Equation~\eqref{554} $C_1\neq 0$ teleparallel $F(T)$ solutions ($p\gg 1$ case):
	\begin{enumerate}
		\item $C_1=0$:
		\small
		\begin{align}\label{554a}
			F(T) =&    -{ \Lambda_0} +(-T)^{(2-b)}\,\Bigg[F_0-\kappa\,p^2\,p_0^2\,\int_{T}\,dT'\,\exp\left(\frac{\sqrt{2}p(1+b)}{c_0\,\left[{ w_{\pm}(T')}\right]^{1/2}}\right)\,(-T')^{(b-3)}\Bigg].
		\end{align}
		\normalsize
		There are solutions for specific values of $b$:
		\begin{itemize}
			\item $b=1$:
			\small
			\begin{align}\label{554aa}
				F(T) \approx & -{ \Lambda_0} -T\,\Bigg[F_0-\,\frac{\kappa\,p_0^2\,p^2}{4\, c_{0} C_{2} }\Bigg[ p  \exp\left({\frac{2 p}{c_{0}}}\right) \mathrm{Ei}_{1}\! \left(-\frac{2 p \left(\sqrt{2}-\sqrt{{ w_{\pm}(T)}}\right)}{c_{0} \sqrt{{ w_{\pm}(T)}}}\right)
				\nonumber\\
				&\,+\frac{\left({ w_{\pm}(T)}\right)}{\left({ 2-w_{\pm}(T)}\right)} \, \exp\left({\frac{2 \sqrt{2}\, p}{c_{0} \sqrt{{ w_{\pm}(T)}}}}\right) c_{0}\Bigg]\Bigg].
			\end{align}
			\normalsize

			\item $b=3$:
			\small
			\begin{align}\label{554ab}
				F(T) \approx &     -{ \Lambda_0} 
				-\Bigg[\frac{F_0}{T}+\frac{64\kappa\,p_0^2p^5C_{2}}{3 c_{0}^{3}}\frac{\sqrt{2}}{T} \sqrt{{ w_{\pm}(T)}} \exp\left({\frac{4 \sqrt{2}\, p}{c_{0} \sqrt{{ w_{\pm}(T)}}}}\right)\Bigg].
			\end{align}
			\normalsize

			\item $b=4$:
			\small
			\begin{align}\label{554ac}
				F(T) \approx &    -{ \Lambda_0} 
				+\Bigg[\frac{F_0}{T^2}+\frac{15625\kappa\,p_0^2\,p^9C_{2}^2}{63\sqrt{2} c_{0}^{7}\,T^2} \, \sqrt{{ w_{\pm}(T)}}\, \exp\left({\frac{5 \sqrt{2}\, p}{c_{0} \sqrt{{ w_{\pm}(T)}}}}\right)\Bigg].
			\end{align}
			\normalsize

		\end{itemize}
		
		\item $C_1=2C_2$:

			\begin{align}\label{554c}
				F(T) =&    -{ \Lambda_0} +\left[-T+C_1\left({ w_{\pm}(T)}\right)\right]^{(2-b)}\left[{ y(T)}\right]^{\frac{(b-2)C_1}{2\left(C_1-2C_2\right)}}
				\,\left[\frac{T+\left(C_1-2C_2\right)\left({ w_{\pm}(T)}\right)}{T+\left(C_1-2C_2\right)\left({ 2-w_{\pm}(T)}\right)}\right]^{\frac{\delta_1\,(b-2)C_1}{2\left(C_1-2C_2\right)}}
				\nonumber\\
				&\,\times\,\Bigg[F_0-\kappa\,p^2\,p_0^2\,\int_{T}\,dT'\,\exp\left(\frac{\sqrt{2}p(1+b)}{c_0\,\left[{ w_{\pm}(T')}\right]^{1/2}}\right)\,\left[-T'+C_1\left({ w_{\pm}(T')}\right)\right]^{(b-3)}
				\nonumber\\
				&\,\times\,\left[{ y(T')}\right]^{-\frac{(b-2)C_1}{2\left(C_1-2C_2\right)}}\,
				\left[\frac{T'+\left(C_1-2C_2\right)\left({ w_{\pm}(T')}\right)}{T'+\left(C_1-2C_2\right)\left({ 2-w_{\pm}(T')}\right)}\right]^{-\frac{\delta_1\,(b-2)C_1}{2\left(C_1-2C_2\right)}}\Bigg].
			\end{align}
		
		There are solutions for specific values of $\delta_1$ and $b$:
		\begin{itemize}
			\item $b=3$:
			\small
			\begin{align}\label{554ca}
				F(T) \approx &    -{ \Lambda_0} +\left[-T+C_1\left({ w_{\pm}(T)}\right)\right]^{-1}
				\nonumber\\
				&\,\times\,\Bigg[F_0+\frac{64\kappa\,p^5\,p_0^2}{3 c_{0}^{3}}\sqrt{2}\, C_{2}\sqrt{{ w_{\pm}(T)}} \exp\left({\frac{4 \sqrt{2}\, p}{c_{0} \sqrt{{ w_{\pm}(T)}}}}\right)\Bigg].
			\end{align}
			\normalsize

			\item $b=4$:
			\small
			\begin{align}\label{554cb}
				F(T) \approx &    -{ \Lambda_0} +\left[-T+C_1\left({ w_{\pm}(T)}\right)\right]^{-2}\,
				\nonumber\\
				&\,\times\,\Bigg[F_0+\frac{15625\kappa\,p^9\,p_0^2}{126\sqrt{2} c_{0}^{7}}\,C_{2}^2\,\sqrt{{ w_{\pm}(T)}}\exp\left({\frac{5 \sqrt{2}\, p}{c_{0} \sqrt{{ w_{\pm}(T)}}}}\right)\Bigg].
			\end{align}
			\normalsize

		\end{itemize}
		
	\end{enumerate} 
	
	\subsection{Exponential Ansatz~Solutions}\label{appenb3}
	
	Equation~\eqref{571} $T_1\neq T_0$ and $b\geq \frac{3c}{2}$ teleparallel $F(T)$ solutions:
	\begin{enumerate}
		\item $b=\frac{3c}{2}$:
		\small
		\begin{align}\label{571b}
			F(T) =&  -{ \Lambda_0} +(T-T_1)^{-1}\Bigg[F_0+\kappa\,p\,p_0^2\,\frac{c(7c+2p)}{2\left(c-p\right)}\,\left(-\frac{2}{c_0^2}\right)^{p/c}\,\left(T-T_0\right)^{1-p/c}\Bigg].
		\end{align}
		\normalsize
		
		\item $b=2c$:
		\small
		\begin{align}\label{571c}
			F(T) =&  -{ \Lambda_0} +(T-T_1)^{-2}\Bigg[F_0+2\kappa\,p\,p_0^2\,c(4c+p)\,\left(-\frac{2}{c_0^2}\right)^{p/c}
			\nonumber\\
			&\,\times\,\frac{\left(T-T_{0}\right)^{\frac{-p+c}{c}} \left(\left(T+T_{0}-2 T_{1}\right) c-p \left(T-T_{1}\right)\right)}{2 c^{2}-3 c p+p^{2}}\Bigg].
		\end{align}
		\normalsize
		
		\item $b=3c$:
		\small
		\begin{align}\label{571d}
			F(T) =&  -{ \Lambda_0} +(T-T_1)^{-4}\Bigg[F_0+\frac{24\kappa\,p\,p_0^2\,c(3c+2c+p)}{\left(24 c^{4}-50 c^{3} p+35 c^{2} p^{2}-10 c \,p^{3}+p^{4}\right)}\,\left(-\frac{2}{c_0^2}\right)^{p/c}
			\nonumber\\
			&\,\times\,\left(T-T_{0}\right)^{\frac{-p+c}{c}} \Bigg(\left(2 T_{1}^{2}+\left(-2 T-2 T_{0}\right) T_{1}+T^{2}+T_{0}^{2}\right) \left(T+T_{0}-2 T_{1}\right) c^{3}
			\nonumber\\
			&-\frac{11}{6} \left(\frac{26 T_{1}^{2}}{11}+\left(-\frac{31 T}{11}-\frac{21 T_{0}}{11}\right) T_{1}+T^{2}+\frac{9 T T_{0}}{11}+\frac{6 T_{0}^{2}}{11}\right) p \left(T-T_{1}\right) c^{2}
			\nonumber\\
			&\,+\left(T+\frac{T_{0}}{2}-\frac{3 T_{1}}{2}\right) p^{2} \left(T-T_{1}\right)^{2} c-\frac{p^{3} \left(T-T_{1}\right)^{3}}{6}\Bigg) \Bigg].
		\end{align}
		\normalsize
	\end{enumerate}
	
	Equation~\eqref{573} $T_1\neq T_0$ and $b\geq \frac{3c}{2}$ teleparallel $F(T)$ solutions ($p\gg 1$ case):
	\begin{enumerate}
		\item $b=\frac{3c}{2}$:
		\small
		\begin{align}\label{573b}
			F(T) \approx &  -{ \Lambda_0} +F_0\,(T-T_1)^{-1}-\kappa\,p\,p_0^2\,c\,\left(-\frac{2}{c_0^2}\right)^{p/c}\,\left(T-T_0\right)^{-p/c}(T-T_1)^{-1}.
		\end{align}
		\normalsize
		
		\item $b=2c$:
		\small
		\begin{align}\label{573c}
			F(T) \approx &  -{ \Lambda_0} +F_0\,(T-T_1)^{-2}-2\kappa\,p\,p_0^2\,c\,\left(-\frac{2}{c_0^2}\right)^{p/c}\,\left(T-T_{0}\right)^{-\frac{p}{c}}\,(T-T_1)^{-1}.
		\end{align}
		\normalsize
		
		\item $b=3c$:
		\small
		\begin{align}\label{573d}
			F(T) \approx &  -{ \Lambda_0} +F_0\,(T-T_1)^{-4}-4\kappa\,p\,p_0^2\,c\,\left(-\frac{2}{c_0^2}\right)^{p/c}
			\left(T-T_{0}\right)^{-\frac{p}{c}} \,\left(T-T_{1}\right)^{-1} .
		\end{align}
		\normalsize
	\end{enumerate}

}
\end{document}